\documentclass[aps,twocolumn,prd,superscriptaddress,nofootinbib]{revtex4-2}

\usepackage[utf8]{inputenc}

\usepackage{array}[=2016-10-06]

\usepackage[utf8]{inputenc}

\usepackage{amsmath}
\usepackage{bbm}
\usepackage{mathtools}
\usepackage{amsfonts}
\usepackage{mathrsfs}
\usepackage{bm}
\usepackage{slashed}
\usepackage{tensor}
\usepackage{bbold}
\usepackage{MnSymbol}
\usepackage{textcomp}
\usepackage[export]{adjustbox}

\usepackage{float}
\usepackage{graphicx}
\usepackage{color}
\usepackage[abs]{overpic}

\usepackage{placeins}

\usepackage{makecell}
\usepackage{subcaption}

\usepackage{xspace}
\usepackage{siunitx}
\usepackage{xfrac}
\usepackage{hyperref}
\usepackage[nameinlink]{cleveref}
\usepackage{appendix}

\Crefname{appsec}{Appendix}{Appendices}

\usepackage{tikz}
\usetikzlibrary{shapes, arrows}
\tikzstyle{roundbox} = [rectangle, draw, text centered, rounded corners, 
minimum height=2em, minimum width=2em, draw=black, fill=blue!10]
\tikzstyle{process} = [rectangle, draw, minimum height=1em, 
minimum width=3em, text centered, draw=black, fill=green!10]
\tikzstyle{integration} = [ellipse, draw, text centered, minimum height=1em, 
minimum width=3em, draw=black, fill=red!10]

\newcommand{\LEGO}{LEGO\textsuperscript{\textregistered}}

\usepackage{xifthen}
\usepackage{xcolor}
\hypersetup{
	colorlinks,
	linkcolor={red!75!black},
	citecolor={blue!75!black},
	urlcolor={blue!75!black}
}

\usepackage{booktabs}
\usepackage{multirow}

\newcolumntype{C}{>{$}c<{$}}
\AtBeginDocument{
	\heavyrulewidth=.08em
	\lightrulewidth=.05em
	\cmidrulewidth=.03em
	\belowrulesep=.65ex
	\belowbottomsep=0pt
	\aboverulesep=.4ex
	\abovetopsep=0pt
	\cmidrulesep=\doublerulesep
	\cmidrulekern=.5em
	\defaultaddspace=.5em
}

\makeatletter
\def\l@subsubsection#1#2{}
\makeatother

\graphicspath{{./figures/}{./figures/equations/}{./figures/results/}}

\newcommand*{\beq}{\begin{equation}}
\newcommand*{\eeq}{\end{equation}}

\newcommand*{\Nc}{N_{\textup{c}}}

\newcommand*{\upB}{\textup{B}}

\newcommand*{\muB}{\mu_{\upB}}

\renewcommand*{\vec}[1]{\bm{#1}}

\DeclareMathOperator{\Tr}{Tr}

\DeclarePairedDelimiterX{\expval}[1]{\langle}{\rangle}{#1}

\usepackage{tikz}
\usetikzlibrary{shapes, arrows}
\tikzstyle{roundbox} = [rectangle, draw, text centered, rounded corners, 
minimum height=2em, minimum width=2em, draw=black, fill=blue!10]
\tikzstyle{process} = [rectangle, draw, minimum height=1em, 
minimum width=3em, text centered, draw=black, fill=green!10]
\tikzstyle{integration} = [ellipse, draw, text centered, minimum height=1em, 
minimum width=3em, draw=black, fill=red!10]

\graphicspath{{figures/}}

\newcommand{\DCSB}{{D$\chi$SB}\xspace}

\allowdisplaybreaks 

\begin{document}

\title{Phase structure and observables at high densities from first principles QCD}

\newcommand{\orcid}[1]{\href{https://orcid.org/#1}{\includegraphics[height=1.9ex,width=1.9ex]{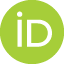}}}

\author{Christian S. Fischer
\orcid{0000-0001-8780-7031}\,}
\email[]{christian.fischer@theo.physik.uni-giessen.de}

\affiliation{Institut f{\"u}r Theoretische Physik, Justus-Liebig-Universit{\"a}t Gießen, 35392 Gießen, Germany} 
\affiliation{Helmholtz Forschungsakademie Hessen für FAIR (HFHF), GSI Helmholtzzentrum für Schwerionenforschung, Campus Gießen, Gießen, 35392  Germany}

\author{Jan~M.~Pawlowski \orcid{0000-0003-0003-7180}\,}
\email[]{J.Pawlowski@thphys.uni-heidelberg.de}
\affiliation{Institut f{\"u}r Theoretische Physik,
	Universit{\"a}t Heidelberg, Philosophenweg 16,
	69120 Heidelberg, Germany
}

\affiliation{ExtreMe Matter Institute EMMI,
	GSI, Planckstr. 1,
	64291 Darmstadt, Germany
}

\begin{abstract}
	We provide a short review of the progress made in the past decade with functional QCD in the description of the phase structure of QCD. 
   We summarise the most important technical aspects of the framework, discuss strategies for truncations and address the problem of 
   systematic error estimates. We detail efforts to gauge the approach systematically with lattice QCD at zero chemical potential,
   also including the physics of the Columbia plot at non-physical quark masses. Our main focus is, however, the high density
   regime of QCD. We address the predictive power of the functional approach for the appearance of new phases beyond the chiral crossover regime for chemical potentials  $\mu_B/T\geq 4.5$.  
   The onset of this regime may be signalled by a critical end point of the crossover line but may also involve a moat regime or the emergence of an instability that indicates an inhomogeneous phase. Respective results include estimates for the location of the onset of new phases, and predictions for their experimental signatures.
\end{abstract}

\maketitle

\section{Introduction}
\label{sec:Introduction}

The physics of QCD at high densities is largely unexplored and harbours some of the most exciting phenomena in strongly correlated QCD. In the past decade, functional QCD at finite temperature and density has developed from a qualitative level, well-suited for exploratory studies, to a first principles quantitative QCD approach. Early milestones are \cite{Braun:2008pi, Braun:2009gm} (functional renormalisation group (fRG) studies, one flavour phase structure, and two-flavour phase structure at imaginary baryon chemical potential). The first work, studying the phase structure of 2+1 flavour QCD at real chemical potential, that passes qualitative finite density benchmarks, is given by the Dyson-Schwinger study (DSE) \cite{Fischer:2014ata}. In \cite{Fu:2019hdw} (fRG), all available lattice benchmarks relevant for the chiral phase structure were met quantitatively for the first time and an estimate for the location of the critical end point (or rather the location of the onset of new physics) was given. 
The result was corroborated by \cite{Gao:2020qsj, Gao:2020fbl} (DSE) and independently by \cite{Gunkel:2021oya} (DSE). All these studies were done with $\mu_S=0$, that is all quark chemical potentials are a third of the baryon chemical potential. The different diagrammatic nature of the approximations provide further non-trivial reliability for this result and the current functional estimate for a chiral critical end point from the combination \cite{Fu:2019hdw, Gao:2020fbl, Gunkel:2021oya} is given by an interval $T(\muB)$ of the chiral transition line with a small width, 
\begin{align}
	(T, \mu_B)_\textrm{CEP} \in \bigl(115-105\,,\, 600-650\bigr)\textrm{MeV}\,, 
\label{eq:FunEstimateCEP}
\end{align} 
which corresponds to $\mu_B/T \approx 5.5 - 6$ for $\mu_S=0$.  
Note that this result does not single out an area but only values in \labelcref{eq:FunEstimateCEP} on a rather well-determined chiral crossover curve $T_\chi(\mu_B)$, see \labelcref{eq:Tchicurvature}. 

The works \cite{Fu:2019hdw, Gao:2020fbl, Gunkel:2021oya}, and the more recent fRG work \cite{Pawlowski:2025jpg}, laid the foundation for further works on phenomenological consequences and observables. In the present mini-review we provide a brief overview on functional QCD, and in particular the systematics behind the computations as well as the systematic error estimates. We also discuss briefly the phenomenological results already achieved in functional QCD. 

At vanishing and finite temperature, the dynamics of QCD is dominantly driven by the phenomena of dynamical chiral symmetry breaking (\DCSB) and confinement that are associated to a momentum scale of the order of 1\,GeV. 
At finite temperature and vanishing density, \DCSB is restored in a relatively sharp crossover at $T_\chi\approx 155$\,MeV. 
Many effects of \DCSB are by now well-understood also on the quantitative level, with non-trivial results and predictions, including a rather comprehensive understanding of the hadron spectrum, see e.g.~\cite{Cloet:2013jya, Eichmann:2016yit, Eichmann:2025wgs, Huber:2025cbd} for results with functional methods. 
While the physics of the hadron spectrum also involves confinement, this phenomenon is more enigmatic as proven by novel unexpected properties discovered recently, \cite{Glozman:2022zpy}. Moreover, the transition from the low energy hadronic phase to the high energy perturbative quark-gluon phase stretches over a large temperature regime between $~T_\chi$ and $~(2-3) T_\chi$, and this regime may include sub-regimes with different dynamics. 

At finite density, the chiral crossover gets increasingly sharper and \textit{assuming} the absence of the emergence of novel dynamics driven by density fluctuations this suggests the existence of a critical end point (CEP). The sharpening of the chiral crossover can be monitored by dedicated observables, as e.g.~fluctuations of conserved charges. Specifically baryon number fluctuations are well suited. The experimental analogue are proton number fluctuations measured in heavy ion (HIC) experiments such as STAR at RHIC (Brookhaven) and future experiments whose data taking starts shortly: most prominently CBM at the FAIR facility (GSI Darmstadt, 2028), MPD at NICA (JINR Dubna, 2026), CEE+ at HIAF (Huizhou, 2028). The sharpening of the crossover leaves its imprint in non-monotonicities of the $\mu_B$-dependence of the fluctuations of conserved charges and in the scenario with a CEP it indicates the approach towards the latter. 
In the literature, these non-monotonicities are sometimes treated as a 'smoking gun' signal for entering the critical regime around the CEP, see e.g.~the recent review \cite{Arslandok:2023utm}. 
This notion might be questioned, since it is readily shown that the latter are also present outside the critical regime and even in the absence of a CEP, see \cite{Fu:2021oaw, Fu:2023lcm} and the reviews \cite{Dupuis:2020fhh, Rennecke:2025bcw}. Thus, we are faced with an intricate situation that requires a comprehensive and unbiased analysis best performed by direct QCD computations in this regime. 

In contrast to extrapolations of any kind, direct QCD computations are also needed to pick up emergent phenomena driven by density fluctuations. 
A well-known example of this phenomenon is the QCD 'condensed matter' regime with colour superconductivity/fluidity and 
competing order effects at low temperature and densities roughly beyond nuclear saturation densities.
In addition, first indications of a so-called 'moat' regime \cite{Pisarski:2021qof} 
have been found with functional QCD \cite{Fu:2019hdw, Fu:2024rto} and have been solidified recently 
in a self-consistent functional QCD computation \cite{Pawlowski:2025jpg}. In the latter work, indications for a true  instability  at even larger density have been found, which suggests the presence of an inhomogeneous phase in this regime. Such inhomogeneous phases have already been studied comprehensively in low-energy effective theories, for a review see \cite{Buballa:2014tba}. Moreover, the path towards applications 
within functional QCD has been prepared by now in \cite{Motta:2023pks,Motta:2024rvk}.
A first analysis of respective experimental signatures of the moat and inhomogeneous phases with Hanbury Brown-Twiss interferometry has been performed in \cite{Rennecke:2023xhc, Fukushima:2023tpv}. Finally, the physics of a potential CEP may be modified by the mixing of the critical scalar massless mode with density fluctuations inducing a sizeable glue component, see \cite{Haensch:2023sig}. 

In the present mini-review we discuss the progress of the past decade in first principles QCD computations at finite temperature and density with the functional QCD approach. 
This approach is based on functional diagrammatic relations of QCD, most prominently functional renormalisation group (fRG) flows and Dyson-Schwinger equations (DSE), potentially augmented with $n$-particle irreducible (nPI) hierarchies for the QCD effective action. 
See \cite{Fischer:2018sdj, Dupuis:2020fhh, Fu:2022gou, Rennecke:2025bcw} for phase structure reviews and \cite{Huber:2025cbd} for a recent introductory review to the functional approach. 
These functional relations  can be cast into the form of an infinite tower of coupled integral (DSE) or integral-differential equations (fRG) for the (one-particle) irreducible parts of $n$-point correlation functions of quarks, gluons and possibly also quark-gluon composites. 
The composites accommodate emergent dynamical low energy degrees of freedom such as pseudoscalar and scalar modes, 
the vector or density channel but also diquarks and further off-shell degrees of freedom that do not have an overlap with asymptotic states in QCD \cite{Gies:2001nw, Gies:2002hq, Pawlowski:2005xe, Fischer:2007ze, Floerchinger:2009uf, Eichmann:2015kfa, Fu:2019hdw, Fukushima:2021ctq}. 

In summary, in the past decade the combination of the different functional methods (fRG, DSE, nPI,...) with systematic and well-controlled expansion schemes has matured into a quantitative functional approach to 1st principles QCD at finite temperatures and density. By now, the finite density regime around the chiral crossover 
line that allows for quantitative predictions with small systematic errors, has been pushed towards $\mu_B/T\lesssim 4$ and recently even to $\mu_B/T\lesssim 4.5$ \cite{Pawlowski:2025jpg}. Beyond this regime, for $4.5 \lesssim \mu_B/T \lesssim 7$, the systematic error for the \textit{chiral dynamics} remains below 10 \%, if the technical uncertainties (truncation artefacts) of the computational expansion schemes in \cite{Fu:2019hdw, Gao:2020fbl, Gunkel:2021oya, Pawlowski:2025jpg} are assessed in combination. This leads to the estimate \labelcref{eq:FunEstimateCEP} for the location of the critical end point and will be discussed
in much more detail below. 

Note, however, that this estimate and the associated systematic error budget 
is only valid under the assumption that potential new phases do not affect the nature and location of the CEP. More specifically this concerns emergent phenomena, driven by density fluctuations, such as a moat regime and/or inhomogeneous phases. This is currently under intense scrutiny, see \cite{Pawlowski:2025jpg} for first results. These indications for a more intricate and exciting phase structure of QCD  suggest to extending the experimental and theoretical search for the CEP to one for the Onset of New Phases (ONP).

We close the introduction with a bird's-eye view  on the review. In \Cref{sec:FunApproach} we introduce functional methods, fRG and DSE, and their systematics as a combined functional approach to first principle QCD, including the highly relevant discussion of the emergence of low-energy effective theories (LEFTs) from functional QCD. This embedding of LEFTs in functional QCD allows for the construction and use of QCD-assisted LEFTs for phenomenological applications in and out of equilibrium that are not yet accessible within 1st principles QCD due to their large numerical costs and the still exploratory state of real-time functional QCD at finite temperature and densities.  In \Cref{sec:FunPhaseStructure} we provide an overview of the functional results for the phase structure of QCD including the current estimates for the location of the onset of new physics/CEP. This includes a brief discussion of the size of the critical region and the importance of soft modes for the access to the potential critical end point. In \Cref{sec:ConservedFlucs+freezeout} we discuss functional results for fluctuations of conserved charges. We shall argue that in combination with experimental measurements of these observables and further theoretical results (lattice QCD and LEFTs) this will allow to pin down the location of the critical end point or rather the onset regime of new phases. In particular, we argue that (only) such a combination allows for an experimental search of smoking gun signals for the CEP or the ONP. In \Cref{sec:TheoryCover} we discuss the embedding of the phase structure of QCD in a higher dimensional parameter space including variations of the quark masses at zero, real and imaginary chemical potential (Columbia plot),  
external magnetic fields and isospin chemical potential, to name a few of these parameters. This provides both, further benchmark tests for functional QCD as well as further constraints and indirect access to the physics at high densities. We conclude this review with a brief summary and outlook in \Cref{sec:ConclusionOutlook}.

\section{Functional approach to QCD}
\label{sec:FunApproach}

This Section is dedicated to a brief introduction to the functional QCD approach. We refrain from repeating derivations (the reader is instead referred to introductory and advanced reviews) but discuss basic properties, expansion schemes and, most importantly, the systematic error analysis. The technically less interested reader, or for a first reading, may very well skip this Section and start with \Cref{sec:FunPhaseStructure}. In \Cref{sec:FunDerivation} we sketch the derivation of functional DSE and fRG equations as well as the derived hierarchies of relations for correlation functions. In \Cref{sec:Expansion+Error} we discuss expansion schemes and combined systematic error analyses.

\subsection{Functional relations in QCD}
\label{sec:FunDerivation} 

Functional relations for QCD can be readily derived from its path or functional integral representation. Both, DSEs and fRG flows can be derived from the simple property  
\begin{align} 
	\int d\hat\Phi \frac{\delta}{\delta \hat\Phi} \left[\Psi[\hat\Phi]\, e^{-S_\textrm{QCD}[\hat\Phi_f] + \int_x J_{\hat\Phi_i} \hat \Phi_i}\right]=0\,. 
\label{eq:TotalDerivativeZ}
\end{align} 
\labelcref{eq:TotalDerivativeZ} entails that integrals of total derivatives vanish. For general functionals or kernels $\Psi[\Phi]$ it comprises general reparametrisations and rescalings of QCD \cite{Wegner:1974sla, Baldazzi:2021ydj, Ihssen:2024ihp, Ihssen:2025cff}. Here, $\hat\Phi$ is the superfield that contains all fields, fundamental and emergent ones, considered in the functional method at hand. For the mean field $\Phi$ we get 
\begin{align} 
	\Phi=\langle \hat\Phi\rangle = (\Phi_f,\phi)\,,\qquad \Phi_f=(A_\mu, c,\bar c, q, \bar q)\,,
\label{eq:Superfield} 
\end{align} 
where $\Phi_f$ is the superfield of the fundamental degrees of freedom, quark, gluons and ghost fields. The composite field $\phi$ typically includes the lowest lying scalar and pseudoscalar resonances, 
\begin{align} 
	\phi=(\sigma, \vec \pi,...)\,,
\label{eq:Composites} 
\end{align} 
 and the dots indicate potential further emergent degrees of freedom. The gluons $A$ and quarks $q$ live in the adjoint and fundamental representation of the colour gauge group SU(3) of QCD. The presence of the ghost and anti-ghost fields $c$ and $\bar{c}$ indicate that functional QCD is typically build upon a gauge-fixed action $S_\textrm{QCD}$ with 
\begin{align}
	S_\textrm{QCD}= S_A[A] +S_\textrm{gf}[A] +S_\textrm{gh}[A,c, \bar c] +S_q[A,q,\bar q] \,,
	\label{eq:SQCD}
\end{align}
and standard gauge fixing parts $S_\textrm{gf}[A]+S_\textrm{gh}[A,c, \bar c]$ (not displayed for brevity).  
In explicit computations, the default choice is Landau gauge, i.e. $\xi\to 0$.\footnote{On the technical side, this choice comes with a significant reduction of the size of the diagrammatic parts of functional relations and the number of dressing functions that have to be computed. The latter originates in the fact that the purely transverse system is closed in the Landau gauge, while for $\xi\neq 0$ the longitudinal system feeds into the transverse one. For a detailed discussion see \cite{Fischer:2008uz, Cyrol:2016tym, Dupuis:2020fhh, Pawlowski:2022oyq}, for functional results (DSE) with $\xi\neq 0$ see \cite{Aguilar:2015nqa, Huber:2015ria, Napetschnig:2021ria}. Therefore, this choice is more than conceptual: it reduces approximation artefacts in the system and therefore also minimises the systematic error.} 
The glue dynamics is carried by the classical Yang-Mills action $S_A$, while the Dirac action $S_q$ contains the quark dynamics. They are given by
\begin{align}\nonumber 
	S_A[A] =   &\, \frac14 \int_x F_{\mu\nu}^a F_{\mu\nu}^a\,,         \\[2ex]
	S_q [A,q,\bar q]  =&\,  \int_x \,\bar q \left( \gamma_\mu D_\mu+m_q-\gamma_0 \mu_q \right)\,q \,. 
	\label{eq:SA+Sq}
\end{align}
The quark field $q$ comprises all quarks used in the computation, in the most general case all three families. 
In the following we shall discuss the cases $N_f=2, N_f=2+1, N_f=2+1+1$, while dropping the 3rd family $b,t$.\footnote{We note in passing, that while the heavier quarks are subleading for the off-shell dynamics, the inclusion of $c,b$ is important for phenomenological purposes, e.g.~the computation of heavy quark transport in heavy-ion collisions. As the $c,b$ off-shell dynamics only gives rise to subleading effects in the glue, $u,d,s$ and mesonic off-shell dynamics, their inclusion can be done in a subsequent (modular) step.} Typically, a flavour diagonal mass matrix is used, 
\begin{align} 
	m_q=~\textrm{diag}(m_u,m_d,m_s,m_c)\,, \qquad m_u=m_d=m_l\,,
	\label{eq:mqall}
\end{align} 
in particular for applications to the QCD phase structure.
The quark chemical potential in \labelcref{eq:SA+Sq} is typically restricted to at most three flavours 
and in this case given by 
\begin{align}
	\mu_q=\textrm{diag}(\mu_l,\mu_l,\mu_s)\,, \qquad  
	\label{eq:muq}
\end{align}
These are related to the chemical potentials of conserved charges, baryon number $\mu_B$, electrical charge $\mu_Q$ 
and Strangeness $\mu_S$ via
\begin{align}\nonumber 
\mu_u &= \frac{1}{3}	\mu_B + \frac{2}{3} \mu_Q\,, \\[1ex]\nonumber 
\mu_d &= \frac{1}{3}	\mu_B - \frac{1}{3} \mu_Q \,,\\[1ex]
\mu_s &= \frac{1}{3}	\mu_B - \frac{1}{3} \mu_Q - \mu_S\,.
\end{align}
In heavy ion collision experiments, all three conserved charges are non-zero. The overall net-strangeness
is zero, resulting in the strangeness neutrality condition 
\begin{align} 
	n_s\bigl(\mu_B,\mu_S(\mu_B) \bigr) = \frac{\partial \Omega}{\partial \mu_s} = 0\,,
	\label{eq:ns=0}
\end{align}
with the Grand Potential $\Omega$ of QCD. Moreover, the experiments take place at a non-zero isospin chemical potential due to proton-neutron
asymmetries and isospin breaking. QCD studies of the chiral transition at finite baryon chemical potential
generically work at zero isospin chemical potential, $\mu_I = \frac{\mu_u-\mu_d}{2} = \frac{\mu_Q}{2}=0$, 
which leads to  
\begin{align}
	(\mu_u, \mu_d, \mu_s) = \left(\frac{\mu_B}{3}, \frac{\mu_B}{3}, \frac{\mu_B}{3} - \mu_S\right)\,. 
	\label{eq:muB-muS}
\end{align}
The effect of finite $\mu_I$ on the location of the CEP remains to be determined in future studies. Furthermore, 
in most functional QCD computations to date, $\mu_S=0$ has been used.  More recently, also strangeness neutrality, 
\labelcref{eq:ns=0}, has been considered, the latter being closer to the experimental situation. 
On the lattice both situations have been studied, with the majority of the more recent and more precise 
computations being done with strangeness neutrality.

\begin{figure*}[t]
	\centering
	\begin{minipage}[t]{.55\linewidth}
		\begin{subfigure}{\linewidth}
			\centering
			\includegraphics[width=\linewidth]{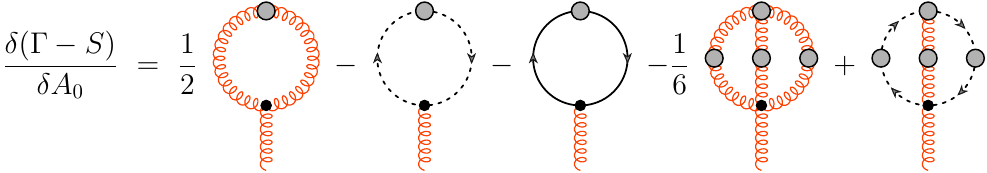} 
			\subcaption{Functional background field DSE.  \hspace*{\fill}}
			\label{fig:DSE-backA0}
		\end{subfigure}%
	\end{minipage}
	\hspace{0.04\linewidth}%
	\begin{minipage}[t]{.38\linewidth}
		\begin{subfigure}{\linewidth}
			\centering
			\includegraphics[width=\linewidth]{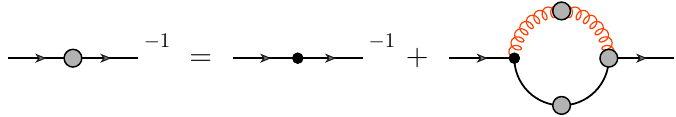} \vspace{-.15cm}
			\subcaption{Quark gap equation.\hspace*{\fill}}
			\label{fig:QuarkGapDSE} 
		\end{subfigure}
	\end{minipage}
	\caption{DSE: Full propagators and vertices are indicated by grey blobs, the classical vertices are indicated by small black blobs. Gluons are represented by red spiral lines, ghosts by back dotted ones, and quark by straight black ones. In \Cref{fig:DSE-backA0} we depict the background field DSE. In contradistinction to the fluctuation field DSE, it also hosts a two-loop ghost-gluon term with the four-point vertex $S^{(4)}_{ac\bar c \bar A_0}$. In \Cref{fig:QuarkGapDSE} we depict the quark DSE. \hspace*{\fill}}
	\label{fig:FunDSE+QuarkDSE}
\end{figure*}
The functional Dyson-Schwinger equation follows from \labelcref{eq:TotalDerivativeZ} with a unity kernel, $\Psi=1$. It  reads for the components of the fundamental superfield $\Phi_f$, 
\begin{align} \nonumber 
   \frac{\delta \Gamma[\Phi]}{\delta A_\mu} &= \left\langle  \frac{\delta S_\textrm{QCD}}{\delta A_\mu}\right\rangle\,, \qquad \frac{\delta \Gamma[\Phi]}{\delta c} = \left\langle  \frac{\delta S_\textrm{QCD}}{\delta c}\right\rangle\,,\nonumber\\[2ex]
     \frac{\delta \Gamma[\Phi]}{\delta q} &= \left\langle  \frac{\delta S_\textrm{QCD}}{\delta q}\right\rangle\,,
	\label{eq:FunDSE}
\end{align} 
and the DSEs for $n$-point correlation functions can be derived by further field derivatives, evaluated on the equations of motion. The first two functional DSEs with gluons and ghosts govern the pure glue sector, and the last functional DSE governs the matter sector. Evidently, DSEs for quark-gluon correlation functions can be derived from either the first or the last DSE in \labelcref{eq:FunDSE}, and similar statements holds true for ghost-gluon and ghost-quark and ghost-gluon-quark correlation functions. 

The right hand side of 	\labelcref{eq:FunDSE} can be rewritten in terms of one- and two-loop terms with full vertices and propagators and one classical one. We illustrate this structure with the DSE for the (background) gauge field, \Cref{fig:DSE-backA0}, that underlies the computation of the Polyakov loop potential.  This DSE is pivotal for the description of the confinement-deconfinement phase transition, see also \Cref{sec:ConfinementDeconfinement}. We also depict the quark gap equation in \Cref{fig:QuarkGapDSE}. Besides being the most important DSE for the present purposes, it also illustrates very clearly the overall structure of functional relations. Apart from the quark propagator, the diagrams contain two further ingredients, the gluon propagator as well as the quark-gluon vertex. Subject to a quantitative grip on these two correlation functions, the quark gap equation is readily solved within an iterative numerical scheme. This already entails, that a systematic error analysis for the gap equation has to survey the systematic error for the gluon propagator as well as the quark-gluon vertex. This is detailed further in \Cref{sec:Expansion+Error}. In general DSEs for $n$-point functions depend on other $m$-point correlation functions with $2\leq m\leq n+2$, i.e. they form an infinite tower of coupled DSEs. 
In approximations this tower has to be truncated, only allowing for a finite set of correlation function with $n\leq N_\textrm{max}$. In \Cref{sec:Expansion+Error} it is discussed how this leads to  well-controlled expansion schemes with quantifiable systematic error estimates. 

The fRG equation for the effective action is obtained by augmenting the classical action with a quadratic infrared cutoff term  $\Delta S_k[\Phi]$, 
\begin{align} 
	S_\textrm{QCD}[\Phi_f]\to& S_\textrm{QCD}[\Phi_f]+\Delta S_k[\Phi]\,,
\label{eq:Sk}
\end{align} 
with 
\begin{align} 
	  \Delta S_k[\Phi]=&\,\int_p \Phi_i(-p) R_k^{ij}(p) \Phi_j(p)\,, 
      \label{eq:Cutoff}
\end{align} 
where $i,j$ sum over all species of fields, and we have suppressed space-time and internal indices. 
The infrared regulator $R_k(p)$ grows large for small momenta with $p^2/k^2 \to 0$ and decays rapidly for $p^2/k^2 \to \infty$. This term suppresses the propagation of small momentum modes and full QCD is obtained in the limit $k\to 0$. By taking a (logarithmic) scale derivative of the path integral (represented by a respective $\Psi$, see e.g.~\cite{Ihssen:2022xjv, Ihssen:2024ihp}, we arrive at the flow equation for the effective action, 
\begin{subequations}
	\label{eq:FunFlow}
	\begin{align}\nonumber 
		\partial_t \Gamma_k [\Phi]  + \dot{\Phi}_a \left(\frac{\delta \Gamma_k[\Phi]}{\delta \Phi_a} + c_{\sigma_i} \delta_{a\mathbf{\sigma}_i} \right) \\[1ex] 
		&\hspace{-2.5cm}=\frac{1}{2} G_{ac} [ \Phi]\,\left(\partial_t\delta^c_b + \frac{\delta \dot{\Phi}_b}{\delta  \Phi_c}\right) R^{ab} \,,
		\label{eq:GenFlow}
	\end{align}
	with $i=l,s$ and $(\sigma_i) = (\sigma_l,\sigma_s)$. In \labelcref{eq:GenFlow}, the parameter $t= \ln (k/\Lambda_\textrm{UV})$ is the RG-time and the effective action $\Gamma$ does not depend on the reference scale $\Lambda_\textrm{UV}$. The propagator $G_k$ is that of all field components in a given background configuration $\Phi$, 
	\begin{align}
		G_k [\Phi] =& \frac{1}{\Gamma_k^{(2)}[\Phi] + R_k } \,. 
			\label{eq:PropPhi}
	\end{align}
\end{subequations}
Here, $\Gamma_k^{(2)}= \delta^2 \Gamma_k/\delta \Phi^2$ and $G_{ab}$ is the propagator of the fields $\Phi_a $ and $\Phi_b$, that is $G_{ab} = (G_k [\Phi])^{\ }_{\Phi_a \Phi_b}$. Importantly, due to the background it also hosts off-diagonal components. We emphasise, that the full propagators are key building blocks in both functional methods, DSE and fRG, and their quantitative nature is one of the important ingredients of a small systematic error budget.

\begin{figure*}
	\centering
	\begin{minipage}[t]{.45\linewidth}
		\begin{subfigure}{\linewidth}
			\centering
			\includegraphics[width=\linewidth]{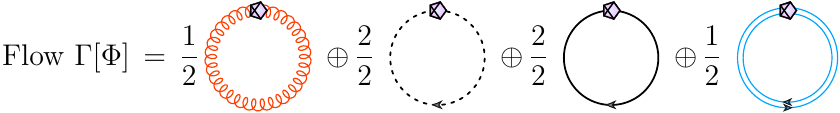} \vspace{.1cm}
			\subcaption{Diagrammatic part of the flow equation \labelcref{eq:FunFlow} for the effective action.  \hspace*{\fill}}
			\label{fig:funfRG}
		\end{subfigure}%
	\end{minipage}
	\hspace{0.02\linewidth}%
	\begin{minipage}[t]{.5\linewidth}
		\begin{subfigure}{\linewidth}
			\centering
			\includegraphics[width=\linewidth]{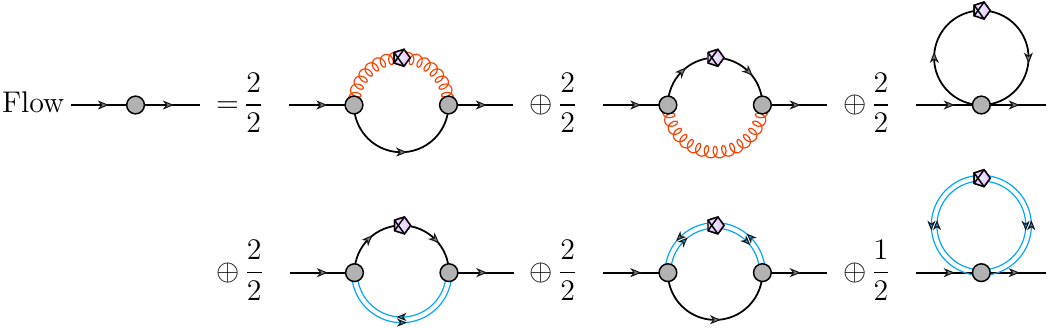} 
			\subcaption{Diagrammatic part of the flow equation for the quark propagator. \hspace*{\fill}}
			\label{fig:QuarkGapfRG} 
		\end{subfigure}
	\end{minipage}
	\caption{fRG: Full propagators and vertices are indicated by grey blobs. Gluons are represented by red spiral lines, ghosts by back dotted ones, and quark by straight black ones, $\bigoplus$ accommodates symmetry factors and relative minus signs. In \Cref{fig:funfRG} we depict the right  hand side of the flow equation \labelcref{eq:GenFlow} of the effective action with $\dot \Phi=0$. A detailed account of the flow can be found in \cite{Ihssen:2024miv}. In \Cref{fig:QuarkGapfRG} we depict the flow equation for the quark propagator, for the respective DSE see \Cref{fig:QuarkGapDSE}. 		
		\hspace*{\fill}}
	\label{fig:FunfRG+QuarkfRG}
\end{figure*}
It follows from the functional flow equation \labelcref{eq:FunFlow}, that the diagrammatic part of the flow of $n$-point function depends on $m$-point functions with $2\leq m\leq n+2$, the same relation as for the DSEs. This leads us to an infinite tower of coupled flows for $n$-point functions that needs to be truncated by well-controlled expansion schemes with quantifiable systematic error estimates. This is discussed in \Cref{sec:Expansion+Error}.

The comparison of the DSE for the (inverse) quark propagator, shown in \Cref{fig:QuarkGapDSE}, and the respective fRG equation, \Cref{fig:QuarkGapfRG}, illustrates very lucidly the different resummation schemes underlying different functional methods. While loops in DSEs always host one renormalised classical vertex stemming from $\delta S_\textrm{QCD}/\delta \Phi_i$, the fRG loops only hosts full correlation functions and is finite by construction as a momentum-shell RG. In the corresponding
DSE for the gluon propagator (not shown for brevity), two-loop diagrams appear, whereas those are absent in fRGs. Also
composite fields appear at different levels. In the fRG these can be made explicit directly on the level of the two-point
functions, included in the second line of \Cref{fig:QuarkGapfRG}. In the DSEs these appear explicitly inside loops on 
the level of the quark-gluon-vertex and are therefore only implicitly contained in \Cref{fig:QuarkGapDSE}. The physics
represented by these diagram is, however, the same in both approaches. We will come back to this point below.  
\footnote{We also remark that the fRG flow can be seen as a differential DSE: to begin with, the functional DSE \labelcref{eq:FunDSE} also holds true in the present of the regulator. Then, one may derive the flow equation by simply hitting the DSEs for $n$-point correlation functions with a $t$-derivative. This is identical to  the flow of $n$-point correlation functions \labelcref{eq:GenFlow}, after using DSEs for the vertices in the diagrams, for further discussions see e.g.~\cite{Fischer:2008uz, Dupuis:2020fhh}.}  

The different diagrammatic structure implies that identical approximations for the effective action $\Gamma$ still lead to different approximations in the functional relations. Consequently, the comparison of results obtained from different functional methods such as DSE and fRG in identical or similar approximations is a rather non-trivial self-consistency check for the functional QCD approach. This is discussed further in the next Section.

\subsection{Systematic expansion schemes, error estimates and apparent convergence}
\label{sec:Expansion+Error}

\begin{figure*}
	\centering
	\begin{minipage}[t]{.55\linewidth}
		\begin{subfigure}{\linewidth}
			\centering
			\includegraphics[width=\linewidth]{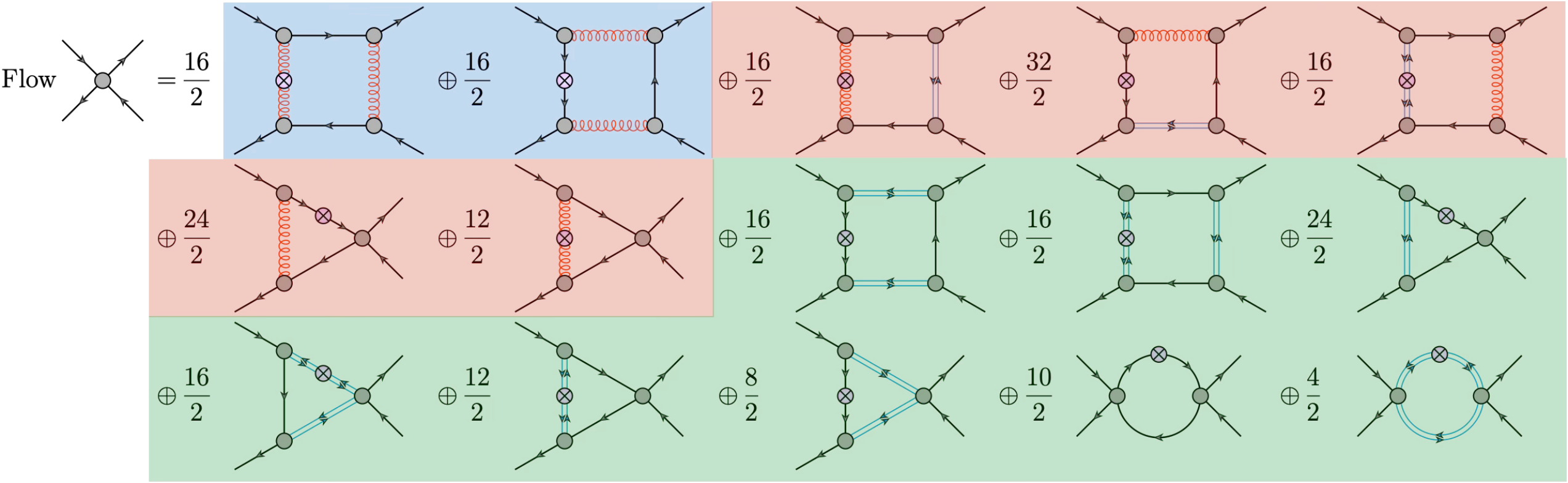} \vspace{.1cm}
			\subcaption{Diagrammatic part of the flow of the four-quark scattering vertex. 
				 \hspace*{\fill}}
			\label{fig:4quarkLego1} 
		\end{subfigure}%
	\end{minipage}
	\hspace{0.02\linewidth}%
	\begin{minipage}[t]{.4\linewidth}
		\begin{subfigure}{\linewidth}
			\centering
			\includegraphics[width=\linewidth]{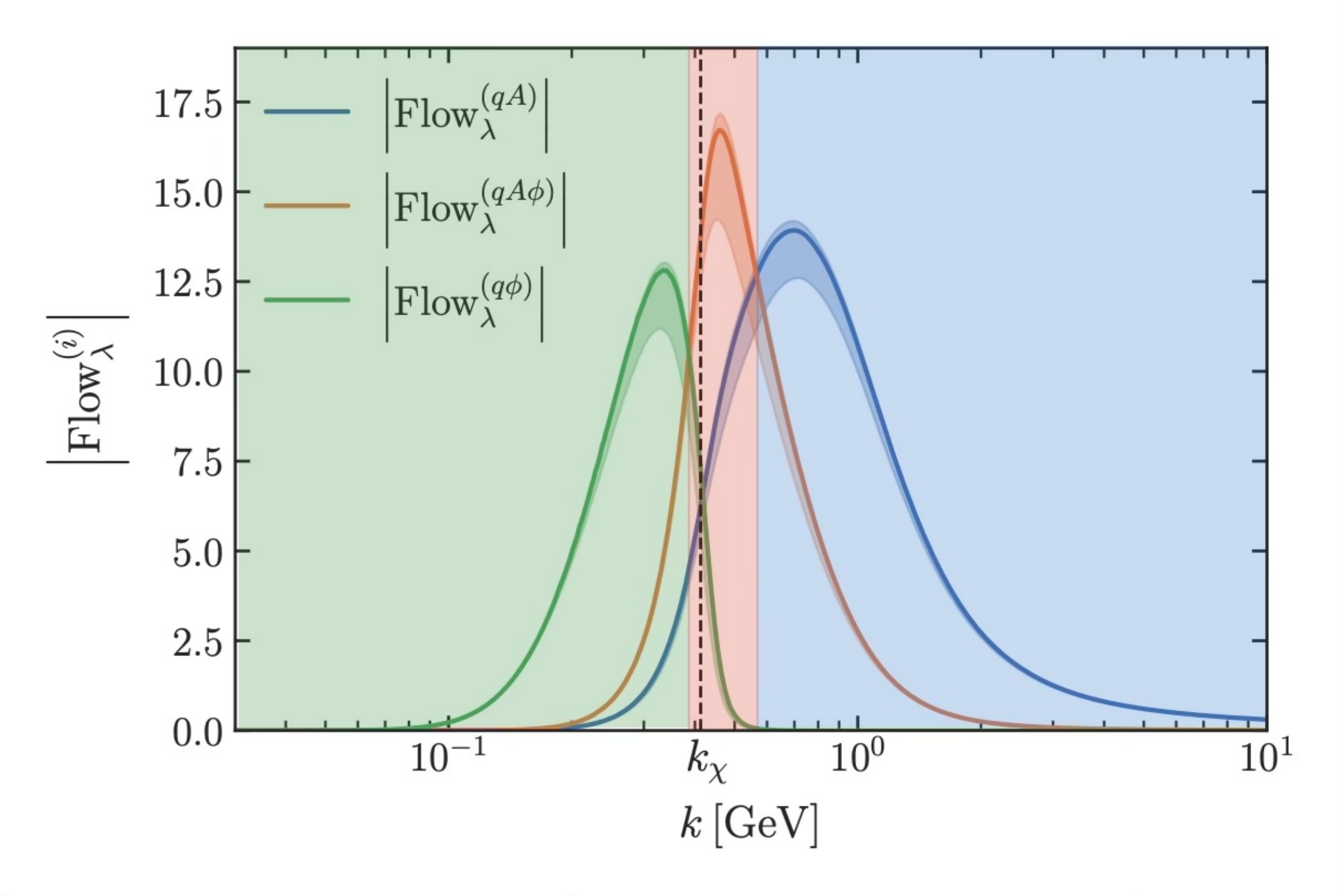} 
			\subcaption{Strength of different diagrammatic subsectors.  \hspace*{\fill}}
			\label{fig:4quarkLego2} 
		\end{subfigure}
	\end{minipage}
	\caption{Flow of the four-quark scattering vertex: Diagrams in the blue-shaded areas in \Cref{fig:4quarkLego1} are quark-gluon diagrams that source the four-quark vertex and dominate in the ultraviolet regime with $k \gtrsim 0.6$\,GeV, see \Cref{fig:4quarkLego2}. Diagrams in the red-shaded area in \Cref{fig:4quarkLego1} are quark-gluon-meson diagrams and dominate in the interface regime with $0.3$\,GeV $\lesssim k\lesssim 0.6$\,GeV. Diagrams in the green-shaded area in \Cref{fig:4quarkLego1} are quark-meson diagrams and dominate the infrared regime with $k \lesssim 0.3$\,GeV, see \Cref{fig:4quarkLego2}. The figures are modified from \cite{Ihssen:2024miv}. The scale $k$ is the infrared cutoff scale in the fRG and is related to an average symmetric-point momentum. The chiral scale  $k_\chi=388$\,MeV is the onset scale of chiral symmetry breaking in the chiral limit, computed in \cite{Ihssen:2024miv}.  \hspace*{\fill}}
	\label{fig:4quarkLego}
\end{figure*}
The predictive power of the functional approach to QCD is in one-to-one correspondence with systematic error estimates. The associated systematics of expansion schemes and approximations (also called truncation) of the full infinite tower of DSEs and FRGs have been the subject of intensive studies over the past decades, see for example the QCD-related fRG reviews \cite{Litim:1998nf, Berges:2000ew, Schaefer:2004en, Pawlowski:2005xe, Gies:2006wv, Braun:2011pp, Dupuis:2020fhh, Fu:2022gou}, the DSE reviews \cite{Roberts:2000aa, Alkofer:2000wg, Fischer:2006ub, Eichmann:2016yit, Fischer:2018sdj, Huber:2018ned} and the recent brief summary of functional QCD, \cite{Rennecke:2025bcw}. The discussion here follows the recent works \cite{Lu:2023mkn, Ihssen:2024miv, Huber:2025kwy} and concentrates on the key aspects. A more detailed description goes beyond the scope of the present review and we refer to these works and the reviews for more details. 

The systematic error analysis in functional QCD relies on two key observations that follow from the \textit{closed} form of the functional equations for any $n$-point function, i.e. the fact that every functional equation only contains a finite amount of diagrams.  
This form entails that diagrams can be divided into subclasses, the two most obvious ones being the class of pure glue diagrams that only involve \textit{internal} gluon and ghost propagators, and the class of pure quark or matter diagrams that only involve quark (and potentially emergent composite) lines. This classification is very apparent in the fRG equations with their one-loop form, but it also persists in the DSEs with their one- and two-loop closed form. The key observations are:
\begin{itemize} 
\item[(i)] \textit{Modular form of functional relations:} Subclasses of diagrams are only connected with each other via very few vertices. The momentum dependence of these vertices is restricted by general considerations (renormalisation, Bose-symmetry, etc.) as well as gauge symmetries (Slavnov-Tayor identities (STIs)). This allows a two-stage error analysis, first of the separate subsystems and then of the combined full system. A complementary hierarchy is induced by the mass gaps $m_i^2$ with $i=A,q,\bar q,\phi$ in the propagators that lead to a very efficient and rapid decoupling of diagrams below the respective mass gap. In combination this modularity has been called the \LEGO-principle and is described in detail in \cite{Ihssen:2024miv}. 
\item[(ii)] \textit{Locality of correlation functions in gauge-fixed QCD:} After including emergent composites, $n$-point correlation functions of quarks and gluons in gauge-fixed QCD are local in momentum and position space, and this locality increases rapidly with $n$. Consequently, diagrams with higher order vertices are parametrically suppressed. 
\end{itemize} 
We proceed with a brief illustration of the two properties within selected examples, for more details we refer the reader to the functional reviews cited above and the works \cite{Lu:2023mkn, Ihssen:2024miv, Huber:2025kwy}.

\subsubsection{(i) Modular form of functional relations:}
\label{sec:ModularFun}

We illustrate property (i) first with the flow of the four-quark scattering vertex in the vacuum. This vertex enters directly the fRG for the quark propagator, see \Cref{fig:QuarkGapfRG}, and governs \DCSB. In a partly different role, it is also present in the quark DSE \Cref{fig:QuarkGapDSE} where it enters via the quark-gluon vertex. As discussed already at the end of \Cref{sec:FunDerivation}, the different occurrence of vertices is related to the different resummation structures and comparing results from different functional methods is part of the systematic error analysis of the functional approach to QCD.

The diagrammatic part of the flow and the absolute strength of the different sectors is depicted in \Cref{fig:4quarkLego}. For the sake of simplicity we have dropped diagrams with quark--two-gluons scattering vertices and higher order ones, there impact has been assessed in \cite{Mitter:2014wpa, Cyrol:2017ewj} and is subleading. 
The flow of the quark-gluon box diagrams completely dominates the momentum regime with $k\gtrsim 1$\,GeV. It is proportional to the forth power of the running strong coupling $g_s(k)$. Importantly, in the non-perturbative regime there are different avatars of the strong coupling as universality only holds up to two-loop. A convenient definition of such a running coupling in the box diagrams is given by the one-gluon exchange coupling $\alpha_{q\bar q A}$, which can be isolated in  \Cref{fig:4quarkLego1} as the contraction of two quark-gluon vertices with a gluon propagator, normalised with the quark dressing squared. The latter normalisation guarantees that the coupling is renormalisation-group invariant. 

\begin{figure}[t]
	\centering
	\includegraphics[width=\linewidth]{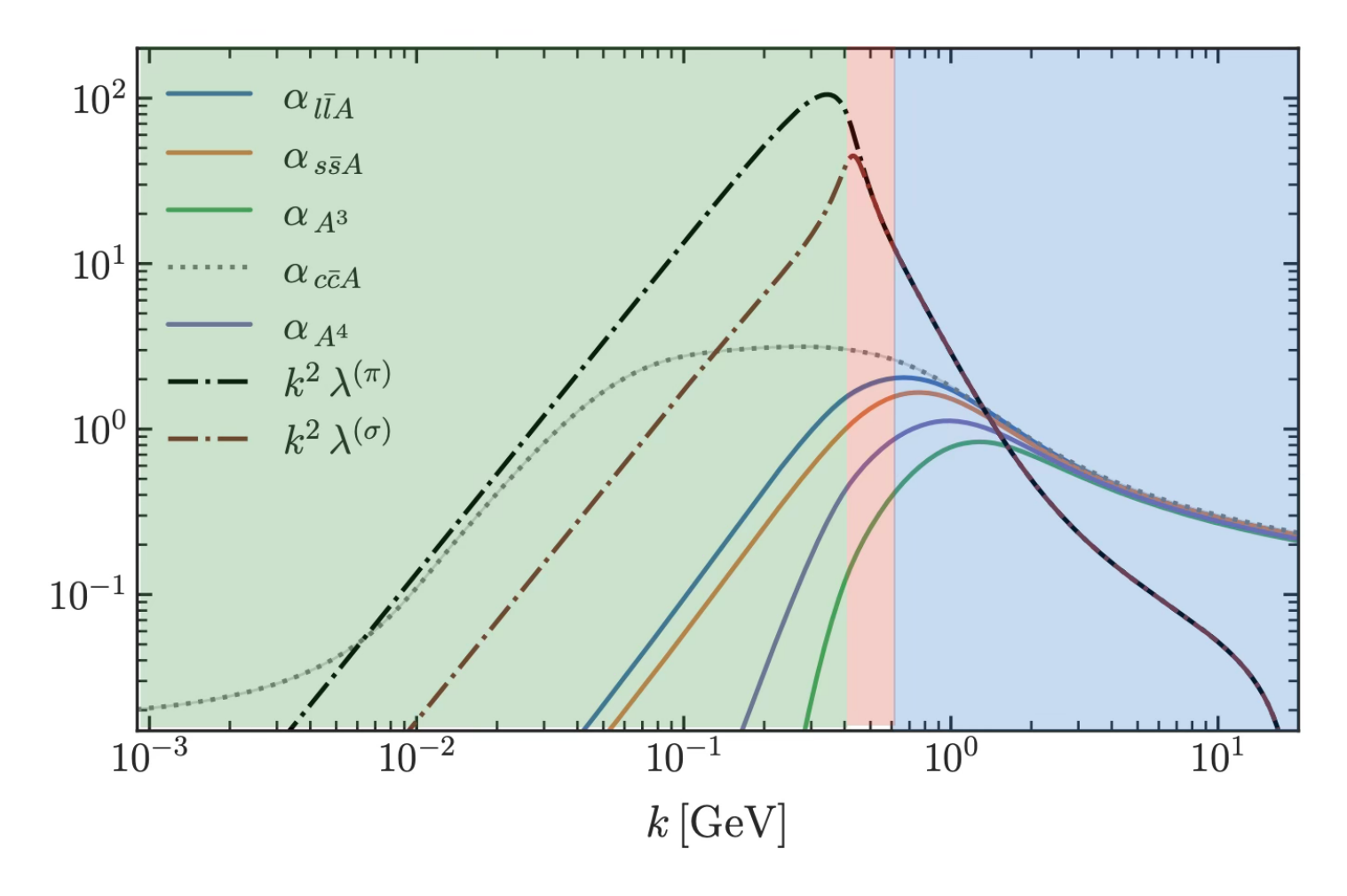}%
	\caption{Momentum scale (RG-scale)dependence of the strong couplings $\alpha_{i}$ for $N_f=2+1$ flavours with $i=l\bar l A, s\bar sA, A^3, A^4$ on a double-logarithmic scale in comparison to the running of the meson exchange couplings  $\lambda^{(\pi)} = (h_\phi^2/2)\, G_\pi$ and $\lambda^{(\sigma)}=(h_\phi^2/2)\, G_\sigma$. The different colours reflect the dominance regimes of the different diagrammatic subsectors, see \Cref{fig:4quarkLego}. Figure modified from \cite{Ihssen:2024miv}. \hspace*{\fill} }
	\label{fig:CouplingStrengths}
\end{figure}
Similar exchange couplings can be defined for purely gluonic vertices as well as meson-exchange couplings. These couplings are depicted in \Cref{fig:CouplingStrengths}. One clearly sees the rapid decoupling of the different degrees of freedom below the respective mass scales,
i.e. the infrared vanishing of their respective couplings. This starts with the decoupling of the purely gluonic exchange coupling at $k\approx 1$\,GeV, quickly followed by the quark-gluon couplings at $k\approx 0.6-0.8$\,GeV and finally the decoupling of the scalar and pseudoscalar exchange couplings below $k\approx 3$\,GeV. While the decoupling of the gluonic and quark-gluon couplings is governed by the gluon mass gap, that of the scalar and pseudoscalar exchange couplings is governed by the mass gap $m_\sigma\approx 0.5$\,GeV of the $\sigma$-mode and that of the pions with $m_\pi\approx 0.14$\,GeV. We also emphasise that the coupling strengths depicted in \Cref{fig:CouplingStrengths}  is only one of many ingredients in the absolute and relative strengths of diagrams as shown in \Cref{fig:4quarkLego2}. This strength also depends on the tensor contraction of the vertices and the remaining propagators in the diagrams. 

\begin{figure}[t]
	\centering
	\includegraphics[width=\linewidth]{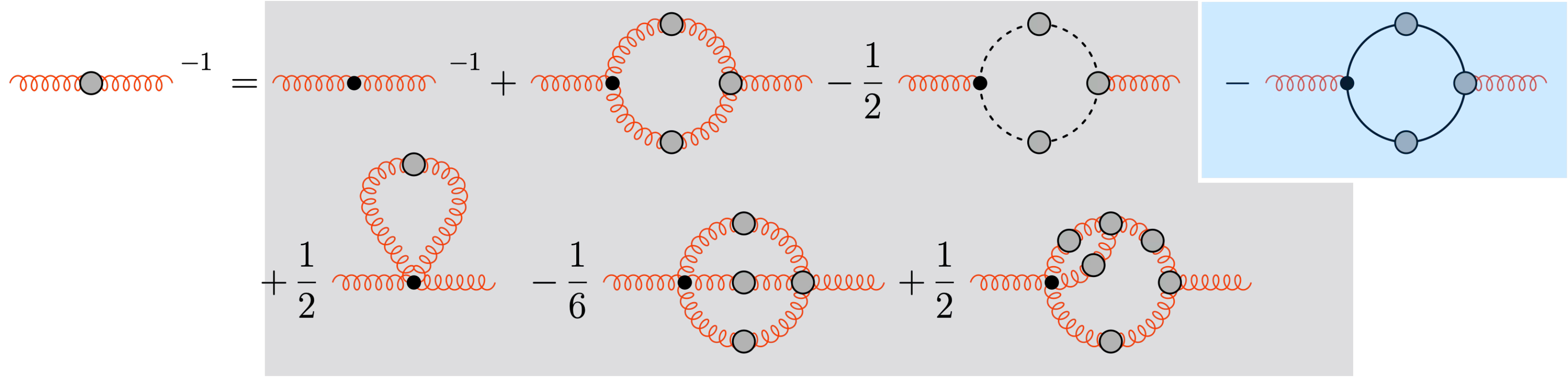}%
	\caption{DSE for the gluon propagator. Diagrams with grey background contain only internal gluon and ghost lines; the quark-loop 
		diagram (blue background) is the explicit matter part. \hspace*{\fill} }
	\label{fig:gluonDSE}
\end{figure}

Our second illustration of property (i) is the DSE for the gluon propagator \Cref{fig:gluonDSE}. The modular structure is readily visible as the matter part only comes from quark loops, and the interface coupling is the quark-gluon vertex. In the corresponding flow equation (not shown) we have an additional connection due the two--quark--two-gluon coupling, which again highlights the different structure of the two towers. Due to the decoupling property of the quark-gluon coupling these loops are parametrically suppressed and their contributions are subleading in the vacuum \cite{Fischer:2003rp} and consequently also at finite temperature. Indeed, their effect on the pure glue results mainly comes from the change of the perturbative momentum running due to the $N_f$-dependent contributions to the UV anomalous dimension of the gluon and to the running coupling. The respective systematic error analysis has been done in \cite{Braun:2014ata}. It suggest that an approximation using quantitative results for the Yang-Mills gluon propagator and simply adding the quark loops works very well in full QCD. This notion has been cross-checked with lattice QCD: In \cite{Fischer:2012vc} the first computation of thermal gluon propagators in QCD has been undertaken, based on pure Yang-Mills precision data from the lattice \cite{Fischer:2010fx}. The results have later been confirmed by lattice simulations in full QCD \cite{Aouane:2012bk}, providing non-trivial evidence for the validity of the functional systematic error analysis. This also illustrates a further advantageous property of the functional approach to QCD: any quantitative results for a subset of correlation functions can be used as consistent input in functional relations. In \cite{Fischer:2012vc} (DSE) this has been done with lattice data for gluon propagators in Yang-Mills theory. In \cite{Gao:2020qsj} (DSE), vacuum QCD input from the fRG has been used.  Finally, one may use a mixed approach: one may compute a subset of correlation functions with the fRG and the complement with the DSE. This has been done in \cite{Fischer:2008uz}, where the ghost propagator in the fRG-computation was computed from the DSE. In our opinion, the examples above illustrate very impressively the flexibility of the functional approach, that can be used without sacrificing the systematics. In short, this flexibility can be used the minimise systematic error budget. 

Moreover, the expansion scheme underlying \cite{Fischer:2012vc} can be used more general: the one- and two-loop closed structure of functional relations can be used to recast them into a difference form, e.g.~2+1 flavour QCD at finite temperature and density can be written as 2+1 flavour vacuum QCD plus the thermal and baryon chemical potential correction of all correlation function. Then one can use that for a large range of temperatures and baryon chemical potentials the difference is but a correction to the vacuum results. Indeed, such an approach also allows to monitor its potential failure if the correction grows large. This has led to the miniDSE scheme \cite{Lu:2023mkn}, and similar developments in the fRG have been put forward in \cite{Braun:2014ata, Rennecke:2015eba, Fu:2019hdw}. For more details we defer the reader to these works. For these schemes one may use lattice input as done in \cite{Fischer:2012vc} but by now also the functional QCD approach offers a plethora of quantitative results for propagators and vertices that can be used as input. Specifically, the vertex input cannot be obtained from the lattice yet as the respective results for vertices still host relatively large error bars. Quantitative state-of-the-art in vacuum Yang-Mills theory and QCD including systematic error analyses can be found in \cite{Cyrol:2016tym, Huber:2020keu} (vacuum Yang-Mills), \cite{Williams:2014iea, Mitter:2014wpa, Williams:2015cvx, Cyrol:2016tym, Cyrol:2017ewj, Gao:2021wun} (vacuum QCD) and \cite{Cyrol:2017qkl} (finite temperature Yang-Mills).

\subsubsection{(ii) Locality of correlation functions in gauge-fixed QCD}
\label{sec:LocalityFun}

This second key property is tantamount to small expansion parameter behind well-controlled approaches. Let us start with an heuristic argument. All functional relations can be derived from the gauge fixed one-particle irreducible effective action $\Gamma$ of QCD. Suppose for the moment that the irreducible parts of scattering vertices of $n$ fields are local in momentum as well as in position space. We can then define a scattering length $\xi_n$ for each of these processes. This scattering length depends on $n$ as well as on the nature of the scattered fields, but as long as the $\xi_n$ are all finite the latter property is not important for our argument. Provided the scattering length is not increasing proportional to $n$ (which is not the case) we readily infer that the probability of such a scattering rapidly decays with $n$. This then leads to a rapid suppression of diagrams with higher order vertices in functional relations of $m$-point correlation functions and suggests 
a specific systematic expansion scheme, the \textit{vertex expansion}. This expansion scheme is indeed used throughout the functional approach to QCD, partially augmented with the \textit{derivative} expansion for emergent composites in the fRG-approach. The latter allows to include all powers of local scatterings of pions and the $\sigma$-modes and gives access to the critical properties of QCD in the critical regimes close to second order phase transitions including the potential CEP in the QCD phase structure.  

But why in the first place should the scattering vertices of $n$ fields be local in momentum as well as in position space? In momentum space the argument is readily made. All $n$-point functions for $n \le 5$ are not primitively divergent and therefore fall off at large momenta with powers of momenta. Thus locality in momentum space is satisfied for $n \le 5$. In turn this means that all truncation schemes that (self-consistently) include all primitively divergent $n$-point function are good candidates for high quality truncations with small systematic error. In Yang-Mills theory, this property has been explicitly demonstrated in functional QCD in \cite{Cyrol:2016tym, Huber:2025kwy}. Position space locality is more intricate. In principle it shows up in momentum space by finite infrared dressing functions of $n$-point functions and corresponding strong running couplings. In particular the latter property is important, since infrared divergences of individual $n$-point functions may be compensated by infrared zeros of others, when combined to RG-invariant running couplings. The corresponding mechanism has been discussed in detail in \cite{Fischer:2006vf, Fischer:2009tn} for a class of gauge completions within the group of Landau gauges. Indeed, a specific property of QCD in covariant gauges, and specifically in the Landau gauge, is the constraint $\alpha_i< \alpha_i^{\textrm{max}}$ with  $\alpha_i^{\textrm{max}} \approx 1-3$, see \Cref{fig:CouplingStrengths}. Thus, in general, there are no divergent couplings that have the potential to undo locality.  

It is worth emphasising that the gauge fixing is a crucial ingredient for this property. It leads to a local covariant quantum field theory and all $n$-point correlation functions are local in the absence of resonances. Gauge fixing is commonly seen as a disadvantage as it only complicates the extraction of specific observables such as the Wilson loop and Polyakov loop and its correlation functions, which are a series of all order correlation functions of the gauge fixed gluons. While this still can be done, see \cite{Herbst:2015ona}, it simply entails that specific observables cannot be easily computed in functional approaches. However, it gives us full access to the phase structure of QCD via functional methods as well as many experimentally accessible observables such as the fluctuations of conserved charges. Indeed, one may say that all experimentally accessible observables are local by definition and hence are in the natural realm of functional computations. 

This leaves us with another possibility: resonant interaction channels in particular in the multi-quark scattering vertices. Indeed, this happens and the most dominant one (in the vacuum and at finite temperature) is the pseudoscalar channel with the pion quantum numbers. These develop poles (and cuts) at time-like momenta, which, however, do not influence the convergence properties of functional relations, since inside loops these correlations are only tested in the space-like momentum domain. Nevertheless, other  interesting phenomena may appear: The pseudoscalar exchange coupling is shown in \Cref{fig:CouplingStrengths}, together with the scalar channel ($\sigma$-mode). Their dimensionless coupling strengths, obtained by the multiplication with momentum squared, peak between 10 and 100 at a scale of roughly the one of chiral symmetry breaking which happens to be quite similar to $\Lambda_\textrm{QCD}$. This maximum is large, but finite and its decay in the infrared is sufficient to not disturb locality. It has been shown in the fRG, that multi-scatterings of these channels become rapidly subdominant, see in particular \cite{Mitter:2014wpa, Braun:2014ata, Rennecke:2015eba, Cyrol:2017ewj, Fu:2019hdw}. In \cite{Ihssen:2024miv} this results have been corroborated within a computation of the full effective potential that takes all order scatterings into account.  Moreover, further four-quark scattering channels are negligible as the respective resonances come with far bigger mass gaps: they can only be relevant close and below the chiral symmetry breaking scale $k_\chi \approx 300 - 400$\,MeV, defined as the onset scale of chiral symmetry breaking in the chiral limit. However,  their mass gaps exceed 500\,MeV and hence these channels can be dropped. This has been explicitly checked in \cite{Mitter:2014wpa, Cyrol:2017ewj} in the vacuum. This property is also present for finite temperature and baryon chemical potential, see \cite{Braun:2019aow} and higher order scatterings only become important in the critical regimes around second order phase transitions including the potential CEP. This will be discussed in \Cref{sec:ChiralSymbreaking}. 

In conclusion, for $\xi_n\leq \xi_\textrm{max}$ for all $n$, the infinite tower of functional relations in the DSE and fRG-approaches can be safely truncated. We are left with a rapidly converging expansion scheme: the vertex expansion, augmented with local scatterings of resonant interaction channels. The systematic error analysis uses the closed form of functional equations and the properties (i) and (ii) as well as the comparison of results from different functional methods, in particular of the DSE and fRG-approaches. In our opinion, in combination this leads to a well-controlled first principles functional approach to QCD that can be applied to QCD at finite temperature and densities beyond the current applicability range of lattice simulations. 

We close with a word of caution concerning an intricacy that is inherent to all expansion schemes. Strictly speaking, the existence of a small parameter, here the rapid decay of diagrams with higher order vertices, is not sufficient to prove convergence. This deficiency is already present in perturbation theory as well as for the continuum limit of lattice QCD. For this reason we cautiously call the rapid convergence, indeed seen in advanced approximations, \textit{apparent convergence}.

\section{QCD phase structure from functional QCD}
\label{sec:FunPhaseStructure}

In this Section we discuss the application of functional QCD to the QCD phase structure. We concentrate on the developments in the past decade which led to quantitative results for $\mu_B/T \lesssim  4.5$ and coherent estimates for the location of the onset of new physics. The accompanying systematic error estimates follow the framework outlined in \Cref{sec:Expansion+Error} and for its importance we shall recall some of its ingredients as well as the additional assessment required at higher densities or chemical potentials. 

We initiate this discussion with a brief review of \DCSB and the thermal chiral phase transition in \Cref{sec:ChiralSymbreaking} as well as the broad transition regime from the confined phase at low temperature to the deconfined phase at large temperatures in \Cref{sec:ConfinementDeconfinement}. Then, in \Cref{sec:PhaseStructure+CEP} we discuss the phase structure of QCD at finite baryon chemical potential including the assessment of the existence of the CEP and its location.

\subsection{Chiral crossover and soft modes}
\label{sec:ChiralSymbreaking} 

In the presence of (small) light current quark masses, the chiral transition is a relatively sharp crossover. Commonly used order parameter are variants of the chiral  condensate which may be used to define the pseudocritical temperature.

In 2+1 flavour QCD in the isospin symmetric limit we have the light quarks $l=(u,d)$ with identical current quark masses $m_l$ of the order of 5\,MeV and one heavy quark $s$ whose current quark mass $m_s$ is of the order of 100\,MeV. This allows for the definition of some variants of the chiral condensate, the most basic ones being the light quark condensate $\Delta_l$ and the strange quark condensate $\Delta_s$. They are given by the respective mass derivatives of the grand potential $\Omega=(T/V)  \Gamma[\Phi_{\textrm{\tiny{EoM}}}]$ of QCD, where $\Phi_{\textrm{\tiny{EoM}}}$ is the solution of the QCD equations of motion.   
\begin{align}\nonumber  
\Delta_l =&\, m_l \frac{\partial \Omega}{\partial m_l}\simeq m_l \frac{T}{V} \int_x \langle \bar l l \rangle \,,\\[1ex] \Delta_s = &\, m_s \frac{\partial \Omega}{\partial m_s}\simeq m_s \frac{T}{V} \int_x \langle \bar s s \rangle \,.
\label{eq:Deltas}
\end{align} 
If the (finite) grand potential is used for the derivation of the chiral condensates, the expressions in \labelcref{eq:Deltas} are finite and their renormalisation follows from that of $\Omega$. These relations are commonly used in the fRG-approach in which a finite effective action $\Gamma$ is computed, for detailed discussions see  e.g.~\cite{Fu:2019hdw, Braun:2023qak} and the reviews \cite{Dupuis:2020fhh, Fu:2022gou}. 

The chiral condensate can also be computed directly from the quark propagators, which requires an explicit renormalisation procedure. This approach is commonly used in DSEs, see e.g.~\cite{Isserstedt:2019pgx, Gao:2020fbl} and the review \cite{Fischer:2018sdj}. 

\begin{figure}[t]
	\begin{center}
\includegraphics[trim=0 2mm 0 0]{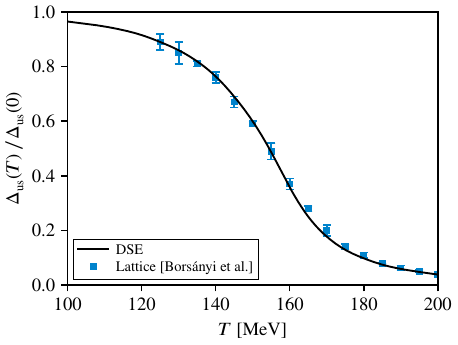}	
	\end{center}
	\caption{Quark condensate from the functional approach \cite{Isserstedt:2019pgx} and lattice QCD \cite{Borsanyi:2010bp}. Plot taken from \cite{Isserstedt:2019pgx}.\hspace*{\fill}}
    \label{fig:condensate}
\end{figure}

We start the discussion of the functional results with ~\Cref{fig:condensate}, which shows the temperature dependence of the \textit{reduced} quark condensate, 
\begin{align}\nonumber 
\Delta_{ls}
&=
\expval{\bar{q} q}_l
-
\frac{m_l}{m_s}  \expval{\bar{q} q}_s \, , \\[1ex]
\expval{\bar{\psi}\psi}_f &= -\Nc Z_2^f Z_m^f \sumint_q \Tr \bigl[ S_f(q) \bigr]
\label{eq:ReducedDelta}
\end{align}
for each quark flavour $f \in {l,s}$ with $l=(u,d)$ in a form which is also used in lattice QCD. Further results of variants of the chiral condensate can be found in \Cref{sec:Flucs+Thermodyn} in the context of the analysis of fluctuations of conserved charges. 

\begin{figure}[t]
	\begin{center}
		\includegraphics[width=0.98\columnwidth]{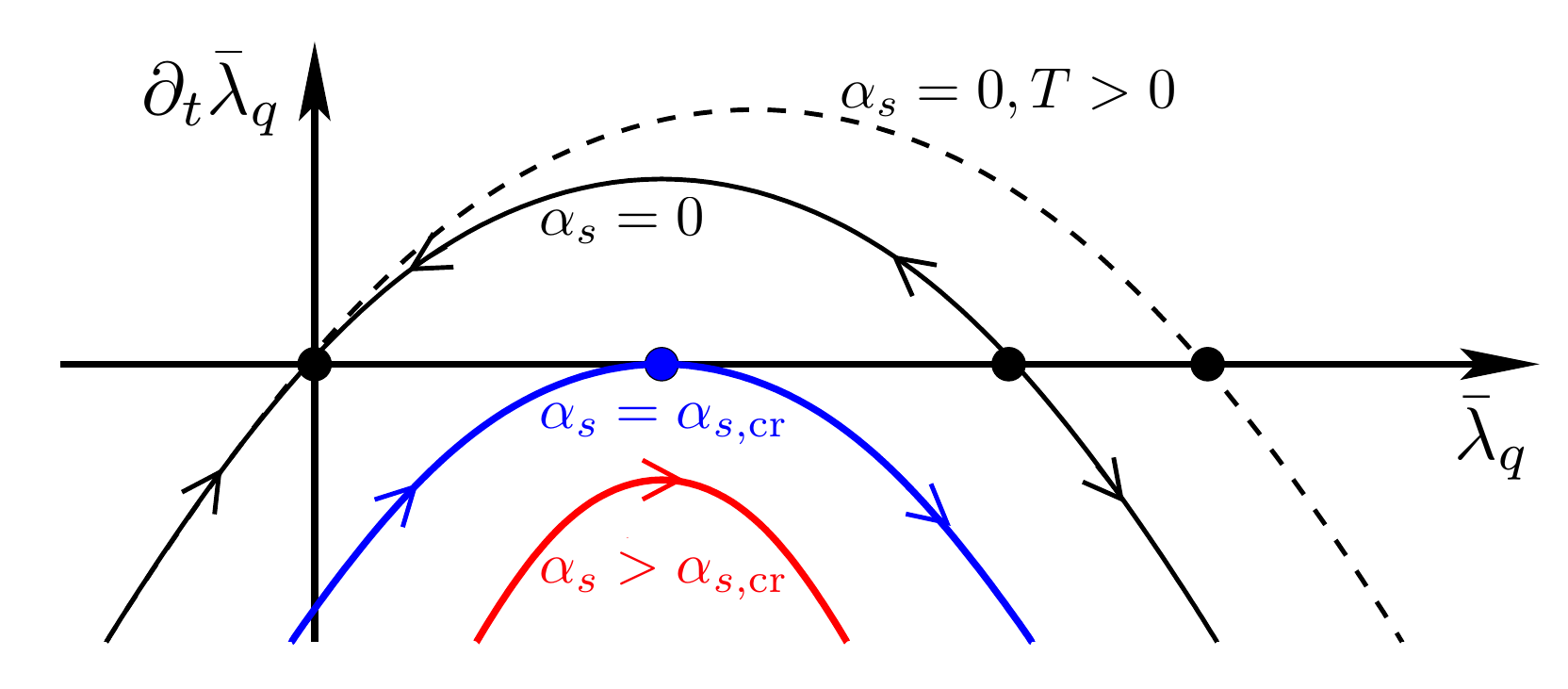}	
	\end{center}
	\caption{Flow $\partial_t\bar{ \lambda}_{q}$ of the dimensionless four-quark coupling $\bar\lambda_q=\lambda_q k^2$ in the scalar-pseudoscalar channel as a function of the four-quark coupling for different values of the strong couplings $\alpha_s$. If the strong coupling is large enough, $\alpha_s>\alpha_{s,\textrm{cr}}$ with the critical coupling $\alpha_{s,\textrm{cr}}$, the flow is negative for all $\bar{ \lambda}_q$ and dynamical chiral symmetry breaking takes place. Figure modified from \cite{PRW2026}.\hspace*{\fill}}
	\label{fig:FlowFourQuarkCoupling}	
\end{figure}

The thermal chiral crossover can be understood with the thermal melting of the exchange couplings displayed in \Cref{fig:CouplingStrengths}. One way to see the mechanism behind this melting is to recall the RG-picture of chiral symmetry breaking in the vacuum. The flow of the dimensionless coupling $\bar \lambda_q= k^2 \lambda_q$ of the scalar-pseudoscalar channel of the four-quark vertex (including the emergent composite channel) has the form 
\begin{align} 
\partial_t \bar \lambda_q = 2\bar\lambda_q -  \bar\lambda_q^2 \,A_k   -   \bar\lambda_q \,\alpha_{s}\, B_k -    \alpha^2_{s}\,C_k  
\end{align}
where we have dropped all other channels for the sake of simplicity and $\alpha_s=\alpha_{l\bar l A}$. 
The first term on the right hand side 
includes the dimensional scaling of the four-quark coupling and originates in the factor $k^2$ in the dimensionless coupling $\bar\lambda_q$.  The other three terms on the right hand side are provided by the blue-shaded diagrams ($\alpha^2_{s} C_k  $), the red-shaded diagrams ($\bar\lambda_q \alpha_{s} B_k$) and the green-shaded ones ($\bar\lambda_q^2 A_k   $). The latter ones also occur in low-energy effective theories such as Nambu--Jona-Lasinio-type models or Quark-Meson models. A cartoon of the $\beta$-function $\partial_t \bar \lambda_q$ is depicted in \Cref{fig:FlowFourQuarkCoupling}. 

We have already discussed that the mixed diagrams proportional to $B_k$ and the quark-gluon boxes proportional to $A_k$ decay rapidly in the infrared but dominate the UV and the interface regime. Then, the four-quark coupling $\bar\lambda_q$, generated by these fluctuations, has to exceed a critical value to generate chiral symmetry breaking in the infrared. Evidently, for large enough $\alpha_{l\bar l A}\geq\alpha_{s,\textrm{cr}}$ this takes place. At finite temperature the quark-gluon scattering coupling melts due to the increasing thermal screening mass, see e.g.~\cite{Cyrol:2017qkl} for a respective study in Yang-Mills theory. Due to the presence of finite current quark masses this happens smoothly within a crossover. 

The physics behind the corresponding mechanism in the DSE-approach is similar but its manifestation is organized differently. The main player
is the self energy in the quark-DSE, which contains the fully dressed gluon propagator and quark-gluon vertex. Together with the renormalization
factors of the second, bare quark-gluon vertex an RG-invariant running coupling emerges that carries enough strength at large, intermediate and
small momenta to generate dynamical chiral symmetry breaking. Note that the same dressings also play a key role for the behaviour of 
the four-quark coupling in the FRG described above. The physics behind chiral symmetry restoration at finite temperature is then, of course,
also the same: the coupling strength melts due to thermal gluon screening masses, see \cite{Fischer:2014ata} for a detailed discussion and 
a comparison to lattice data.

By now, the functional approach to QCD leads to a chiral crossover temperature of $T_\chi \approx 155$\,MeV with a small systematic error of less 
than 5\% in agreement with the lattice results, cf.~\Cref{fig:condensate}.

\subsection{Confinement-deconfinement crossover and strong correlations}
\label{sec:ConfinementDeconfinement} 

Confinement in Yang-Mills theory can be related to the breaking of centre symmetry of the gauge group SU(3) and the respective order parameter for the thermal phase transition is the expectation value of the traced Polyakov loop $L$ with 
\begin{align} 
{L}(\boldsymbol{x})= \frac{1}{3} \textrm{tr}_\textrm{f}\, P(\boldsymbol{x})\,,\qquad P(\boldsymbol{x}) =  \mathcal{P}\, e^{ -i\, g_s \int\limits_0^\beta d \tau \,A_0(\tau,\boldsymbol{x}) }\,.
\label{eq:PolloopP+L}
\end{align} 
Loosely speaking, it measures the free energy $F_q$ of a single quark, $L\propto \exp \{ - F_q\}$. In the confined phase, the free energy of such a state is infinite and the Polyakov loop vanishes. In turn, in the deconfinement phase at high energies or temperatures, the temporal gauge field takes values close to zero and $\langle L\rangle \to 1$. In SU(3) with the centre $Z_3$ this phase transition is a first order transition. 

\begin{figure}[t]
	\centering
	\includegraphics[width=.98\columnwidth]{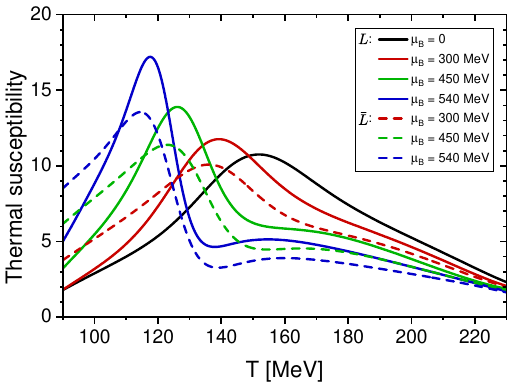} 
	\caption{Thermal susceptibilities of the Polyakov loops $L(\langle \varphi\rangle),\bar L(\langle\varphi\rangle)$. They are shown as functions of temperature $T$ for different values of baryon chemical potential $\mu_B$. Plot taken from \cite{Lu:2025cls}.\hspace*{\fill}}
	\label{fig:suscL-app}
\end{figure}

In physical QCD with its very light current quark masses, centre symmetry is not even an approximate symmetry any more and the dynamics changes a lot. While it is common to label the respective transition as a (broad) crossover, it is more appropriate to map out a wide temperature range where the system undergoes the transition from the confined hadronic phase to the deconfined phase. As already discussed in the introduction, in physical QCD this range can be located between $T_\chi $ and $2-3 T_\chi$, where $T_\chi$ is the chiral crossover temperature. This makes the measurement of the confining physics so elusive and the Polyakov loop is not a good probe for confinement any more. 

For this reason and for their experimental accessibility fluctuations of baryon charge have been used as a measure of confinement: specifically the kurtosis $\kappa$ (see e.g.~\cite{Stephanov:1999zu} for a definition) offers an observable that simply measures the number of degrees of freedom. In the deconfinement phase at asymptotically large temperatures, $\kappa \to 1/9$, measuring the 3 colours (squared) of free quarks, while for the limit of vanishing temperature, $\kappa \to 1$, measuring a baryon degree of freedom. In functional QCD as well as low energy EFTs, this observable has to be computed in the self-consistent $A_0$-background that solves the equations of motion, for a detailed discussion of the dynamics see \cite{Fu:2015naa}. While this background seemingly is a gauge variant quantity it can be directly related to the expectation value of the algebra element $\varphi(A_0)\propto A_0$ of the Polyakov loop $P(\boldsymbol{x})$ in \labelcref{eq:PolloopP+L}, whose eigenvalues are gauge invariant. This has been introduced in \cite{Braun:2007bx, Fister:2013bh, Herbst:2015ona} within functional methods both for the fRG- and DSE-approaches, for related advances in the Curci-Ferrari model see e.g.~\cite{Pelaez:2021tpq, Reinosa:2024njc, MariSurkau:2026irs}. Applications to the phase structure of QCD can be found in \cite{Braun:2009gm, Fischer:2013eca, Fischer:2014ata, Fu:2019hdw, Lu:2025cls, Lu:2026ezr}. There it has been shown that the onset of the transition regime, measured by the peak of the $\varphi$-susceptibility is at about $T_\chi$ at $\mu_B=0$, see the black curve in \Cref{fig:suscL-app}. One also observes that the susceptibility has a broad shoulder for $T\gtrsim T_\chi$, indicating a strongly correlated phase above $T_c$. For larger baryon chemical potentials this shoulder develops into a second peak, clearly indicating non-trivial underlying physics that may be closely related to a region with emerging chiral spin symmetry reviewed in~\cite{Glozman:2022zpy}.

\subsection{Phase structure of QCD}
\label{sec:PhaseStructure+CEP} 

\begin{figure*}[t]
   \begin{minipage}{0.98\columnwidth}
			\begin{tabular}{|c|c|c|}\hline\hline
				                                									& $\kappa_2$ 	& $\kappa_4$ 	\\\hline\hline
				Lattice \cite{Bonati:2015bha}	($\mu_S=0$ and $\mu_s=0$)			& 0.0135 (20)	& not computed	\\
				Lattice \cite{Bellwied:2015rza} ($n_s=0$)							& 0.0149 (21)	& not computed	\\
				Lattice \cite{Bonati:2018nut}	($\mu_s=0$)	 						& 0.0145 (25)	& not computed	\\
				Lattice \cite{Bazavov:2018mes}	($n_s=0$)	 						& 0.0120 (40)	& 0.000(4)		\\
				Lattice \cite{Borsanyi:2020fev}	($n_s=0$)	 						& 0.0153 (18)	& 0.00037(66)	\\
				Lattice \cite{Ding:2024sux}     ($n_s=0$)                           & 0.0134 (14)	& not computed	\\
				Lattice \cite{Ding:2024sux}     ($\mu_S=0$)                         & 0.0150 (10)	& not computed \\
				Lattice \cite{Ding:2024sux}     ($\mu_s=0$)                         & 0.0148 (13)	& not computed	\\[.5ex]\hline\hline
				DSE     \cite{Fischer:2014ata}	($\mu_s=0$)	 						& 0.0238 (100)	& not fitted    \\				
				fRG     \cite{Fu:2019hdw}	    ($\mu_S=0$)	 						& 0.0142 (2)	& 0.00029(2)	\\
				DSE \cite{Gao:2020qsj}	    ($\mu_S=0$)	 						& 0.0150 (7)	& 0.00029 (2)   \\	
				DSE \cite{Gao:2020fbl}	    ($\mu_S=0$)	 						& 0.0147 (5)	& 0.00028(2)    \\				
				DSE     \cite{Gunkel:2021oya}	($\mu_s=0$)	 						& 0.0167 (5)	& 0.00005(5)    
				\\				
			fRG    \cite{Pawlowski:2025jpg}	($\mu_S=0$)	 						& 0.0148 (4)	& 0.00047(5)    	\\\hline											
			\end{tabular}
   \end{minipage}
   \begin{minipage}{0.98\columnwidth}
	\includegraphics[width=.98\columnwidth]{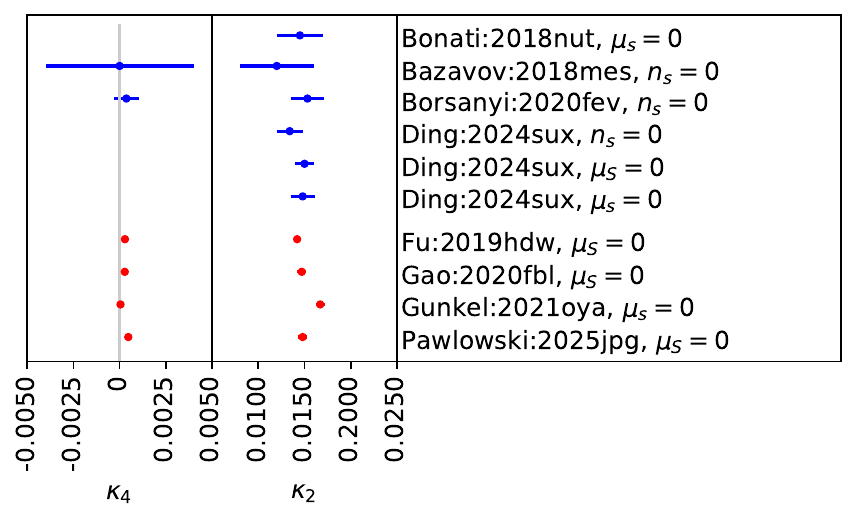} 		
   \end{minipage}
		\caption{Left: Results for the curvature coefficients $\kappa_{2,4}$ of the transition line from different approaches.
			     Whereas results with ($n_s=0$) impose strict strangeness neutrality, results with ($\mu_s=0$) come close to 
			     this condition at small baryon chemical potential. We also quote the ratios $\kappa_2(n_s)/\kappa_s(\mu_S=0)=0.893(35)$ and $\kappa_2(\mu_s=0)/\kappa_s(\mu_S=0)=0.968(23)$ from \cite{Ding:2024sux}. They can be used to compute estimates for all the $\kappa_s$'s from the given results. The functional QCD results are obtained by \textit{global} fits with \labelcref{eq:Tchicurvature}  with $\kappa_{2n}=0$ for $n\geq 3$. The error for the functional QCD results for $\kappa_{2,4}$ is the numerical error, the $\chi^2$-error per degree of freedom is $\approx 0.1$. The full combined systematic and computational error estimate for $\kappa_2$ amounts to 10\%, that for $\kappa_4$ is far higher. Note that $\mu_S=0$ implies $\mu_s=\mu_l$ in the isospin symmetric limit $\mu_Q=0$ chosen in all approaches.  Right: graphical representation of recent lattice results and the state-of-the-art functional results \cite{Fu:2019hdw, Gao:2020fbl, Gunkel:2021oya,Pawlowski:2025jpg}. 
		         \hspace*{\fill}
			}
			\label{tab:curvature}
\end{figure*}

As a preparation for the following discussion we first supply some general remarks on the curvature of the crossover line. 
At small chemical potential, this line can be parametrised by an expansion quadratic in the dimensionless ratio of chemical 
potential to temperature:
\begin{align} 
	\frac{T_\chi(\mu_B)}{T_\chi}=  1 - \kappa_2 \frac{\mu_B^2}{T_\chi^2} - \kappa_4\left( \frac{\mu_B^2}{T_\chi^2}\right)^2+\cdots \,, 
	\label{eq:Tchicurvature}  
\end{align} 
with baryon chemical potential $\mu_B$, pseudo-critical temperature $T_\chi(\mu_B)$ and $T_\chi=T_\chi(0)$. 
The expansion is quadratic, since the grand canonical QCD partition function $Z$ is symmetric with respect to 
a change of sign in $\mu_B/T$ and therefore all odd powers of $\mu_B/T$ in the expansion have to vanish. 
As has been discussed in \cite{Fischer:2018sdj}, it is interesting to consider an ellipse as a suitable 
functional form for the entire transition line including the crossover and first-order regimes, i.e.
\begin{equation}\label{eq:ellipse}
	\left(\frac{T_\chi(\mu_B)}{T_\chi}\right)^2 = 1 - 2\kappa\left(\frac{\mu_B}{T_\chi}\right)^2\,.
\end{equation}
This form features \labelcref{eq:Tchicurvature} as a low $\mu$ expansion and naturally captures all aspects of the 
$\pm \mu_B/T$-symmetry as well as constraints from the Clausius-Clapeyron relation at zero temperature. 
Provided it is at least approximately correct, it predicts that $\kappa_4 \approx \frac{1}{2} \kappa_2^2$,
which is indeed the case as can be seen from the explicit results collected in \Cref{tab:curvature}. Note that
the expansion coefficients $\kappa_{2,4,6,..}$ could be made of roughly equal magnitude if in the denominators on
the right hand side of \labelcref{eq:Tchicurvature} the more natural scale $(2-3) \pi T_\chi$ would be used instead of $T_\chi$. Indeed, such a ratio is that of the density and temperature energy scales. The further discussion of this very interesting observation is deferred to \Cref{app:kappaExpansion}, as it is outside the core topics of this review.

The above observations and the systematic error estimate discussed below in this Section and more comprehensively in \Cref{app:CEP-Estimates} suggest the following global determination of the curvature coefficients: We shall use a \textit{global} $\chi^2$-fit of $T_\chi(\mu_B)$ in the baryon-chemical potential regime with quantitative reliability for the state-of-the-art functional QCD works discussed below. In the vicinity of the chiral crossover line,  this regime extends from $\mu_B=0$ to $\mu_B/T\lesssim 4$, see \labelcref{eq:muBT4} and the respective evaluation. Even though this bound has recently been extended in \cite{Pawlowski:2025jpg} to $\mu_B/T\lesssim 4.5$, \labelcref{eq:muBT4.5}, we shall use universally $\mu_B/T\lesssim 4$ for all global fits. Moreover, \labelcref{eq:Tchicurvature} with $\kappa_{2n}=0$ for $n\geq 3$ provides already quantitative fits with a 'minimal' Ansatz. 

With the preparations of \Cref{sec:ChiralSymbreaking,sec:ConfinementDeconfinement}, as   well as the approach to the systematic error analysis summarized in \Cref{sec:Expansion+Error},  we can comprehensively review the advances in the mapping out of the phase structure with functional QCD. 

A first significant milestone was achieved in 2014 with the DSE work \cite{Fischer:2014ata} with $\mu_S=0$, see \labelcref{eq:muB-muS}. This work is based on a solution of the coupled DSEs of quark and gluon propagators. Here, the DSE for the gluon propagator ('gluon gap equation') was expanded about the Yang-Mills gluon propagator at finite temperature, neglecting the back coupling of the quark effects in the pure glue diagrams. We shall discuss this work in more details as it allows us to introduce some further notation and observables, as well as also emphasising the importance of benchmarks and self-consistent error estimates as well as the vertex expansion scheme at work.
 
As discussed in \Cref{sec:Expansion+Error}, the neglected back reaction is subleading on the very few percent level. This has been confirmed by respective lattice results at finite temperature and vanishing density and from the explicit check within functional approaches in \cite{Braun:2014ata} in the same year. The subleading nature also holds true for the density fluctuations induced by the quark loop only. The small (linear) density effects observed in the gluon propagator further support such an approximation a posteriori. 

This setup leaves us with only one further ingredient, the quark-gluon vertex. It has been shown in the vacuum, \cite{Williams:2014iea, Mitter:2014wpa, Williams:2015cvx, Cyrol:2017ewj, Gao:2021wun}, that only three of the eight transverse tensor structure are sizeable. These are denoted by $\tau_{1,4,7}$ in the notation of \cite{Gao:2021wun}, see also \cite{Eichmann:2016yit, Ihssen:2024miv} for vertex definitions with full crossing symmetry. These tensor structures can be derived from the four gauge-invariant local operators $\bar q \slashed{D}^n\,q $ with $n=1,2,3,4$. It has been shown in \cite{Mitter:2014wpa, Cyrol:2017ewj}, that the vertex dressings at the symmetric momentum point (i.e. with equal external scales at all legs) are related to exactly the tensor structures derived from the corresponding gauge-invariant local operator. This emphasises the strength of gauge-consistent constructions.
It turns out that the two most important structures are the chirally symmetric classical tensor structure $\gamma_\mu$ ($\tau_1$) and a chirally breaking one $\sigma_{\mu\nu}$, ($\tau_4$), related to the spin term $\sigma_{\mu\nu} F_{\mu\nu}$. The third important tensor structure already contributes less than 10\% and is a further chirally symmetric one ($\tau_7$), derived from $\bar q \slashed{D}^3\,q $, see e.g.~\cite{Mitter:2014wpa, Cyrol:2017ewj, Ihssen:2024miv, Gao:2021wun}. 

In \cite{Fischer:2014ata} only $\tau_1\propto \gamma_\mu$ has been considered whose dressing $\lambda_1$ has been extracted from the corresponding STI. The effects of the other two vertex structures has been accommodated by an additional infrared enhancement of $\lambda_1$ adjusted such that the chiral crossover temperature agrees with the lattice value $T_\chi\approx 155$\,MeV. Moreover, the chiral condensate at $\mu_B=0$ agrees quantitatively with the lattice predictions, see ~\Cref{fig:condensate}, which is a non-trivial benchmark test for the quantum and thermal fluctuation effects that are taken into account in \cite{Fischer:2014ata}. In turn, this justifies the computation of the density effects on the chiral crossover lines with $T_\chi(\mu_B)$, and the result allows for a comparison with the lattice benchmarks at $\mu_B=0$. 
The STI construction guarantees that the chemical potential modifications of the quark-gluon vertex are captured. In this aspect, the construction qualitatively supersedes previous ones based on simple gluon models. The respective chiral and confinement phase structure captures the qualitative aspects well. Its curvature coefficient $\kappa_2$ in \labelcref{eq:Tchicurvature} is given
in \Cref{tab:curvature}. Finally, the chiral crossover was ending in a critical end point with the location 
\begin{align}
(T, \mu_B)_\textrm{CEP} = \bigl(115 \, ,\, 504\bigr)\,\textrm{MeV}\,,  
\label{eq:Fun-CEP-2014}
\end{align}
for $N_f=2+1+1$ flavours and very similar for $N_f=2+1$. These results, and in particular that for $\kappa_2$ have to be 
confronted with the curvature results from the lattice, see table \Cref{tab:curvature}. 
Evidently, the curvature coefficient from \cite{Fischer:2014ata} exceeds the lattice curvature by far. Thus, in the end,
we will not include the results of \cite{Fischer:2014ata} in our selection of works with highest quality truncations. 

%
\begin{figure}[t]
	\begin{center}
		\includegraphics[width=0.48\textwidth]{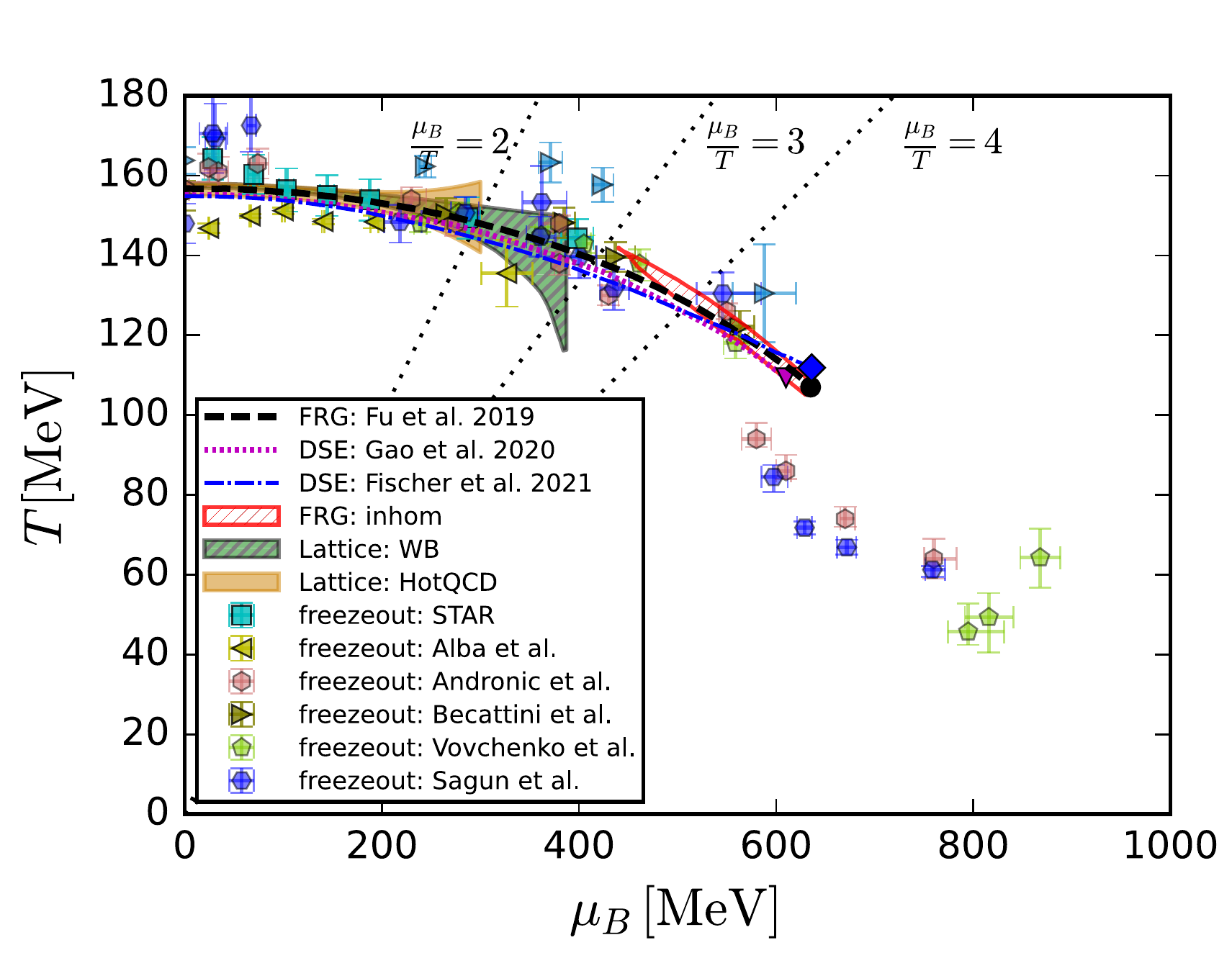} 
		\caption{Crossover lines from functional QCD and lattice QCD. FRG: Fu \textit{et al.} \cite{Fu:2019hdw}, DSE: Gao \textit{et al.} \cite{Gao:2020fbl}, DSE: Gunkel \textit{et al.} \cite{Gunkel:2021oya}. Lattice: Wuppertal-Budapest Collaboration
			\cite{Bellwied:2015rza} (WB) and HotQCD Collaboration \cite{Bazavov:2018mes} (HotQCD). The red hatched area denotes a region of inhomogeneous instability obtained in fRG \cite{Fu:2019hdw}, known also as the moat regime \cite{Pisarski:2021qof}. The estimates of the CEP from functional methods are in the regime \labelcref{eq:FunEstimateCEP-Recall}. While all come from functional approaches, the respective approximations and resummations are different, which further improves the respective reliability. We have also added freeze out points: STAR \cite{Adamczyk:2017iwn}, Alba \textit{et al.} \cite{Alba:2014eba}, Andronic \textit{et al.} \cite{Andronic:2017pug}, Becattini \textit{et al.} \cite{Becattini:2016xct}, Vovchenko \textit{et al.} \cite{Vovchenko:2015idt}, Sagun \textit{et al.} \cite{Sagun:2017eye}. Note that freeze-out data from Becattini \textit{et al.} are shown in two different colours, corresponding to with (light blue) and without (dark green) afterburner-corrections, respectively. \hspace*{\fill}}
		\label{fig:QCDPhasediagram_2025} 
	\end{center}
\end{figure}
%

The next milestone has been achieved in the fRG work \cite{Fu:2019hdw}. In this work the fRG-approach of \cite{Braun:2014ata, Rennecke:2015eba} was extended to finite temperature and density for a computation of the chiral and confinement phase structure for $\mu_S=0$ in \labelcref{eq:muq}, see \Cref{fig:QCDPhasediagram_2025}. 
In \cite{Braun:2014ata, Rennecke:2015eba}, different avatars of the strong coupling have been computed in a self-consistent expansion about the Yang-Mills results, including the feed-back of the quark effects in the pure glue sector. The tensor structures of the primitively divergent vertices have been reduced to the classical ones and the dressings are not taken fully momentum dependent but carry an average momentum dependence in the RG-cutoff scale. It has been shown that this resolves effectively the momentum dependence of the vertices at the symmetric point. The quantitative nature of the approximation scheme in the vacuum has been tested thoroughly, in particular within a comparison with fully momentum-dependent computations in very advanced approximations showing apparent convergence, \cite{Mitter:2014wpa, Cyrol:2017ewj}. Very recently, this test has been extended in a series of papers to the full momentum dependence of four-quark vertices, including the dependence on all Mandelstam variables as well as a thorough test of the angular momentum, see \cite{Fu:2022uow, Fu:2024ysj, Fu:2025hcm}. The computation in \cite{Braun:2014ata} also allowed to estimate self-consistently the effects of neglecting the (higher order) feedback of quark loops onto pure Yang-Mills diagrams by comparing the computations with and without feedback. In short, the effects were negligible if the RG-point $\mu_R$ was taken at sufficiently small momenta $\mu_R \lesssim 10$\,GeV, providing independent support for the approximation used in \cite{Fischer:2014ata}. It is worth emphasising that due the \LEGO-principle, the systematic error estimate of such an approximation estimated in one approach (here the fRG computation) can directly be applied as well in the other approach (here the DSEs). 

The direct comparison of the approximations in \cite{Fischer:2014ata} and \cite{Braun:2014ata, Rennecke:2015eba, Fu:2019hdw} faces some intricacies, both due to the fRG vs DSE diagrammatics but also due to the fact that \cite{Braun:2014ata, Rennecke:2015eba, Fu:2019hdw} have used the fRG-approach with emergent composites, \cite{Gies:2001nw, Gies:2002hq, Pawlowski:2005xe, Floerchinger:2009uf, Fu:2019hdw, Fukushima:2021ctq}. Specifically this entails that the quark-meson loop in the flow of the quark propagator in \Cref{fig:QuarkGapfRG} covers part of the spin part of the quark-gluon vertex. Indeed, while the computation in \cite{Fu:2019hdw} still contains an infrared enhancement factor for precisely matching the full functional QCD results in the vacuum, \cite{Mitter:2014wpa, Cyrol:2017ewj, Gao:2021wun, Fu:2025hcm}, it is reduced to $\lesssim 3\%$. This leads to a prediction of the chiral crossover temperature at vanishing density with 
\begin{align} 
	    T_\chi= 156\,\textrm{MeV}\,, 
\end{align}
in quantitative agreement with lattice results, for recent ones see e.g.~\cite{Borsanyi:2020fev, Ding:2024sux}. Moreover, the prediction of the curvature coefficient, also shown in \Cref{tab:curvature}, was 
\begin{align} 
	 \kappa_2(\mu_S=0) = 0.0142(2)\,,
	 \label{eq:Fun-kappa-2019}
\end{align}
where the error is a purely numerical one, and the very conservative systematic error estimate 10\% in \cite{Fu:2019hdw} has to be added to the numerical error. This leads us to the final results with the full error 
\begin{align} 
	\kappa_2(\mu_S=0) = 0.0142(14)\,,
	\label{eq:Fun-kappa-2019FullError}
\end{align}
\Cref{eq:Fun-kappa-2019FullError} is compatible with the lattice results in \Cref{tab:curvature}, and the results for the chiral condensate agree quantitatively with the lattice results as for \cite{Fischer:2014ata}. The chiral crossover line in \cite{Fu:2019hdw} ends in a critical end point, whose location is at 
\begin{align}
	(T, \mu_B)_\textrm{CEP} = \bigl(107\,,\,635\bigr)\,\textrm{MeV}\,, 
	\label{eq:Fun-CEP-2019}
\end{align}
with the ratio $\mu_B/{T}\approx 5.9$. 

\begin{figure}[t]
	\centering
	\includegraphics[width=0.98\columnwidth]{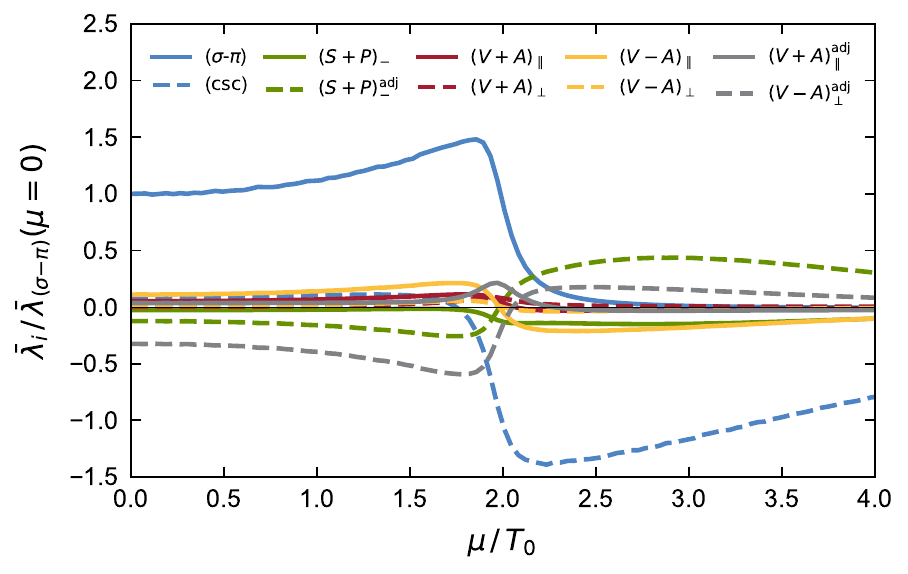}
	\caption{strength of the dressings in a Fierz-complete $N_f=2$ flavour computation, reprinted from \cite{Braun:2019aow}.\hspace*{\fill}}
	\label{fig:FierzcompleteDominance}
\end{figure}
\Cref{eq:Fun-CEP-2019} has been assessed in \cite{Fu:2019hdw} as an estimate and not a bona fide prediction due to the following systematic error analysis: As discussed above, in the fRG-approach the driving element for dynamical chiral symmetry breaking and its restoration along the phase boundary is the four-quark interaction. The relative relevance of its different components has been assessed in a Fierz-complete tensor basis in \cite{Braun:2019aow}, see \Cref{fig:FierzcompleteDominance}. This analysis entails that for baryon chemical potentials $\mu_B/T \lesssim 7-8$ the scalar-pseudoscalar channel (straight blue line) is the only relevant one. This can be already seen from the strength comparison depicted in \Cref{fig:FierzcompleteDominance}. In fact, the analysis in \cite{Braun:2019aow} reveals that in integrated flows all other tensor structures can be dropped completely without changing the results. This changes only for larger $\mu_B$, where a diquark channel (dashed blue line) takes over. Thus the leading order approximation of the four-quark interaction used in \cite{Fu:2019hdw} only introduced a very small systematic error. 

This leaves us with only one major error source, the potential occurrence of a moat regime for $\mu_B/T \gtrsim 4$. In \cite{Fu:2019hdw} an early sign of such a possibility showed up computationally: the momentum slope of the pion and $\sigma$ wave functions turned negative for larger chemical potentials and sufficiently large temperatures, however without instability. The latter is signalled by a emergence of a zero in the inverse pion and $\sigma$ propagator, see e.g.~\cite{Tripolt:2017zgc} for an LEFT-study. The full physics understanding of the moat regime was achieved in the seminal work \cite{Pisarski:2021qof}, and the analysis in \cite{Fu:2019hdw} was extended with analysis of the spectral properties in the moat regime in \cite{Fu:2024rto}. 

In general, we consider the fact, that the moat regime was first detected numerically, as yet another proof of the capacity of functional QCD to unravel novel features in the phase structure. The moat regime intersects with the chiral crossover line at $	\mu_B/T\approx 4$, 
see \Cref{fig:QCDPhasediagram_2025}. Beyond this point a fully quantitative analysis of the phase structure requires the full momentum resolution of propagators and vertices. This was not included in \cite{Fu:2019hdw}, which led to the following conservative systematic error assessment: for 
\begin{align} 
	\mu_B/T\lesssim  4\,,
	\label{eq:muBT4} 
\end{align} 
the results are quantitatively reliable. In turn, for larger baryon chemical potentials the non-trivial momentum dependence of pion and $\sigma$-propagators has to be fully feed-back in order to achieve self-consistency required for a small systematic error budget. 

\begin{figure}[t]
	\begin{center}
		\includegraphics[width=0.48\textwidth]{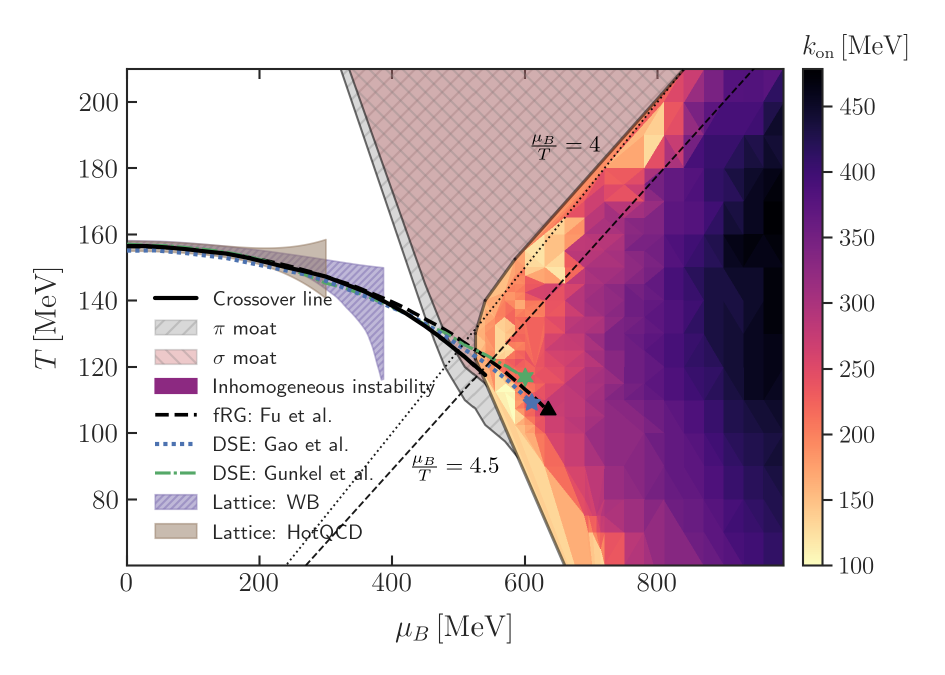} 
		\caption{Update of the QCD Phase structure of 2+1 flavour QCD from \cite{Pawlowski:2025jpg}: In comparison to \Cref{fig:QCDPhasediagram_2025} we also display the moat regimes of the pions (hatched grey) and the sigma mode (hatched red). Moreover, the region with signatures of inhomogeneous condensation is shown with a heatmap whose colour indicate value of $k_\textrm{on}$ which is the lowest value of the RG-scale that can be reached before the instability terminates the flow. The current computation pushes the quantitative reliability bound of functional QCD to $\mu_B/T\approx 4.5$ (dashed black line) with 10\% accuracy. We also show the previous bound $\mu_B/T\approx 4$ (dotted black line). Further results: Black dashed line \cite{Fu:2019hdw} (fRG, fQCD), \cite{Gao:2020fbl} (DSE, fQCD), dashed green line \cite{Gunkel:2021oya} (DSE). Violet area \cite{Bellwied:2015rza} (lattice, WB), brown area \cite{Bazavov:2018mes} (lattice, HotQCD). Figure reprinted from \cite{Pawlowski:2025jpg}. \hspace*{\fill}}
		\label{fig:PhaseDiagramMoat} 
	\end{center}
\end{figure}

Such a self-consistent computation has been done very recently in \cite{Pawlowski:2025jpg}, were a successively deeper moat with increasing baryon chemical potential was reported. For $\mu_B/T \gtrsim 4.5$ this deepening even led to potential instabilities and hence a potential  inhomogeneous regime. The resolution of this regime with a potential inhomogeneity requires further attention and is subject of ongoing investigations. In summary, the results in \cite{Pawlowski:2025jpg} push  the regime with quantitative reliability of functional QCD to 
\begin{align} 
	\mu_B/T\lesssim  4.5\,, 
	\label{eq:muBT4.5} 
\end{align} 
see \Cref{fig:PhaseDiagramMoat}. There, also the separate moat regimes for pion and $\sigma$-mode are shown. The regime with the potential instability is shown as a heatmap which indicates the emergence of the instability at the respective cutoff scale. For more details we refer to \cite{Pawlowski:2025jpg}.
%
%

Before we discuss the range for the combined fRG/DSE estimates, we discuss the progress of the DSE works \cite{Gao:2020qsj, Gao:2020fbl, Gunkel:2021oya} (DSE) in comparison to the early milestone \cite{Fischer:2014ata}. 

In \cite{Gao:2020qsj}, difference DSEs where solved for vertices and gluon propagator. The base point used where the $N_f=2$ flavour fRG vacuum results from \cite{Cyrol:2017ewj}, hence the name fRG-assisted DSE. In short, both the fluctuations of the strange quark as well as the thermal and density fluctuations of the vertices and gluon propagator where computed. In contradistinction, the quark gap equation was solved without using the $N_f=2$ flavour base point. It is worth noting in this context that the computation in \cite{Gao:2020qsj} is a DSE computation and its diagrammatic systematics is that of the DSE-approach. Indeed, in later works in the miniDSE scheme set up in \cite{Lu:2023mkn} the $N_f=2+1$ flavour DSE results from \cite{Gao:2021wun} where used as the base point.  The qualitatively non-trivial extension of the previous approximations in \cite{Gao:2021wun} of the vertex expansion used in fRG and DSE works at finite temperature and/or density, was the inclusion of all transverse tensor structures of the quark-gluon vertex instead of only the classical one, e.g.~\cite{Fischer:2014ata} or the classical one and part of the chirally breaking tensor structures in \cite{Fu:2019hdw}. This extension leads to quantitative results in the vacuum as the input from \cite{Cyrol:2017ewj} used in \cite{Gao:2020qsj} was obtained with all tensor structures of the quark-gluon vertex. It also leads to predictions of the chiral crossover temperature $T_\chi=154 $\,MeV at vanishing $\mu_B$ as well as the curvature $\kappa_2 = 0.0150(7)$ in agreement with the lattice benchmarks and \cite{Fu:2019hdw}. The corrections of the quark-gluon vertex at finite temperature and baryon chemical potential where determined with a combination of STIs and RG-scaling arguments. The ensuing phase structure agrees with that from \cite{Fu:2019hdw} except for the location of the critical end point at $(T, \mu_B)_\textrm{CEP} = (93\,,\,672)\textrm{MeV}$, that is at $\mu_B/T =7.23$. The computation in \cite{Gao:2020qsj} also revealed an inherent and remarkable stability of the location of the chiral crossover line, and the only volatile result was that of the location of the CEP on the fixed crossover line. 

These results triggered a follow-up work, \cite{Gao:2020fbl}, where the STI and RG-scaling determination of the thermal and density corrections of the dressings of the quark-gluon vertex was dropped and these corrections where computed. The predictions for the chiral crossover temperature $T_\chi=155.1$\,MeV at vanishing $\mu_B$ as well as the curvature $\kappa_2 = 0.0147(5)$ are basically unchanged from \cite{Gao:2020qsj} and are in agreement with the lattice benchmarks and \cite{Fu:2019hdw}. The respective location of the CEP shifted on the fixed chiral crossover curve to smaller values with $(T, \mu_B)_\textrm{CEP} = (109\,,\,610)\textrm{MeV}$, that is $\mu_B/T\approx 5.6$. 

Finally, the DSE work \cite{Gunkel:2021oya} improved upon the earlier milestone \cite{Fischer:2014ata} by taking into account composite contributions in the pseudoscalar and scalar exchange channel using Bethe-Salpeter amplitudes for the pion and $\sigma$-mode \cite{Fischer:2007ze}. As in the fRG with emergent composites, this accommodates for all chirally breaking tensor structures in the quark-gluon vertex. Moreover, it explicitly introduces the soft pion mode into the system, which is missing in the approximation underlying earlier DSE works \cite{Fischer:2014ata, Gao:2020fbl, Gao:2020qsj}. 

To summarise: the most advanced truncation schemes in the fRG \cite{Fu:2019hdw, Pawlowski:2025jpg}, fRG-assisted DSE \cite{Gao:2020fbl} and lattice-assisted DSE
approaches \cite{Gunkel:2021oya} all have different systematic errors due to (slightly) different truncation schemes. They all share the common feature that they systematically take into account the back-reaction of the quarks onto the Yang-Mills sector and they incorporate the most important parts of the quark-gluon interaction including pseudoscalar and scalar composites. Moreover, they all share the general caveat that a possible inhomogeneous phase is not yet properly taken into account, while the moat regime is only fully taken into account in \cite{Pawlowski:2025jpg}. Taken together, they serve as cross-checks for each other and it is a remarkable and non-trivial result, that the estimates from \cite{Fu:2019hdw, Gao:2020fbl, Gunkel:2021oya} cluster in a very narrow regime,  
\begin{align}
	(T, \mu_B)_\textrm{CEP} \in \bigl(115-105\,,\, 600-650\bigr)\textrm{MeV}\,, 
	\label{eq:FunEstimateCEP-Recall}
\end{align} 
already quoted in the introduction, \labelcref{eq:FunEstimateCEP}. \Cref{eq:FunEstimateCEP-Recall} lies in the regime with a potential instability revealed in \cite{Pawlowski:2025jpg}. However, if the CEP survives in this regime, the results in \cite{Pawlowski:2025jpg} corroborate \labelcref{eq:FunEstimateCEP-Recall}. Finally,
while it is difficult to nail down this combined reduction of the systematic error on a quantitative basis, it certainly is very reassuring.

\subsection{Critical end point: Estimates and predictions}
\label{app:CEP-Estimates}

In the previous Section we have discussed the development of the functional QCD analysis of the (chiral) phase structure of QCD over the past decade, cumulating in \Cref{fig:QCDPhasediagram_2025} with the state-of-the-art results from fRG and DSE, \cite{Fu:2019hdw, Gao:2020fbl, Gunkel:2021oya} as well as the estimate for the location of the critical end point \labelcref{eq:FunEstimateCEP-Recall}. In the present Section we put these results in the context of results from other methods. While \labelcref{eq:FunEstimateCEP-Recall} may also be the location of new phases such as an inhomogeneous regime in QCD, in the following we concentrate on the CEP. 

`Scatter' plots of the QCD phase diagram often include estimates and predictions of potential CEP positions from many different sources on a democratic level. This equal-importance evaluation ignores qualitative differences between different predictions. In order to provide some guidance here, the following grouping may be helpful:
\begin{itemize}
	\item[(A)] Functional QCD,\\[-1ex]
	\item[(B)] Analytic continuations of lattice QCD results at $\mu_B=0$,\\[-1ex]
	\item[(C)] Low-energy effective theories (LEFTs).
\end{itemize}
As already explained in great detail in this mini-review, high quality estimates from functional methods, class (A), share the unique feature that they include quark and glue dynamics at finite temperature and chemical potential. Functional QCD does this at the cost of approximation of the full diagrammatic system, but it allows for a systematic improvement and the evaluation of apparent convergence of the truncation scheme \cite{Fischer:2018sdj, Dupuis:2020fhh, Fu:2022gou, Rennecke:2025bcw}. On the other hand, analytic continuations, class(B), of lattice QCD results at $\mu_B=0$ provide high quality results at small chemical potential with rapidly increasing systematic error for $\mu_B/T > 3.5$. In principle, analytic continuations are not able to pick up new dynamical information. Thus regions with new physics such as a moat regime or inhomogeneous phases cannot be detected by analytic continuation. Taking into account the singularity structure of the complex chemical potential plane, recent studies of Lee-Yang zeros made considerable progress towards the identification of a possible CEP location \cite{Basar:2023nkp, Clarke:2024ugt}, but it seems fair to say that much more work is needed until these results can be transformed into reliable predictions. Finally, there are low-energy effective theories, i.e. theories that are designed to capture important properties of QCD without invoking its full dynamics. Examples for this class of theories are simplified truncations of functional QCD  such as \cite{Fu:2023lcm} (QCD-assisted LEFTs), low energy effective models such as the (Polyakov-)quark-meson model (PQM) or the (Polyakov-)Nambu-Jona-Lasignio model (PNJL) (see e.g.~\cite{Fukushima:2017csk} and references therein), both with restrictive dynamics of the Yang-Mills sector. At the other end of this spectrum are black-hole engineering approaches \cite{Hippert:2023bel} that rely on chiral potentials deduced from lattice input at vanishing baryon chemical potentials. Similarly to investigations of Lee-Yang zeroes these approaches are insensitive to emergent novel chiral dynamics at finite chemical potential that is not put in at $\mu_B=0$.

In short, LEFTs may be qualitatively and even quantitatively successful in certain applications and serve as important toy models in others. But, ultimately, by definition they lack the rigorousness of the full functional approach. Their inherent systematic error is therefore always larger. For a more detailed discussion see also~\cite{Fu:2023lcm}.

\section{Pinning down the CEP}
\label{sec:ConservedFlucs+freezeout}

The QCD phase structure, derived from functional QCD, results in the exciting situation, that QCD does not exhibit a chiral crossover for $\mu_B> 700$\,MeV for $n_s=0$. Moreover, the regime with $\mu_B=600 - 700$\,MeV contains a rapid change from the chiral crossover to either a first order regime with a critical end point or Lifshitz point or a regime with inhomogeneous instabilities. In the following we shall call this regime the onset regime of new phases (ONP). In all these cases it is of paramount importance to assess the nature and size of this transition regime with tailor-made observables. In this Section we discuss the dominant dynamical modes and effects in this regime and the size of regimes with critical scaling, \Cref{sec:SoftModes+Criticality}, observables such as the fluctuation of conserved charges and thermodynamics that either carry direct signatures of this onset regime of new phases or are an important input in the transport or hydrodynamical phase in this regime, \Cref{sec:Flucs+Thermodyn}. The final step concerns the translation of observables in equilibrium QCD to their experimental  signatures that carry the physics at chemical freeze-out.

\subsection{Soft modes and the size of QCD critical regions }
\label{sec:SoftModes+Criticality}

The enhancement of experimental signals in scaling regimes (critical regimes) around second order phase transitions, and in particular the potential critical endpoint in QCD, has triggered a prolonged interest in these experimental signals. One of these signals is provided by a non-monotonic behaviour of the fluctuations of conserved charges to be discussed in \Cref{sec:Flucs+Thermodyn}. 

Commonly these non-monotonicities are  viewed as a smoking gun for the critical endpoint. It has been argued, though, that these non-monotonicities also occur in the absence of a critical end point. Indeed, there are many studies in LEFTs that explicitly accommodate the latter scenario. While these low-energy effective theories are not QCD, some of these studies are self-consistent QFT-studies including quantum, thermal and density fluctuations, for an  important early fRG-work see \cite{Schaefer:2006ds}. In particular the fRG as a renormalisation group method allows for a quantitative grip on the anomalous scaling and its absence, for a comprehensive overview see \cite{Dupuis:2020fhh}. In short, the results from these analyses seem to invalidate the smoking gun property of the non-monotonicities or other properties that occur in critical scaling regimes, as they also can occur outside these regimes. In turn, a smoking gun for a critical regime would be the proof of anomalous scaling in these regimes. However, this scaling is even exceedingly difficult to establish in clean table top experiments designed for the study of critical systems. 

It has been pointed out that one may devise observables that are more sensitive to the critical scaling. One of these observables are $n$th order fluctuations of conserved charges whose divergence at the critical end point increases with $n$, each further order enhancing the singularity by order one. However, this better resolution of the singularity is going hand in hand with a dramatic rise of the statistics needed (increasing with powers of the volume) which probably makes this a hopeless task, both for the experimental measurements as well as for lattice simulations. 

We stress, that the situation is qualitatively different for the functional approach to QCD. It is built on analytic diagrammatic methods and is not based on Monte Carlo sampling or experimental statistics. It is important to keep these differences in mind if results from and  challenges for different QCD approaches are compared: In order to understand these differences, one has to appreciate the fact that in functional QCD, thermodynamic observables such as the pressure $p$, entropy $s$, energy density $\epsilon$ and the trace anomaly $I$, as well as fluctuations of conserved charges are derived directly from analytic expressions for the grand potential $\Omega$ or its first chemical potential derivatives. In most cases the baryon density $n$ is used, but in general derivatives with respect to single quark chemical potentials are the starting points. These first derivatives are taken analytically, and also further ones can be taken analytically in a recursive way, see \cite{Fu:2015naa, Faigle-Cedzich:2023rxd}. However, typically further ones are taken numerically as the analytic expressions grow rapidly. Consequently, the computation of the grand potential (containing the effective potentials of the order parameters) has to be done in a numerically accurate way that allows for numerical high order derivatives, see e.g.~\cite{Wagner:2009pm, Karsch:2010hm}. For example, the evaluation of the scaling regime in the chiral limit required the relative numerical error of $10^{-8}-10^{-10}$, see the fRG study \cite{Braun:2023qak}. Only with this accuracy conclusive statements  about the size of the critical scaling regime were possible from the scaling behaviour of the effective potential of the chiral order parameter. This accuracy also allows for the extraction of hyper-fluctuations of conserved charges to a relatively high order. 

However, it has been also shown in \cite{Braun:2020ada, Braun:2023qak} and the accompanying DSE work \cite{Gao:2021vsf} that, while the critical scaling regime around the second order chiral phase transition is tiny, the regime with \textit{soft modes} (pions) is rather large. Here, soft modes are dynamical modes that are quasi-massless if compared with the QCD scale $\Lambda_\textrm{QCD}$ and the pion decay constant or chiral condensate.  Naturally, soft modes can and do give rise to large signals such as the rise of fluctuations observables or non-monotonicities without showing critical behaviour. 

In a broader sense, one may also call the regime with these signals 'critical' and more recently this notion has been used. We emphasise, thought, that this is not only an abuse of the well-defined notion of the technical term critical regime in the theory of critical phenomena; it also diffuses the fact that quantitatively reliable expansions and extrapolations in regimes with critical and non-critical components require a far larger statistics for convergence than \textit{regular} expansions in  non-critical regimes. Accordingly, these two regimes should not be mixed up. This suggests to use the term \textit{soft mode regime} (SMR) for the regimes in the phase structure dominated by the dynamics of soft modes. For small and medium size baryon chemical potential this is the hadronic regime with pions being the soft modes including the validity regime of chiral perturbation theory, see \cite{Gao:2021vsf}. Sufficiently close to the critical end point or rather the onset of new phases (ONP)  the lowest lying scalar mode ($\sigma$) is taking over and this change of mass order is heralding the approach to the ONP. Indeed this has been seen in \cite{Pawlowski:2025jpg}, where the $\sigma$-mode is turning lighter than the pion already within the moat regime, close to the regime with a potential instability. 

Finally, we would like to stress again that the distinction discussed in the previous paragraph is not mere semantics but also has important consequences for the systematic error estimates for functional QCD: In the \textit{proven} absence of critical scaling in a given regime, quantitatively reliable approximation in functional methods do \textit{not} have to accommodate this potential scaling. As indicated above, this not only significantly lowers the need for numerical accuracy but also the need for all-order scatterings of the critical modes. We shall use this property in the studies of fluctuation observables in \Cref{sec:ConservedFlucs+freezeout} as well as for the discussion of the Columbia plot in \Cref{sec:TheoryCover}.  

In summary, 
it may be a stretch of imagination to hope for the experimental proof of critical scaling  in heavy ion collision with its intricate non-equilibrium and equilibrium dynamics along its time line. While this may look discouraging at first, the absence of humongous regimes with critical scaling together with the presence of a rather large soft mode regime is rather a benefit for the task of pinning down the onset of new phases:\\[-2ex]
 
If we assume that the regime with critical scaling is exceedingly small, the location of the ONP can be inferred from the extrapolation of experimental observables within a simple Taylor expansion or rational expansion schemes. In particular, in this scenario there is no necessity to accommodate both, a critical part with \textit{anomalous} scaling and a regular part. The latter scenario would necessitate an accurate determination of the relative strengths of these parts, which is getting exponentially difficult with the increasing distance to the critical regime which is dominated by the critical scaling part. 
In short, the accuracy needed to pin down the critical scaling regime with anomalous scaling is in one-to-one correspondence to the accuracy needed for the computation of higher orders, which is related to the higher statistics needed for their experimental measurement or their determination within lattice simulations. 

This leaves us with a clear task for combined experimental and theoretical studies: the location of the ONP will only be revealed by salvaging a combination of all observables throughout a large temperature and chemical potential regime around the critical end point and the related experimental signatures for a respective large $\sqrt{s}$-regime on the freeze-out line. Moreover, on top of the equilibrium QCD computations on the theory side, the whole timeline of HICs has to be resolved accurately. The latter part requires kinetic, transport and hydrodynamical analyses that are firmly anchored in QCD, built upon the thermodynamic and fluctuations observables as well as the order parameter potentials and dispersion relations. 

%
\begin{figure}[t]
	\includegraphics[width=0.5\textwidth]{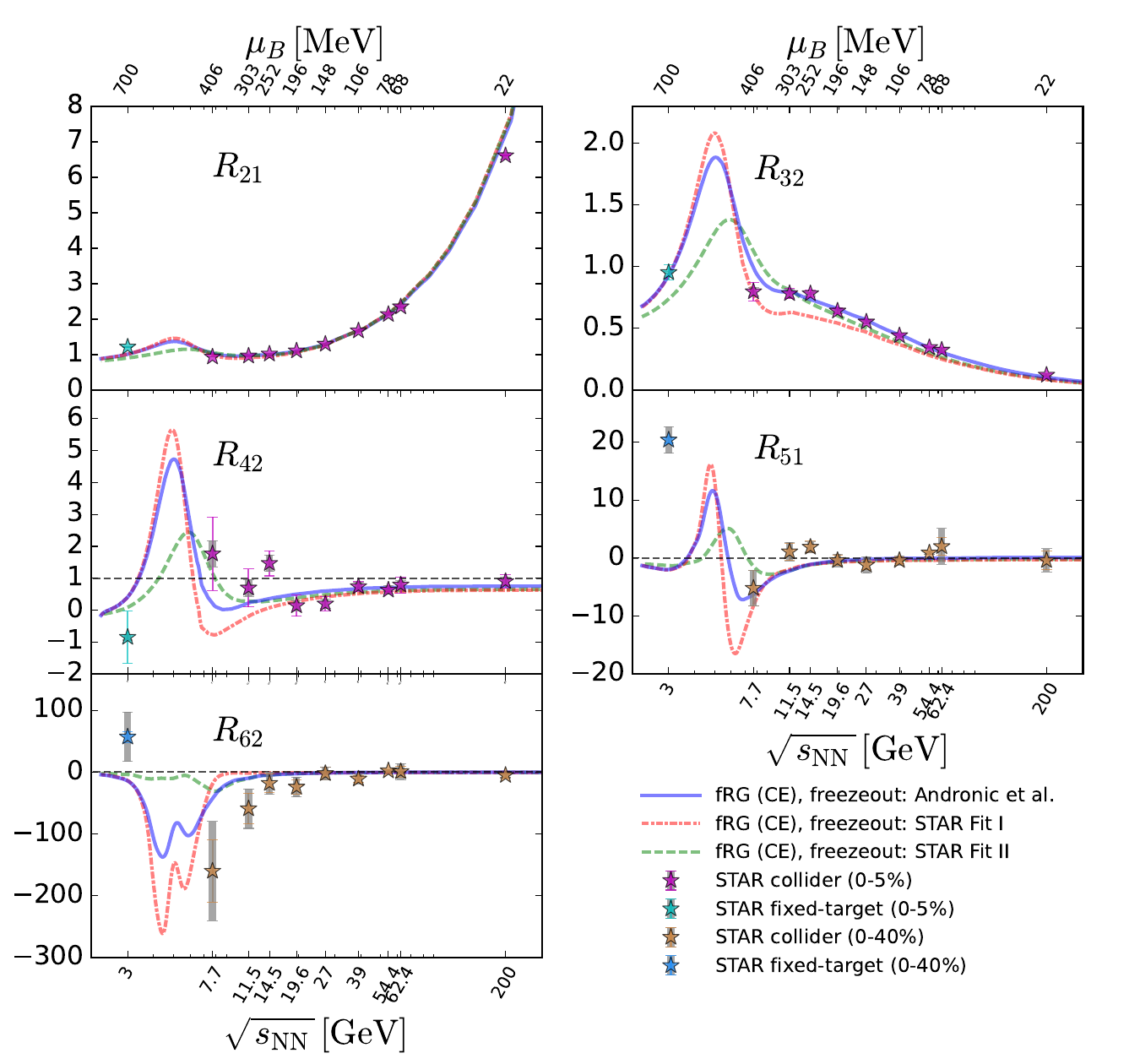}
	\caption{Baryon number fluctuations of different orders as functions of the collision energy, calculated in the QCD-assisted LEFT with the fRG on three different freeze-out curves \cite{Fu:2021oaw}, where the effects of global baryon number conservation are taken into account in the range of $\sqrt{s_{\mathrm{NN}}}\lesssim$\,11.5\,GeV through SAM \cite{Vovchenko:2020tsr}, see text for more details. Experimental data measured by STAR are also presented for comparison~\cite{STAR:2020tga, STAR:2021rls, STAR:2021fge, STAR:2021iop, STAR:2022vlo}. Plot taken from \cite{Fu:2023lcm}. \hspace*{\fill}}
	\label{fig:chi-CE}
\end{figure}
%

\subsection{Fluctuations of conserved charges and thermodynamics}
\label{sec:Flucs+Thermodyn}

Based on the functional QCD results on the phase structure of QCD and specifically the location of the crossover line, thermodynamics and fluctuations of conserved charges, in particular the baryon charge, are readily computed. While thermodynamic observables  are computed from the grand potential $\Omega$ and its $T$- and $\mu_B$-derivatives, $n$th order fluctuations of the baryon charge require $n$th order derivatives with respect to $\mu_B$ and are increasingly sensitive to the dynamics of soft modes. With the discussion of the last Section this entails that both functional approaches, fRG and DSE, can be used for the computation of thermodynamic observables. For the hyper-fluctuation of the baryon charge the current approximations of the DSE do not fully accommodate the dynamics of soft modes, for a discussion see \cite{Gao:2021vsf}. However, the importance of the latter for hyper-fluctuations can be assessed within fRG computations and has been found to be sub-leading. In conclusion, this supports the quantitative reliability of the DSE results. 
\begin{figure}[t]
	\centering
	\includegraphics[width=0.86\columnwidth]{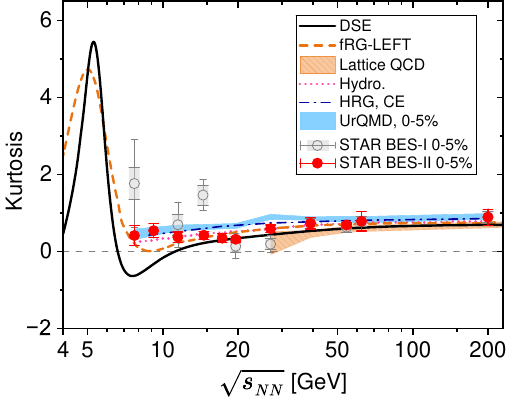}
	\caption{Kurtosis $\chi_4^B/\chi_2^B$ along the freeze-out line~\cite{Andronic:2017pug} as a function of collision energy $\sqrt{s_{NN}}$. We also show results with a canonical ensemble within the QCD-assisted fRG-LEFT~\cite{Fu:2023lcm}, the lattice QCD result~\cite{Bazavov:2020bjn}, the STAR BES-I~\cite{STAR:2020tga} and BES-II~\cite{STAR:2025zdq} data for the net-proton cummulant ratio $C_4/C_2$, together with baselines from HRG~\cite{Braun-Munzinger:2020jbk}, UrQMD~\cite{STAR:2021iop} and hydrodynamic simulations~\cite{Vovchenko:2021kxx}. Plot taken from \cite{Fu:2023lcm}. \hspace*{\fill}}
	\label{fig:R42-Frz}
\end{figure}

Moreover, the computation of the observables under discussion, in particular that of the fluctuations of conserved charges, is best done in a self-consistent gluonic background $\langle A_0\rangle \neq 0$: here, \textit{self-consistent} refers to an $A_0$-background that solves the gluon equations of motion and hence is the expectation value of $A_0$. We rush to add that this $A_0$-background is gauge-invariant: it is defined via the eigenvalues $\nu_i$ of the algebra element of the Polyakov loop $P(\vec x)$ and can be directly mapped to $\nu(A_0)$ in specific gauges. For more details see the recent work \cite{Lu:2025cls}. We also note that the expectation value of the traced Polyakov loop $L=1/N_c \textrm{tr} P(\vec x)$  or rather its susceptibility is commonly related to the confinement-deconfinement crossover. However, the Polyakov loop is only an order parameter or pseudo-order parameter in the heavy quark limit of QCD, and physical current quark masses are far away from this limit. In our opinion the combination of different observables and investigations rather hints at a strongly correlated temperature regime between the chiral crossover temperature $T_\chi$ and $(2-3)T_\chi$ without (complete) deconfinement. 

With these intricacies in mind, we note that the self-consistent gluon background $\langle \nu\rangle = \nu(\langle A_0\rangle )$ or $L(\langle \nu\rangle)$ is instrumental for results for the fluctuations of conserved charges that even qualitatively match the full ones. For instance, the kurtosis of baryon charge, $\kappa_B$, measures the transition from a regime with quark-dynamics at high temperature to one with baryon dynamics at low regimes. The kurtosis counts dynamical degrees of freedom and the transition is signalled by the limits $\kappa_B(T\to \infty) =1 /9$ ($1/3^2$ for quarks with $3$ colours)and  $\kappa_B(T\to 0) =1$ (baryon). The transition happens at about $T_\chi$ and it goes hand in hand with a peak of the susceptibility of the Polyakov loop  $L(\langle \nu\rangle)$ at about this temperature.  

State of the art results from fRG and DSE can be found in \cite{Fu:2023lcm} (fRG) and \cite{Lu:2025cls} (DSE), relevant preparatory work is provided in \cite{Fu:2015amv, Fu:2016tey, Fu:2021oaw, Isserstedt:2019pgx}. While the DSE computations are performed directly in functional QCD, the fRG computations have been performed in \textit{QCD-assisted} low-energy effective theories: The modularity of functional equations (the \LEGO-principle in \Cref{sec:Expansion+Error}) allows us to insert couplings from full functional QCD computations into QCD low-energy effective theories for the computation of observables such as fluctuations of conserved charges which are dominated by the matter fluctuations fully taken into account in the LEFT itself. These LEFTs are called \textit{QCD-assisted}: the allow for quantitatively reliable computations with a significantly reduced numerical cost. 

We proceed with the discussion of selected results, for a comprehensive discussion we refer to the works \cite{Fu:2021oaw, Fu:2023lcm, Lu:2025cls}. In \Cref{fig:chi-CE} we depict baryon number fluctuations of different orders on the freeze-out line based on the canonical ensemble taken from \cite{Fu:2023lcm}. These results are in surprisingly good agreement with the experimental results for proton number fluctuations, given all the caveats: baryon vs proton number, lack of non-equilibrium effects and the ambiguities in the determination of the freeze-out line, and more. These results are corroborated by functional QCD computations with DSEs in \cite{Lu:2025cls}, see \Cref{fig:R42-Frz}. 

\begin{figure}[t]
\includegraphics[width=0.35\textwidth]{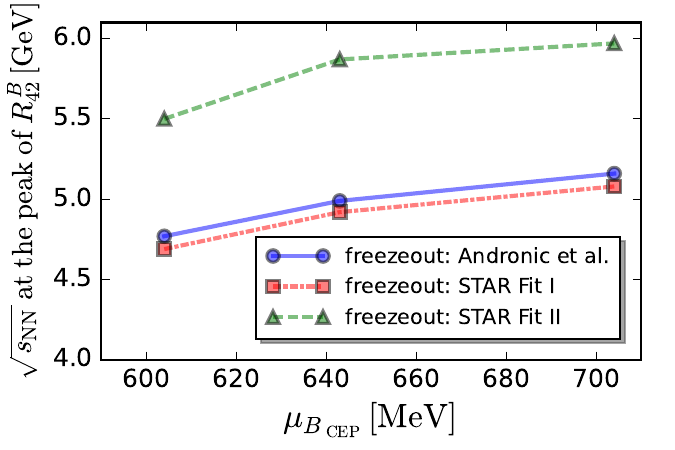}\\ \vspace{-0.14cm}
\includegraphics[width=0.35\textwidth]{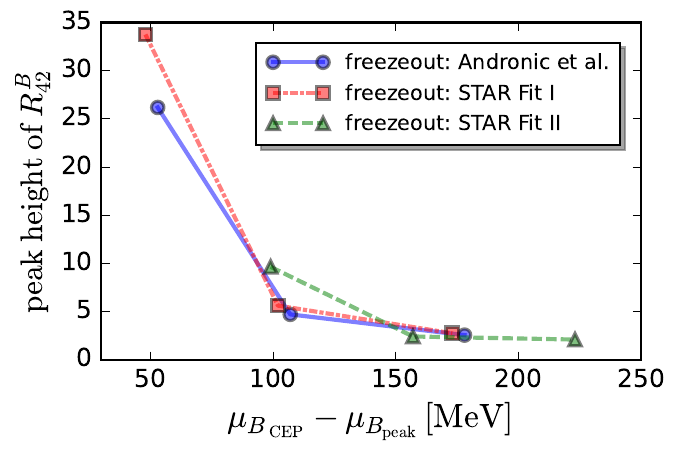}
\caption{Top: Dependence of the position of the peak in $R_{42}^{B}$ on the location of the CEP and the freeze-out curve \cite{Fu:2021oaw}. Bottom: Height of the peak in $R_{42}^{B}$ as a function of the difference between the ${\mu_B}_{_{\mathrm{CEP}}}$ and  ${\mu_B}_{_{\mathrm{peak}}}$, where the latter corresponds to $\mu_B$ related to the $\sqrt{s_{\mathrm{NN}}}$ of the peak in $R_{42}^{B}$. Plot taken from \cite{Fu:2023lcm}. \hspace*{\fill}}
\label{fig:peak-muBCEP}
\end{figure}

The results in \Cref{fig:chi-CE,fig:R42-Frz} show a clear peak structure of (hyper) fluctuations of conserved charges related to the critical end point in QCD \labelcref{eq:FunEstimateCEP} found in functional QCD. These findings beg the question whether one can infer the location of the critical end point from specific features of these observables. In \cite{Fu:2023lcm} a first step in this direction was taken: there the relation between the location of the CEP and the location of the peak on the freeze-out line as well as its height was investigated, see \Cref{fig:peak-muBCEP}. In short, the location of the peak only marginally depends on that of the CEP while the peak height is directly related to it.

%
%

%
\begin{figure}[t!]
	\centering
	\includegraphics[width=0.91\columnwidth]{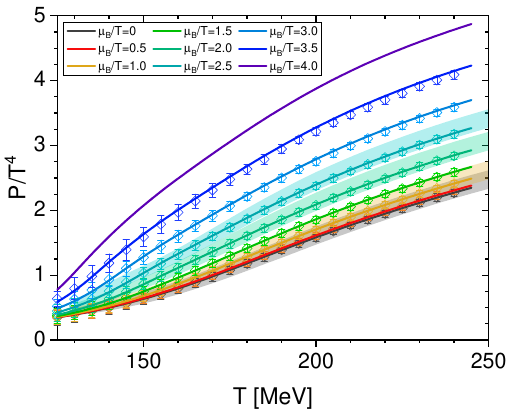}\\[2ex]
	\includegraphics[width=0.91\columnwidth]{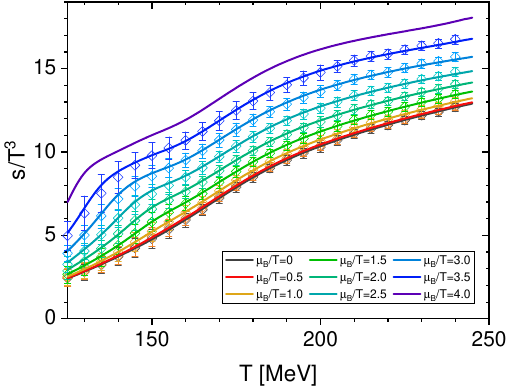}\\[2ex]
	\includegraphics[width=0.91\columnwidth]{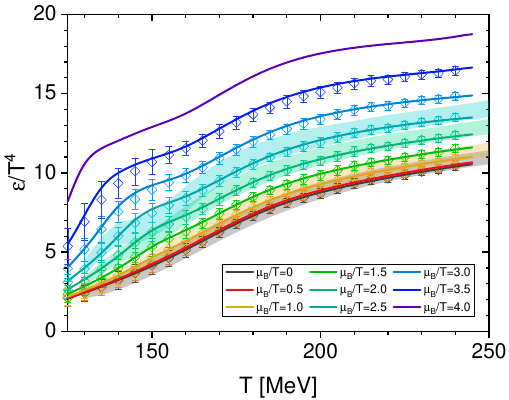}
	\caption{QCD equations of state at finite temperature and density: pressure ($P/T^4$), entropy density ($s/T^3$) and energy density ($\epsilon/T^4$) as  functions of temperature with $T \in (100,250)\,\textrm{MeV}$ for baryon chemical potentials $\mu_B/T = 0,0.5,\cdots,4.0$. We also show lattice QCD data from the $T^{\prime}$ expansion scheme~\cite{Borsanyi:2021sxv} (open diamonds) and the Taylor expansion~\cite{Bazavov:2017dus} (coloured bands). Plot taken from \cite{Lu:2025cls}. \hspace*{\fill}}
\label{fig:EoST+Entropy+EnergyDensitymuB}
\end{figure}

These first results are very promising and suggest a more refined analysis and also indicates that these observables have to be measured rather accurately in order to allow for a determination of the location of the CEP or, more precisely, that of the location of the ONP. 

We close this Section with some remarks on the next steps towards a quantitative comparison of the theoretical results with the experimental results for proton number fluctuations. To begin with, this requires a theoretical computation of proton number fluctuations which is currently under way in functional QCD. Moreover, non-equilibrium effects have to be taken into account, first steps towards functional QCD computations in the transport regime have been taken in \cite{Bluhm:2018qkf, Tan:2025bsv}. QCD transport and hydrodynamics requires the access to thermodynamic quantities, for state-of-the-art functional results see \Cref{fig:EoST+Entropy+EnergyDensitymuB}, taken from \cite{Lu:2025cls}. A more comprehensive discussion can be found there. Finally, we remark that first steps in this programme have been taken very recently in \cite{Lu:2026ezr}, where a functional freeze-out curve is discussed.

\section{Columbia plot, isospin chemical potential, magnetic fields and more}
\label{sec:TheoryCover}

The phase diagram of QCD at finite temperature and baryon chemical potential can be embedded into a higher-dimensional one with further external parameters. 
Particularly interesting in the present context are variations of the current quark masses (Columbia plot) and the extension to complex chemical potentials (Lee-Yang singularities). Physical parameters are isospin (ion species and neutron stars) and magnetic fields (triggered in the initial phase of heavy ion collisions).

\subsection{Columbia plot and Lee-Yang singularities}
\label{sec:columbia}

\begin{figure}[t]
	\centering
	\includegraphics[width=0.7\columnwidth]{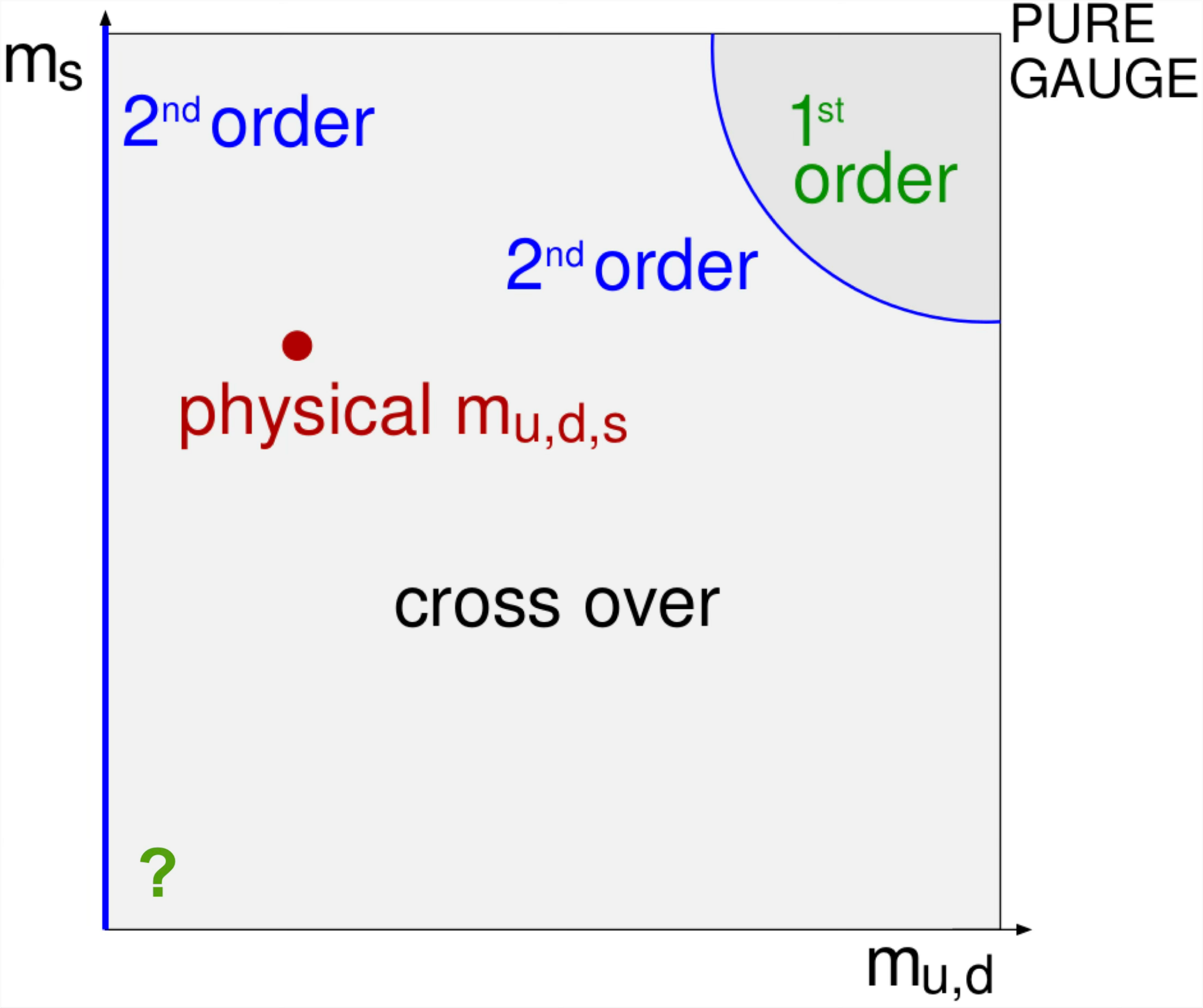} 
	\caption{Phase structure of three-flavour QCD in the light and strange quark plane. Figure modified from \cite{Bernhardt:2025fvk}. The green question mark in the lower left corner indicates the open question of a first or second order phase transition. A 3d-version of this plot is shown in \Cref{fig:columbia}. 
	For a possible 1st order scenario see \Cref{fig:columbiafull}. \hspace*{\fill}}
	\label{fig:Columbia}
\end{figure}

The physics of the Columbia plot is extremely interesting by itself, for a cartoon see \Cref{fig:Columbia}. Specifically the lower left corner with small and vanishing light and strange quark masses is subject to an ongoing open debate. This is indicated by the green question mark in \Cref{fig:Columbia}. 

In addition, the quark mass dependence of the QCD phase structure  provides 
ample opportunity for systematic cross-checks between different ab initio approaches to the QCD phase diagram such as lattice QCD
and functional methods. Lattice QCD is fully operational at not too small quark masses and zero as well as imaginary chemical
potential and provides extrapolations toward the various chiral limits, $m_{u,d} \rightarrow 0$ with $m_s \rightarrow \infty$,  
$m_{u,d} \rightarrow 0$ with $m_s$ fixed at the physical strange quark mass and $m_{u,d,s} \rightarrow 0$. Functional methods
can be cross-checked in these regions with lattice QCD and, provided these cross-checks are successful, may deliver reliable 
results in all regions inaccessible by lattice QCD. For real chemical potential at the physical point, we have extensively 
discussed corresponding results above and provided arguments for a reasonable assessment of their systematic error. In this 
section, we extend our discussion to the other regions of the Columbia plot.    

The physics of the pure gauge/heavy quark region of the Columbia plot is dominated by the gauge theory, the
associated deconfinement phase transition and the Roberge-Weiss point at imaginary chemical potential. In order 
to address these issues within functional QCD it is mandatory to take the Yang-Mills sector explicitly into account. 
In functional QCD this has been done in~\cite{Fischer:2009wc, Fischer:2009gk, Fischer:2014vxa}, using the truncation schemes  
discussed at the beginning of section \Cref{sec:PhaseStructure+CEP}. The results have already been reviewed in~\cite{Fischer:2018sdj} and shall not be discussed in detail here for brevity. The main results of these studies 
is the reproduction of the second order critical surface emerging from the plane of the first Roberge-Weiss transition
including the expected tricritical scaling from the Roberge-Weiss endpoint in agreement with previous and later findings from 
lattice gauge theory \cite{deForcrand:2010he, Saito:2011fs, Fromm:2011qi, Ejiri:2019csa, Cuteri:2020yke, Kiyohara:2021smr}.

At the physical point and in the chiral limit, the direction of imaginary chemical potential has been explored within functional QCD in~\cite{Bernhardt:2023ezo} and \cite{Braun:2009gm} respectively. The corresponding pseudo-critical transition temperatures have been determined in the 
region between zero chemical potential and the first Roberge-Weiss transition. The purpose of the study in~\cite{Bernhardt:2023ezo} has been two-fold.
On the one hand it served as a cross-check for the functional approach by direct comparison with the corresponding lattice 
results of~\cite{Borsanyi:2020fev} for $T_c(i \mu_B)$. On the other hand, it served as a cross-check for the extrapolation
method used in~\cite{Borsanyi:2020fev} to determine $T_c(\mu_B)$ at real baryon chemical potential from $T_c(i \mu_B)$.
Both cross-checks have been successful. The functional approach reproduced $T_c(i \mu_B)$ with only small systematic error 
at large $i \mu_B$, presumably originating from neglecting the effect of the $A_0$-background that proved to be indispensable
in the heavy quark studies \cite{Fischer:2014vxa}. Moreover, using the same extrapolation method as in~\cite{Borsanyi:2020fev},
functional QCD results for $T_c(\mu_B)$ at real baryon chemical potential could be reproduced: up to large chemical potentials 
not very much smaller ($\approx 20 \%$) than the one of the critical endpoint, the extrapolation worked extremely well. 
For larger chemical potential, the extrapolated transition line undershooted the calculated one; at the critical
chemical potential $\muB^{\textup{CEP}} \approx \SI{636}{\MeV}$, the resulting temperature of the extrapolation was about
$\SI{13}{\MeV}$ too small. Also, of course, it was not possible to extract the location of the CEP from the extrapolation 
procedure, cf. the discussion in \Cref{app:CEP-Estimates} above. 

This changes, when one takes into account not only strictly imaginary and real chemical potential, but extends to the 
complex chemical potential plane, where the Lee-Yang edge singularities can be found, see e.g.~\cite{Connelly:2020gwa, Mukherjee:2021tyg, Rennecke:2022ohx, Johnson:2022cqv}. This has been explored directly 
(i.e. without any extrapolations) within functional QCD in \cite{Wan:2024xeu, Wan:2025wdg}. 

The results explicitly 
confirm the scenario in which the CEP coincides with the edge singularities on the real baryon chemical potential axis. 
Since all results are explicitly available, one can also test extrapolation procedures that are involved in corresponding
lattice calculations. These tests underline our comment made in  \Cref{app:CEP-Estimates}: it seems fair to say that much 
more work is needed until the lattice extrapolations performed in \cite{Basar:2023nkp,Clarke:2024ugt} can be transformed 
into reliable predictions.

\begin{figure}[t]
	\centering
	\includegraphics[width=0.75\columnwidth]{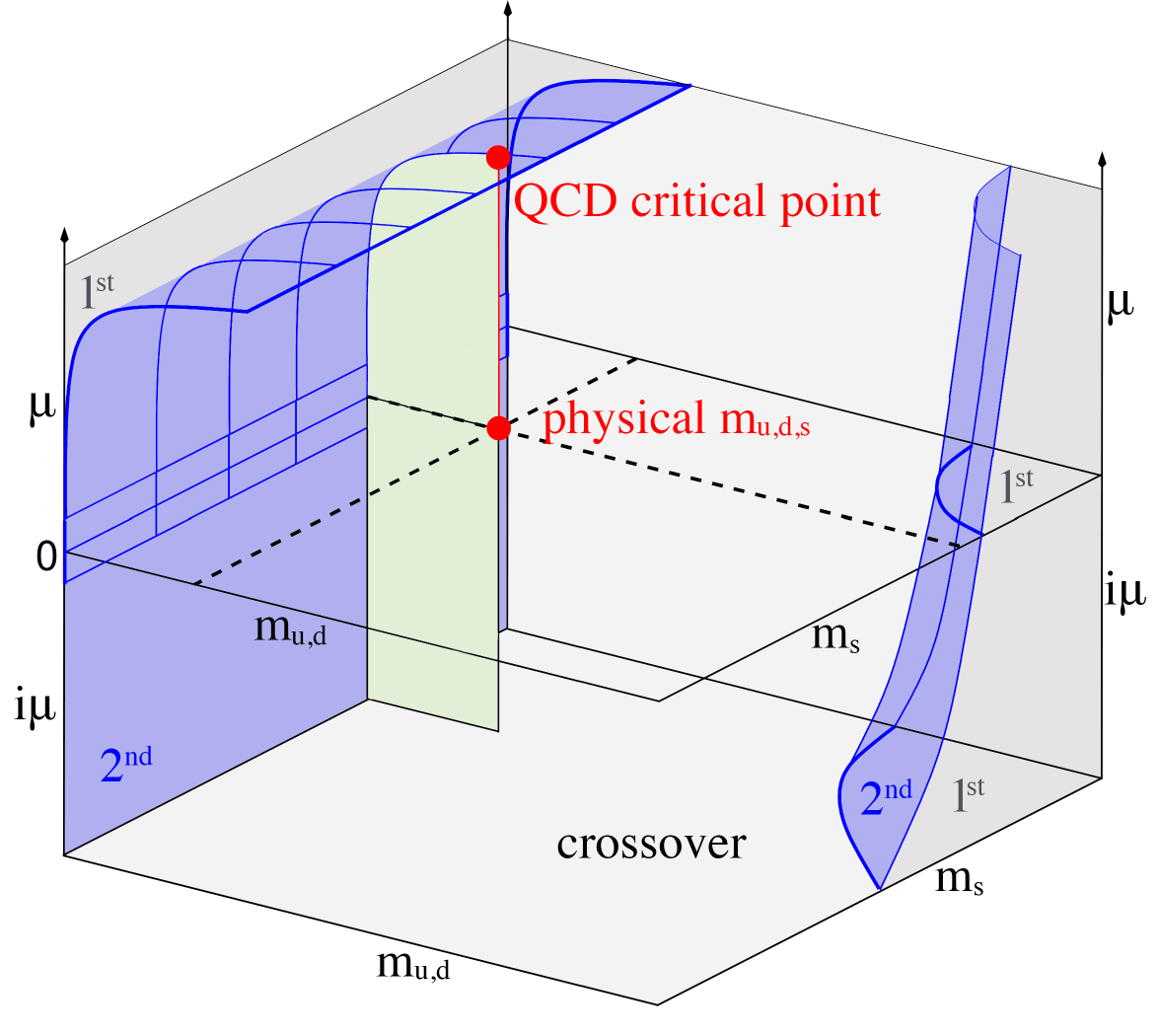}
	\caption{
		3d-Columbia plot with real and imaginary chemical potential as third axis. The second order critical surface
		of the confinement-deconfinement transition in the heavy quark limit has been explored with functional methods
		in Ref.~\cite{Fischer:2014vxa}. The chiral critical surface connecting the QCD critical point discussed in 
		previous sections with the second order transition line at zero chemical potential has been studied in
		Ref.~\cite{Bernhardt:2025fvk}.
		Moreover the cross-over nature of the green area in the imaginary chemical potential region is confirmed by both,
		lattice calculations \cite{DAmbrosio:2022kig, Cuteri:2022vwk, DAmbrosio:2025ldv} and functional methods \cite{Bernhardt:2025fvk}.   \hspace*{\fill}}
	\label{fig:columbia}
\end{figure}

A further interesting, and not yet fully resolved, situation emerged in the past years for the chiral limit, i.e. the left hand
side of the 2d Columbia plot of \Cref{fig:Columbia} and the zero chemical potential plane of \Cref{fig:columbia}. Starting from the physical point and approaching the chiral limit
$m_{u,d} \rightarrow 0$ with fixed strange quark mass, strong indications for the second-order nature of this point have 
been reported from lattice QCD \cite{HotQCD:2019xnw,Cuteri:2021ikv} and functional methods
\cite{Braun:2020ada, Gao:2021vsf, Bernhardt:2023hpr, Braun:2023qak}. The pseudocritical transition temperatures on the way towards this limit
have been compared between the two approaches at pion masses of $m_\pi = 140, 110, 80$ MeV for $N_\tau=12$ lattice data and
$m_\pi =  140, 110, 80, 55$ MeV for $N_\tau=8$ and found to agree within error bars at least for the first set, i.e. close
to the continuum limit. Deviations occur for the second set and, interestingly, directly in the chiral limit where the extrapolated
lattice result and the result from direct computations within the fRG and DSE-approach deviate by about 10 percent. This quantitative discrepancy clearly warrants further study. The $N_f=2$ chiral limit with infinite strange quark mass has been explored already very early in the functional approach \cite{Braun:2009gm} with clear indications of an O(4) second order transition and corresponding
critical behaviour. Also in this limit, lattice extrapolations of the corresponding transition temperature \cite{Bornyakov:2009qh} and results from functional methods \cite{Braun:2009gm, Bernhardt:2023hpr} match on 
the ten percent level. 

%
\onecolumngrid

\begin{figure}[h!]
	\begin{center}
		\vspace{.3cm}
		\includegraphics[scale=0.22]{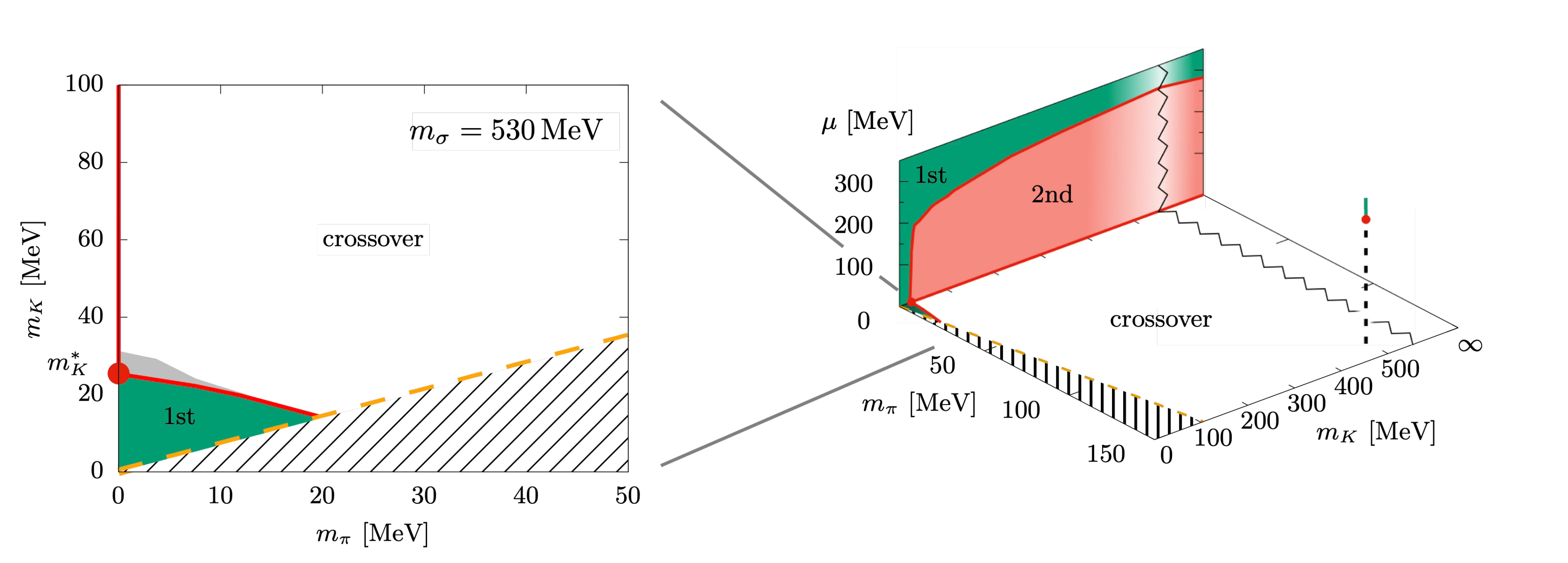}
	\end{center}
	\caption{Chiral phase structure for 2+1 flavour QCD in the $(m_\pi,m_K)$- and $(\mu,m_K)$-planes (Columbia plot) obtained in \cite{Resch:2017vjs}. At $\muB=0$ there may a small first-order region  around the chiral limit. The anomalous $U_A(1)$-breaking is chiefly important for this results, without the anomaly the phase structure is significantly changed, see \cite{Resch:2017vjs}. This question is a hot topic of ongoing research, see e.g.~\cite{deForcrand:2017cgb, Cuteri:2021ikv, Dini:2021hug,DAmbrosio:2025ldv} (lattice), \cite{Bernhardt:2023hpr,Bernhardt:2025fvk} (DSE), and \cite{Resch:2017vjs, Fejos:2022mso, Pisarski:2024esv, Giacosa:2024orp} (low energy effective models). Figure modified from \cite{Dupuis:2020fhh}.\hspace*{\fill}}
	\label{fig:columbiafull}
\end{figure}
\twocolumngrid
%
\ 
\newpage
\ 

\begin{figure*}[t!]
	\centering
	\includegraphics[width=0.9\columnwidth]{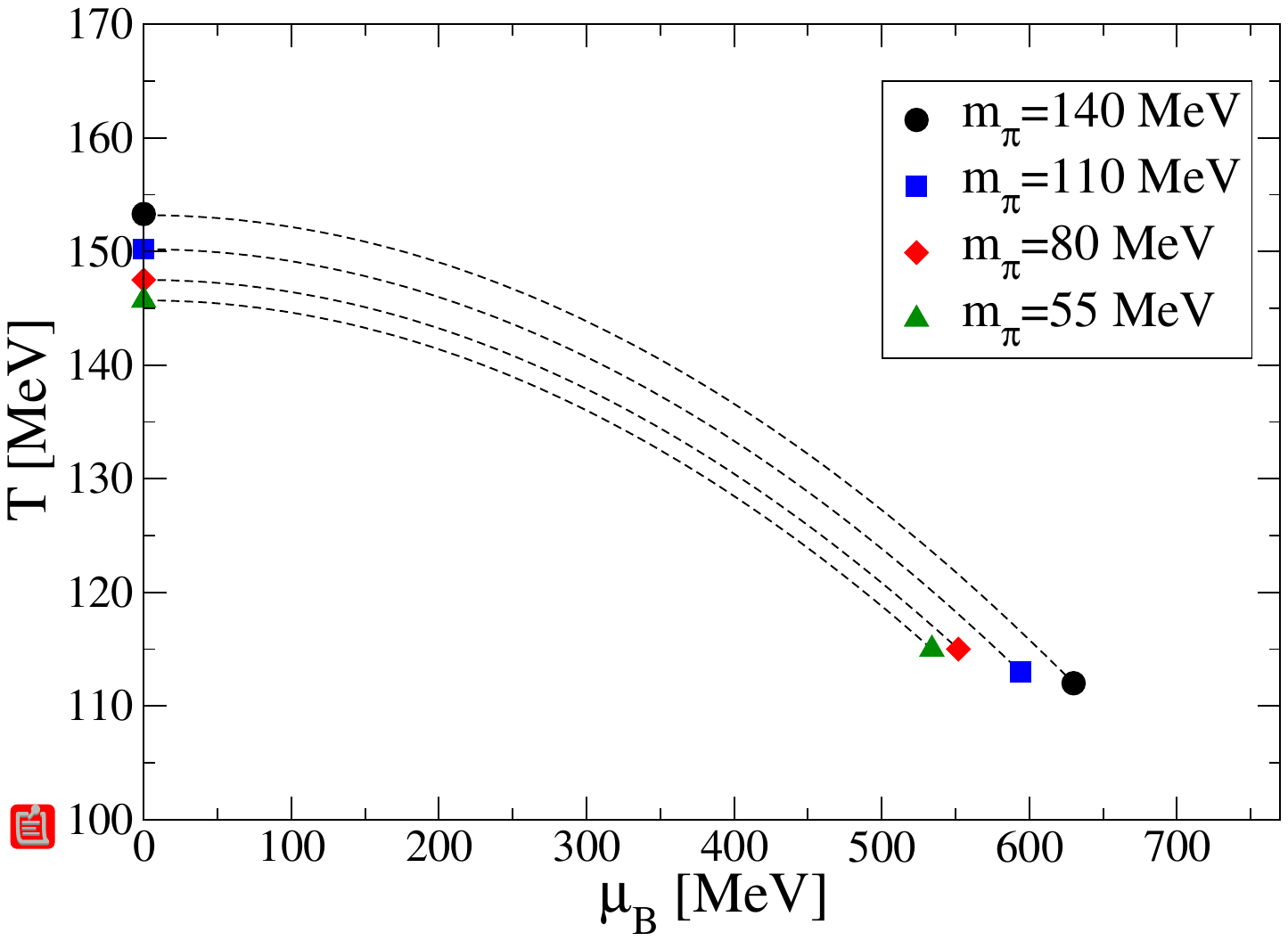}\hspace{2cm}
	\includegraphics[width=0.7\columnwidth]{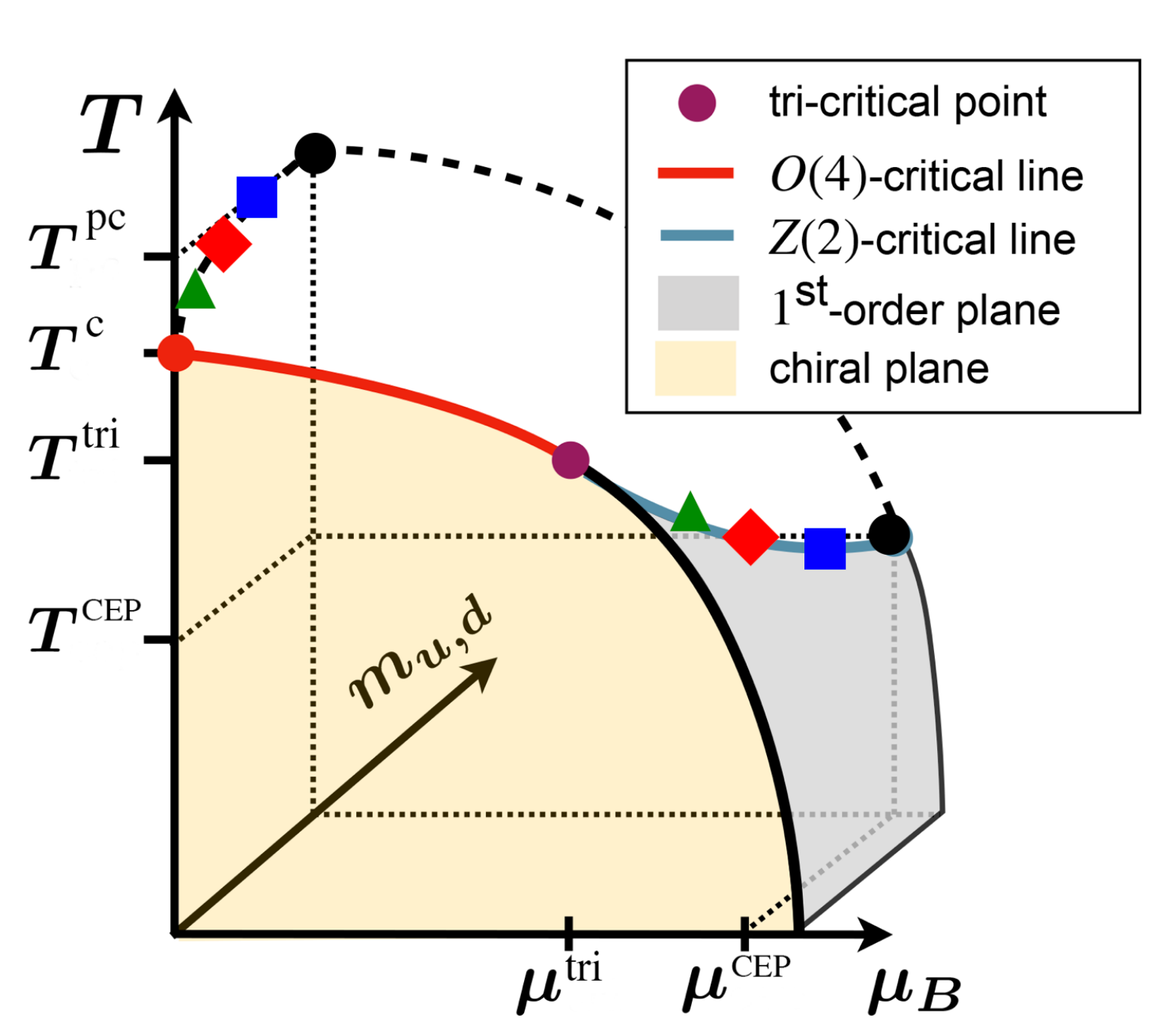}
	\caption{Left: Variations of the location of the crossover line and the CEP under change of the light quark masses towards the chiral limit \cite{Bernhardt:2025fvk}. Plot taken from \cite{Bernhardt:2025fvk}. Right: Sketch of the QCD phase diagram in temperature $T$ and baryon chemical
		potential $\muB$ for varying degenerate light quark masses. Adapted from~\cite{Ding:2024sux}. The symbols
		indicate the quantitative results from~\cite{Bernhardt:2025fvk} for $m_\pi =  140, 110, 80, 55$ MeV also shown 
		in the left diagram. The qualitative behaviour under variation of the quark mass matches general expectations, 
		see text for details.\hspace*{\fill}}
	\label{fig:Tofm}
\end{figure*}
Furthermore, it is not clear, 
whether the SU(3)-symmetric limit $m_{u,d,s} \rightarrow 0$ features 
a second order or a (weak) first order  transition with a very
small extension into the finite mass region of the Columbia plot. 
Here, the fate of the $U_A(1)$-symmetry is expected
to affect the order of the chiral $SU(3)$ transition, see e.g.~\cite{Pisarski:1983ms, Mitter:2013fxa, Resch:2017vjs, Pisarski:2024esv, Giacosa:2024orp}
and references therein. In \Cref{fig:columbiafull} we show a possible scenario with a small first order regime, for more details see \cite{Resch:2017vjs}, which hosts a rather comprehensive discussion of possible scenarios as well as an assessment of the sensitivity to the strength of the $U_A(1)$-anomaly. Notably, the possible scenario of a second-order transition for all non-zero masses of the strange 
quark has emerged in the past years. Recent lattice results are compatible with a second-order nature of this limit,  
\cite{deForcrand:2017cgb, Cuteri:2021ikv, Dini:2021hug}, and this is in line with a recent study in the DSE-approach \cite{Bernhardt:2023hpr}. How such a scenario can be accommodated and explained, is currently scrutinised within low energy effective theories, \cite{Resch:2017vjs, Fejos:2022mso, Pisarski:2024esv, Giacosa:2024orp}. While these models lack the full flue dynamics of QCD, the allow for a comprehensive study of the matter dynamics of QCD that is the relevant one for this question.

Finally, we comment on the shape of the critical surface connecting the critical end point at real baryon chemical potential 
to the critical physics in the zero chemical potential chiral limit. A very recent work in the DSE-approach has addressed this issue, 
\cite{Bernhardt:2025fvk}. There, a qualitative study of the critical surface  has been performed and the nature of the transition in the 
green area indicated in \Cref{fig:columbia}. Interestingly, a flat chiral critical surface at large
chemical potential has been found, that presumably ends in a line of tricritical points along the chiral left-hand side of the 3$d$-Columbia 
plot. Furthermore, inside the green area in this figure, a crossover down to the plane of the first 
Roberge-Weiss transition has been found in he results in \cite{Bernhardt:2025fvk}. These findings are in agreement with and confirm previous notions from lattice gauge theory for 
the structure of the Columbia plot \cite{DAmbrosio:2022kig, Cuteri:2022vwk, DAmbrosio:2025ldv} and for 
temperature bounds for the 
transition temperature of the critical endpoint \cite{Halasz:1998qr, Karsch:2019mbv}. The latter scenario is summarised in
\Cref{fig:Tofm} with explicit results from functional QCD denoted by coloured stars. These results support the existence of 
a tri-critical point, whose precise location still needs to be determined.

\subsection{Isospin chemical potential and magnetic fields}
\label{sec:iso_mag}

A further physically interesting parameter in the QCD phase diagram is finite isospin chemical 
potential $\mu_I$, see \Cref{sec:FunDerivation}. It creates an imbalance between 
up- and down-quarks \cite{Son:2000xc}, which is realised in heavy ion collisions (mainly
 due to different numbers of protons and neutrons in the colliding nuclei), in 
compact stars due to $\beta$-equilibrium and charge neutrality and is also relevant for the early universe, see 
e.g.~\cite{Vovchenko:2020crk} and references therein. In contradistinction to baryon chemical potential, lattice QCD at finite isospin chemical
potential is not hampered by the sign problem, see e.g.~\cite{Kogut:2004zg, deForcrand:2007uz}. This allows for systematic cross-checks with functional QCD as
well as with mean-field models, see e.g.~\cite{Kamikado:2012bt, Brandt:2025tkg}. 

The physics of finite isospin chemical 
potential is related to several phase transitions. For $\mu_I < \mu_\pi/2$ the silver-blaze phenomenon takes place, 
similar to the situation at finite baryon chemical potential. At $\mu_I = \mu_\pi/2$, the system undergoes a second order transition to a phase 
with pion condensation, which may even play a role in compact stars, see e.g.~\cite{Brandt:2018bwq, Brandt:2022hwy, Basta:2025svw} 
and references therein. The pion condensation phase extends to finite temperature and the system may exhibit another first order 
transition accompanied with a CEP, see e.g.~\cite{Son:2000xc, Kamikado:2012bt}. Further  interesting phenomena, like a Fulde-Ferrell-Larkin-Ovchinnikov (FFLO) phase \cite{Fulde:1964zz, Larkin:1964wok, Son:2000xc} may occur in a regime with finite baryon chemical and isospin chemical potential. 

In view of the main topic of this review, one of the most interesting questions is perhaps the 
fate of the CEP or the ONP and the associated fluctuations at finite $\mu_I$. This has not been studied yet in
functional QCD. Lattice results on the fate of the chiral crossover at zero baryon chemical potential
and the onset of the pion condensed phase provide strong indications that the transition temperatures 
do not change much, see~\cite{Brandt:2022hwy}. Thus one may expect a similarly small effect for the location
of the CEP, at least for small values of $\mu_I$. Whether this is also true for the fluctuations, 
is completely open and should be studied in the near future. In any case, the potential of further systematic 
cross-checks between lattice QCD and functional QCD at finite isospin chemical potential is very promising. 

A huge amount of literature is available that deals with the effects of non-zero magnetic field onto the various transitions
discussed above. Magnetic fields are interesting in connection with important physics applications: (i) in heavy 
ion-collisions huge (but short-lived) magnetic fields are created by the nuclei moving rapidly in opposite directions;
(ii) some compact stars, so called magnetars, are characterised by extremely high magnetic fields that may have 
a profound impact on the equation of state; (iii) magnetic fields may have played an important  role in the electroweak 
phase transition of the early universe. Comprehensive reviews on the effects of magnetic field are e.g.~\cite{Andersen:2014xxa, Miransky:2015ava}. From a fundamental point of view, one of the most interesting 
effects of a non-vanishing magnetic field is the generation of magnetic catalysis, i.e. the fact that an external 
magnetic field initiates (or enhances) the dynamical generation of fermion masses even in theories with only weak 
interaction~\cite{Klevansky:1989vi, Suganuma:1990nn}. This effect has been studied in detail over the years and
similarities and discrepancies with Cooper pairing effects have been worked out, see \cite{Miransky:2015ava} for
an overview. Perhaps surprisingly, results from lattice QCD also demonstrate that the opposite effect ('inverse catalysis')
happens for large enough temperatures: the dynamical mass generation is reduced and consequently the transition temperatures
for the crossover to the chirally restored phase is decreased 
\cite{Bali:2011qj, Bali:2012zg, Bruckmann:2013oba, Bali:2013esa, Ilgenfritz:2013ara}.
This effect, not present in simple models, can be traced back on the back-reaction of the quarks onto the Yang-Mills 
sector, and has also been seen also in functional QCD \cite{Braun:2014fua, Mueller:2015fka}. Latest developments include lattice studies
of the chiral magnetic effect and non-uniform magnetic fields, see e.g.~\cite{Brandt:2023dir, Brandt:2024wlw, Brandt:2024fpc} and references therein.

\section{Summary and outlook}
\label{sec:ConclusionOutlook}

We have provided a short review of important aspects of the functional QCD approach to the phase structure of QCD. We have mainly focused on works mapping out the phase structure as well as computing respective observables. At present, functional QCD estimates the location of the critical end point, or rather that of the onset of new phases, at roughly $\mu_B\approx 600 - 650$\,MeV for $\mu_S=0$ or $\mu_s=0$, see \labelcref{eq:FunEstimateCEP-Recall}. The details are provided in 
\Cref{sec:PhaseStructure+CEP}. Respective observables have been discussed in \Cref{sec:ConservedFlucs+freezeout}, concentrating on thermodynamics and fluctuations of conserved charges. Finally, we have discussed changes of the phase structure under the variation of external parameters such as the quark masses, external magnetic fields, isospin chemical potential as well as complex chemical potentials, see \Cref{sec:TheoryCover}. 

The results reported here lay the foundation for the quantitative resolution of the phase structure for  $\mu_B/T\lesssim  4.5$, see \labelcref{eq:muBT4.5}.   
To our mind, a central task in the upcoming years is to map out quantitatively the regime with $\mu_B/T\gtrsim  4.5$ with its potential moat regimes and regions with inhomogeneous condensate(s). This will allow us to pin down the location and nature of the onset regime of new phases. In particular this concerns the very existence of the CEP and more generally, the signals of the onset regime of new phases. We hope to have convinced the reader, that functional QCD is well suited for this task.

\begin{acknowledgments}

We thank Julian Bernhardt, Szabolcs Borsanyi,Jens Braun, Michael Buballa, Yong-rui Chen, Wei-jie Fu, Kenji Fukushima, Fei Gao, Leonid Glozman, Jana Guenther, Pascal Gunkel, Chuang Huang, Friederike Ihssen, Philipp Isserstedt, Keiwan Jamaly, Yi Lu, Frithjof Karsch, Volker Koch, Konrad Kockler, Xiaofeng Luo, Larry McLerran, Theo Motta, Jorge Noronha, Owe Philipsen, Rob Pisarski, Fabian Rennecke, Dirk Rischke, Franz Sattler, Bernd-Jochen Schaefer, Lorenz von Smekal, Misha Stephanov, Yang-yang Tan, Shi Yin, Rui Wen, Nicolas Wink, Nu Xu and the members of the QCD collaboration \cite{fQCD} for enlightening discussions and work related to topics of this review. 

This work is funded by the Deutsche Forschungsgemeinschaft (DFG, German Research Foundation) under Germany’s Excellence Strategy EXC 2181/1 - 390900948 (the Heidelberg STRUCTURES Excellence Cluster) and the Collaborative Research Centre SFB 1225 - 273811115 (ISOQUANT) as well as the Collaborative Research Centre TransRegio CRC-TR 211 “Strong-interaction matter under extreme conditions” and the individual grant FI 970/16-1.

\end{acknowledgments}

\ 
\newpage 
\ 
\appendix 

\section{Approximation schemes and crosschecks in functional QCD}
\label{app:ApproximationsandChecks+Balances}

In this Appendix we provide a brief overview on the general strategy and the detailed approximations used in the three different 
but related works \cite{Fu:2019hdw, Gao:2020fbl, Gunkel:2021oya} that led to the estimate \labelcref{eq:FunEstimateCEP}. This analysis also extends to the very recent work \cite{Pawlowski:2025jpg}, that is used for the determination of the state-of-he-art quantitative reliability regime $\mu_B/T\lesssim 4.5$, see \labelcref{eq:muBT4.5}. We also comment on the internal checks that have been done within each of the  approaches (in previous and later 
publications) and on the systematic cross-check provided by the combination of all three approaches. The non-trivial fact
that all approaches with their different expansion schemes lead to similar predictions despite their differences may hint at a small overall systematic error even for baryon chemical potentials beyond $\mu_B/T\lesssim 4.5$, as we already argued in various places in the main text of this review. Here we provide more details supporting this argument.

\subsection{Summary of approximations and difference relations}
\label{app:SummaryofApprox}

\begin{figure*}
	\centering
	\begin{minipage}[t]{1\linewidth}
		\includegraphics[width=0.98\textwidth]{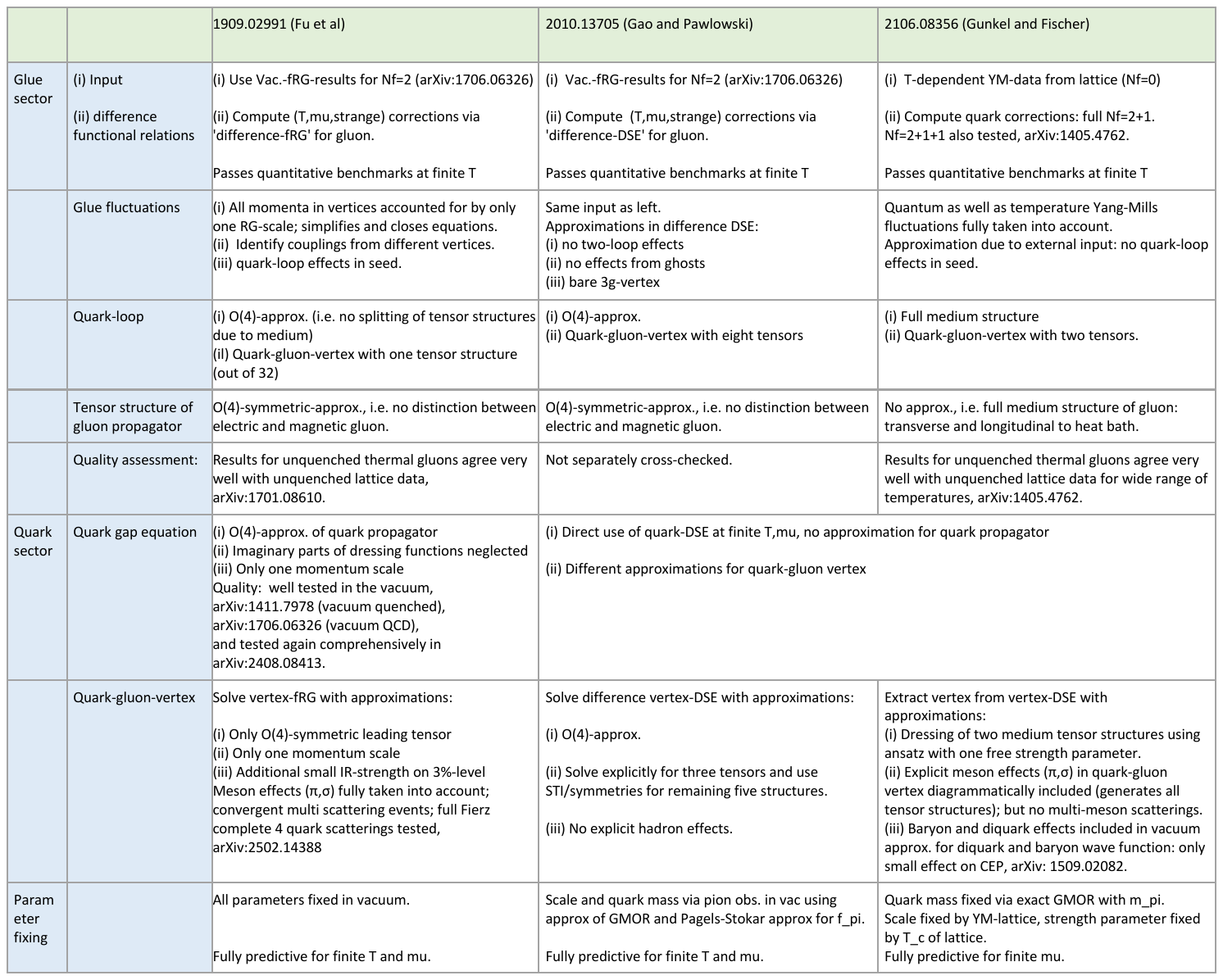} 
		\caption{Overview of approximations in the three works \cite{Fu:2019hdw, Gao:2020fbl, Gunkel:2021oya} predicting 
		 the location of the CEP. The very recent work \cite{Pawlowski:2025jpg} has not been included in the above list for the sake of readability. It is based on the approximation scheme in \cite{Ihssen:2024miv}, which itself is a considerable upgrade of that used in \cite{Fu:2019hdw}. Summary of the upgrades in \cite{Pawlowski:2025jpg}: full effective potential of the light and strange condensates , $\rho_l$ and $\rho_s$ respectively. This takes into account all order scatterings of pions, $\sigma,\sigma_s$-modes. Moreover,  momentum-dependent (light and strange) quark , pion and $\sigma,\sigma_s$-mode propagators, and respective Yukawa couplings, are included with thermal splits parallel and perpendicular to the heat bath.  \hspace*{\fill}}
		\label{fig:Approximations19-20-21} 
	\end{minipage} 
\end{figure*}

The relevant approximations are summarised in \Cref{fig:Approximations19-20-21} and are discussed in the following.
The three works \cite{Fu:2019hdw, Gao:2020fbl, Gunkel:2021oya} are arranged in order of publication date from left to right, the recent update \cite{Pawlowski:2025jpg} is discussed separately in \Cref{app:SystematicErrorBudget}. All works make use of an important structural property of functional equations: assume one has access to a set of QCD correlation functions in a specific regime or setup. Prominent cases, for which quantitatively accurate correlation functions are available, are Yang-Mills correlation functions in the vacuum and at finite temperature (both functional QCD results and lattice results) as well as full vacuum QCD with $N_f=2\,,\, 2+1$ flavours (mostly functional QCD results for general correlation functions and lattice gluon and quark propagators). Then, functional methods can be re-organised in terms of \textit{difference} functional relations, 
\begin{subequations} 
\label{eq:DiffFunRel}
\begin{align} 
\Gamma^{(n)}= \Gamma^{(n)}_\textrm{input} + \Delta \Gamma^{(n)}\,,\quad 
\Delta \Gamma^{(n)} = \Gamma^{(n)}-\Gamma^{(n)}_\textrm{input} \,.
\label{eq:DeltaGamma} 
\end{align}
Importantly, the difference correlations $\Delta \Gamma^{(n)}$ satisfy \textit{exact} difference functional relation, 
\begin{align} 
\Delta \Gamma^{(n)} =\Delta \textrm{FunRel}_n\left[\left\{\Delta \Gamma^{(m)}\right\}\,,\, \left\{\Gamma^{(m)}_\textrm{input}\right\}\right]\,,
\label{eq:DeltaFunRel}
\end{align} 
with $m\leq n+2$. 
\end{subequations} 
This bound on the order $m$ of the correlation functions on the right hand side of \labelcref{eq:DeltaFunRel} can be readily used for estimates on the systematic error budget: In a given order of the vertex expansion, some of the higher order correlations on the right hand side \labelcref{eq:DeltaFunRel} of are not computed  and their impact can be estimated by a conservative survey of their potential strength.

These functional difference relations have the same loop structure as the underlying functional method: they are one loop-exact for the fRG and two-loop exact for the DSE. In practise these relations are solved within approximations and are subject to the same systematic error analysis already discussed before. 

The main advantage of such a setup is twofold: 
\begin{itemize} 
\item[(i)] For quantitatively accurate input $\{\Gamma^{(n)}_\textrm{input}\}$, only the difference is subject to systematic errors caused by the approximation.   
\item[(ii)] Thermal and density fluctuations cause mild effects in pure glue correlation functions. This also holds true for general quark-gluon and meson correlation functions for a large part of the phase diagram. Accordingly, the difference functional relations already provide significantly smaller systematic errors in relatively simple approximations.     
\end{itemize} 
The two advantageous properties (i,ii) of difference functional relations have been used in all three works \cite{Fu:2019hdw, Gao:2020fbl, Gunkel:2021oya} and we proceed with a brief discussion of the respective setups. 
In \Cref{fig:Approximations19-20-21} we have aimed for a concise overview of the different approximation schemes. We emphasise that such an overview cannot be comprehensive and a full appreciation is only possibly within the works themselves.

%
%

\subsection{Glue sector}
\label{app:GlueSector}

Let us start the detailed discussion with the glue sector. Here, the \LEGO-principle, already discussed in \Cref{sec:Expansion+Error}, unfolds its full power as well as the complementarity of the different approaches.

\subsubsection{Input correlation functions}
\label{app:InputGlueSector}

The input used in the glue sector in \cite{Fu:2019hdw, Gao:2020fbl} is the same, namely the vacuum results  from the fRG for full 
$N_f=2$ QCD \cite{Cyrol:2017ewj} as the input $\{\Gamma^{(n)}_\textrm{input}\}$ and solve the difference functional relations \labelcref{eq:DiffFunRel} for the additional strange quark as well as temperature and chemical 
potential. In the case of \cite{Fu:2019hdw} this is  a difference-fRG 
equation, for \cite{Gao:2020fbl} this is a difference-DSE. 

These difference functional relations are now solved within approximations: For example, in \cite{Fu:2019hdw} the back-reaction of the strange quark onto the gluon and quark-gluon sector has been taken into account and was found to be sub-leading. This has informed the approximation in \cite{Gao:2020fbl}, where the back-reaction 
effects of the strange quark onto pure glue  vertices have been  neglected. Further approximations are listed in \Cref{fig:Approximations19-20-21}. We also remark that an additional systematic error is that of the $N_f=2$ flavour vacuum QCD input from \cite{Cyrol:2017ewj} and this error propagates through the system. We note in this context that the correlation functions in \cite{Cyrol:2017ewj} match lattice QCD benchmarks where available (mostly quark and gluon propagators) quantitatively (percent level) and we consider the respective systematic error negligible in comparison to that stemming from the approximations in the difference fRG flows and the DSEs.   

The input used in the glue sector in \cite{Gunkel:2021oya} is lattice data for the temperature-dependent gluon propagator in pure Yang-Mills theory \cite{Fischer:2010fx, Maas:2011ez, Eichmann:2015kfa}. We emphasise that the use of lattice data is merely convenience and availability of data. Then, the gluon propagator at finite temperature and chemical potential is computed with its difference-DSE \labelcref{eq:DiffFunRel}. In practise, the $T,\mu_B$-dependent 
quark loops are added to the Yang-Mills input. This neglects the back-reaction effects of the quarks onto Yang-Mills vertices. Moreover, it picks 
up uncertainties of the initial lattice results but in turn does not rely on truncations in generating the seed input. 

In short, both setups with their different input are complementary to each other. The solutions of the functional difference equations in the glue sector provide in particular the gluon propagator at finite temperature and density. Together with the light and strange quark-gluon vertices this is the key input into the matter sector of QCD, and its systematic error estimate propagates to that of the matter correlations.

\subsubsection{Systematic error budget glue sector}
\label{app:ErrorGlueSector}

The quark-gluon vertex will be discussed below and we proceed with a discussion of the systematic error of the gluon propagator, which, apart from being a key input in the matter sector, contains interesting physics in itself both related to confinement as well as chiral symmetry breaking. To begin with we note that the different inputs used in \cite{Fu:2019hdw, Gao:2020fbl, Gunkel:2021oya} naturally lead to different approximation schemes listed in the second row of the table ('Truncations of glue fluctuations') for the computation of the temperature
corrections to the glue correlation functions: 

In the fRG work \cite{Fu:2019hdw}, mainly momentum dependencies of propagators and vertices have been simplified due to the presence of the average momentum scale given by the RG-scale. This approximation and its convergence has been well studied beyond the difference equations used here, for a Yang-Mills study see \cite{Cyrol:2016tym}, for a recent assessment see \cite{Ihssen:2024miv}. Moreover, its reliability has been checked in a plethora of theories ranging from condensed matter systems to quantum gravity, see \cite{Dupuis:2020fhh}. 

In the DSE works \cite{Gao:2020fbl, Gunkel:2021oya} main approximations in the difference DSE concern the quark-gluon vertex discussed separately below. Moreover, as discussed above, the inputs used in  
\cite{Gao:2020fbl} and \cite{Gunkel:2021oya} differ and so do the fluctuations that are taken into account by the difference DSEs. For example, the difference DSE in \cite{Gao:2020fbl} accommodates all thermal and density fluctuations and only the vacuum fluctuations of the $s$-quark; while the difference DSE in \cite{Gunkel:2021oya} accommodates all quark vacuum and thermal and density fluctuations but the input already accommodates the thermal fluctuations of Yang-Mills theory. This different structure already implies that the pure glue vertices are treated differently: In 
\cite{Gao:2020fbl} the bare three-gluon vertex is used in the difference DSE, while in \cite{Gunkel:2021oya} the Yang-Mills finite temperature three-gluon vertex is used (implicitly) in full QCD. 

All the above approximations are well-motivated and cross-checked. The decisive cross-check for the present use as input in the matter system is the comparison of the $N_f=2+1$ flavour gluon propagator in the vacuum and at finite temperature with benchmark data from the lattice (vacuum and finite temperature) and functional QCD in more sophisticated approximations (vacuum): this crucial benchmark is passed with flying colours by the propagators from \cite{Fu:2019hdw, Gao:2020fbl, Gunkel:2021oya}. Further explicit checks for \cite{Gunkel:2021oya} can be found in~\cite{Fischer:2014ata}, for further checks for  \cite{Fu:2019hdw} we refer to \cite{Ihssen:2024miv}. 

We conclude that the glue sector is under quantitative control also for $\mu_B/T\lesssim 4$: the density modifications of the glue sector in this regime are only triggered indirectly via the quark parts of the functional relations. In particular the gluon propagator only received subleading corrections via the quark loop. In summary, the respective systematic error budget is very small and negligible for the regime of interest, $\mu_B/T\lesssim 4$.

\subsection{Quark-gluon interface}
\label{app:QuarkGluonInterface}

A crucial part in all functional QCD methods is the quark-gluon interface that steers the feedback from the glue to the matter sector and vice versa. This interface was discussed in detail in \Cref{sec:ModularFun} at the example of the four-quark scattering vertex, see in particular \Cref{fig:CouplingStrengths,fig:4quarkLego} and the related evaluation. The interface is controlled by the gluon propagator,  comprehensively discussed in \Cref{app:GlueSector}, and the quark-gluon vertex.
For example, the accurate determination of the latter is crucial for the physics of the Columbia plot: In the fRG setup it is tightly related to the quantitative determination of the dynamics of the four-quark vertex which governs the details of chiral symmetry breaking, see \Cref{fig:FlowFourQuarkCoupling} in \Cref{sec:ChiralSymbreaking} and the related discussion. In the DSE-approach, the quark loop in the gluon DSE generates the global frame in the Columbia plot discussed in section \Cref{sec:columbia} and therefore requires 
careful attention.

\subsubsection{Systematic error budget quark-gluon vertex}
\label{app:ErrorQuarkGluon}

For the systematic error budget, induced by the quark-gluon vertex, the complementarity of all three approaches is most helpful: All three approaches use different
approximations for the (dressed) quark-gluon vertex. Guided by vacuum results \cite{Williams:2014iea, Williams:2015cvx, Cyrol:2017ewj, Gao:2021wun}, all three 
approaches take into account the leading tensor structures of this vertex but neglect a different amount of subleading ones. In comparison, \cite{Fu:2019hdw, Gao:2020fbl} apply simplifications of the momentum dependencies of the vertex dressings accounted for, whereas \cite{Gunkel:2021oya} relies 
on an ansatz for the dressing functions that is guided by Slavnov-Taylor identities and vacuum results, but contains an additional 
parameter (see below). Moreover, \cite{Gunkel:2021oya} works with the full medium structure of the gluon propagator (i.e. takes into
account the split into electric and magnetic parts), whereas \cite{Fu:2019hdw, Gao:2020fbl} use an O(4)-symmetric approximation with
equal electric and magnetic parts. Note that in the fRG this goes hand in hand with an RG-invariant expansion scheme that takes advantage of the fact that propagator dressings and parts of vertex dressings cancel out and leave us with RG-invariant dressings  that are less sensitive to thermal and density effects.  

We emphasise in this context that the quark-gluon vertex in the fRG plays a different rôle in comparison to the DSE. This is very apparent in the comparison of the gap equation \Cref{fig:QuarkGapDSE} and its flow \Cref{fig:QuarkGapfRG}: the latter hosts a four-quark tadpole which carries part of the dynamics encoded in the non-classical tensor structures of the quark-gluon vertex in the gap equation. 

This different structure naturally also leads to a different treatment of mesonic effects. Most complete in this respect is the fRG work \cite{Fu:2019hdw}. There, all multi-scatterings of mesons are  
taken into account, though in a point-like approximation. In \cite{Gao:2020fbl}, this sector is only included implicitly (corresponding to non-resonant effects in the 
(pseudo-)scalar part of the quark four-point function). In \cite{Gunkel:2021oya}, implicit and explicit meson effects have been included (i.e. 
also resonant contributions), but multi-scattering effects have been neglected. Additionally, resonant diquark
and baryon effects have been included and found to be almost negligible in the high temperature/chemical potential region of the
QCD phase diagram with the estimated  CEP. This DSE-analysis corroborates the results of \cite{Braun:2019aow}, where a Fierz-complete four-quark tensor structure was used in the fRG, see \Cref{fig:FierzcompleteDominance}. 

\subsubsection{Induced error budget from quark-gluon vertex}
\label{app:InducedErrorQuarkGluon}

This leads us to chiefly important combined systematic error estimates: while the computations in \cite{Fu:2019hdw} show the subleading nature of higher order meson-scatterings except for a small vicinity of the CEP (a few MeV), the computations in \cite{Braun:2019aow, Gunkel:2021oya} show the subleading nature of diquark and baryon fluctuations in the relevant regime. We note in passing that such contributions may be expected to take over and be dominant in the low 
temperature finite density region of the phase diagram. 

Importantly, in combination these works offer respective support of the approximations used in \cite{Fu:2019hdw, Gao:2020fbl, Gunkel:2021oya} and reduces significantly the combined systematic error estimate. Indeed, by varying the different correlation functions, including the pure matter ones, within their own systematic error estimate, the ensuing changes for the location of the crossover line are significantly less than 10\%. In summary, for $\mu_B/T \lesssim 4$, this allows for the combined systematic error budget of the quark-gluon interface of 10\%.

\subsection{Matter sector}
\label{app:MatterSector}

In the quark sector, the situation is again complementary between all three approaches. In the DSE works \cite{Gao:2020fbl, Gunkel:2021oya},  the quark gap equation is solved
for the full quark propagator, and the systematic error estimate only concerns the approximation of the quark-gluon vertex. In the fRG work \cite{Fu:2019hdw}, an O(4)-approximation for the quark propagator is used within the RG-invariant expansion scheme discussed in \cite{Ihssen:2024miv}. This entails that two of the three tensor structures\footnote{Strictly speaking there are four structures,
but the fourth one has been shown in the DSE-framework to be completely negligible, see \cite{Fischer:2018sdj} for a review.}
 in the medium are two are assumed to be degenerate. 
 
In the fRG work \cite{Fu:2019hdw}, the potential soft modes, pions and $\sigma$-mode, are taken into account including their higher order scattering events. However, the thermal split for quark and mesonic modes, that is the difference between modes longitudinal and perpendicular to the heat bath, was not taken into account. The quantitative nature of this approximation has been checked in many LEFT-applications that include the full quantum, thermal and density fluctuations of the quark and mesonic modes, see \cite{Dupuis:2020fhh} for a comprehensive survey of the literature. A direct check in functional QCD has been performed recently in \cite{Pawlowski:2025jpg}, and the respective upgrade of the approximation scheme in comparison to \cite{Fu:2019hdw} is done below in \Cref{app:SystematicErrorBudget}. 
 
Finally, we emphasise that the matter sector is also used for fixing the parameters of all functional approaches, see last line of \Cref{fig:Approximations19-20-21}. Moreover, the extra parameter in the ansatz function of the third approach is fixed by the crossover
 temperature taken from lattice QCD. Therefore we only obtain predictive power for the transition temperatures and the order of the
 transitions away from this point, i.e. for variations in the quark masses (as done for the Columbia plot) and in general 
 at finite chemical potential.
 
In combination, these considerations sustain the systematic error budget discussed at the end of \Cref{app:ErrorQuarkGluon}.

 \subsection{Systematic error budget for the location of the chiral crossover line and that of the CEP/ONP}
 \label{app:SystematicErrorBudget}
 
 With the discussion of the systematic error estimates of the glue sector, glue-matter interface and the matter sector we are in the position of discuss the systematic error budget for the location of the chiral crossover line and the location of the critical end point or the onset of new phases.

  For the combined systematic error estimate as well as going beyond $\mu_B/T \approx 4$, we also include the very recent work \cite{Pawlowski:2025jpg} into this discussion. It relies on the same expansion scheme as \cite{Fu:2019hdw}, and mainly improves on this work in two aspects: First, it includes a full resolution of the momentum dependence of the pion,  $\sigma,\sigma_s$-modes and quark propagators as well as that of the Yukawa couplings in the quark momentum channel. These momentum dependences also resolve thermal splits parallel and perpendicular to the heat bath. The full momentum dependence is key to the self-consistent resolution of the moat regime and allows us to push the quantitative reliability bound to $\mu_B/T\lesssim 4.5$, see \labelcref{eq:muBT4.5}. Moreover, all-order scatterings of pions and $\sigma,\sigma_s$-modes are included in a full effective potential instead of using a convergent Taylor expansion as in \cite{Fu:2019hdw}. This generalisation allows for a resolution of the critical regime around a CEP as well as the full access to the first order regime. It is also mandatory for the resolution of potential instabilities. In view of the present discussion, \cite{Pawlowski:2025jpg} corroborates the results of \cite{Fu:2019hdw} as well as allowing us to push the quantitative reliability bound towards larger baryon chemical potential.

 \subsection{Combined error budget in functional QCD}
 \label{app:CombinedErrorBudget}
 
 Here we wrap up our analysis of the systematic error. We start this summary with emphasising, that the current analysis was not done on a technical level and the respective computational tests are scattered across roughly twenty works. We are very well aware of the fact that this complicates the access and understanding of these error estimates. This asks for a compilation of these tests in an independent work which is work in progress.  
 
 In short, the works \cite{Fu:2019hdw, Gao:2020fbl, Gunkel:2021oya, Pawlowski:2025jpg} allows for combined systematic error estimates of the correlation functions involved. One of the most important take home message for the reader is that there is not a single approximation involved that is employed in all works \cite{Fu:2019hdw, Gao:2020fbl, Gunkel:2021oya}. Roughly speaking, \cite{Fu:2019hdw, Pawlowski:2025jpg} (fRG) and \cite{Gao:2020fbl} (DSE) use the same input ($N_f=2$ flavour vacuum QCD) and distinguish themselves mainly by the 
 use of different functional equations. Both neglect full momentum dependencies and medium effects in most 
 tensor structures to a different degree and differ in the treatment of the dressings of the quark-gluon vertex and the higher order scatterings. In turn, in \cite{Gunkel:2021oya} (DSE) the input is Yang-Mills theory at finite temperature. It genuinely includes medium effects for the propagators and the meson exchange part of the quark-gluon vertex, which is even dropped completely in \cite{Gao:2020fbl}. The non-resonant contributions to the quark-gluon vertex is motivated by the STI and relies on an ansatz function. As discussed before, the respective approximations in one of the works are confirmed by the subleading nature of the effects computed explicitly in either one of the other two works, but also in further functional QCD computations in the literature. 
 
 Specifically important is the impact of variations of correlation functions within this error margin on the location of the chiral crossover line as well as for the critical end point or the onset of new physics. To begin with, varying the correlation functions within their error margin triggers changes for the location of the crossover line for $\mu_B/T \lesssim 4$, that are significantly smaller than 10\%. The inclusion of the recent work \cite{Pawlowski:2025jpg} allows us to extend this regime with quantitative reliability to $\mu_B/T \lesssim 4.5$, see the discussion around \labelcref{eq:muBT4.5}. This leads to our combined very conservative systematic error budget for the location of the crossover line of 10\% for $\mu_B/T \lesssim 4.5$. 
 
In summary, the overall complementarity of \cite{Fu:2019hdw, Gao:2020fbl, Gunkel:2021oya, Pawlowski:2025jpg}, and its embedding in the rich functional QCD literature in the past decade, provides for small systematic error bars for $\mu_B/T \lesssim 4.5$. Finally, the highly non-trivial and contingent fact that the resulting end-points from \cite{Fu:2019hdw, Gao:2020fbl, Gunkel:2021oya} are very close together, motivated our discussion
and the resulting error budget at the end of section \Cref{sec:PhaseStructure+CEP}.

\section{Natural expansion parameter}
\label{app:kappaExpansion}

In this Appendix we discuss the convergence radius of the series in terms of an optimised or appropriate expansion parameter. The commonly chosen expansion parameters is $\mu_B/T$, both quantities being external parameters that are linked to an energy scale. While $\mu_B=3 \mu_q$ is directly related to the excitation energy of a baryon in the medium, the respective thermal energy scale is the lowest lying Matsubara frequency $\pi T$ of the quarks. Indeed, they both enter quark correlation functions in the combination 
\begin{align} 
	2 \pi T + i\, \mu_q\,. 
	\label{eq:T+mu} 
\end{align} 
which is sourced in the Dirac term with $\gamma_0(2 \pi T + i\, \mu_q)$. This suggests an expansion in $\mu_q /(\pi T)$ or, in terms of the baryon chemical potential, $\mu_B/(3 \pi T)$.  Further corrections come from the thermal fluctuations that deform this proportionality, typically lowering the relative importance of the $T$-corrections. We rewrite \labelcref{eq:Tchicurvature} in this spirits, leading to 
\begin{align} 
	\frac{T_\chi(\mu_B)}{T_\chi}=  1 - \bar \kappa_2 \frac{\mu_B^2}{( 3 \pi T_\chi)^2} - \bar \kappa_4\left( \frac{\mu_B^2}{(3 \pi T_\chi)^2}\right)^2+\cdots \,, 
	\label{eq:Tchicurvature3PiT}  
\end{align} 
with the expansion coefficients $\bar\kappa_{2n}$ with 
\begin{align} 
	\bar \kappa_n = \left(3\pi\right)^{2n}\kappa_n
\label{eq:barkappa-kappa}
\end{align} 
As an explicit example we take the results from \cite{Fu:2019hdw} with $\kappa_2$ in \labelcref{eq:Fun-kappa-2019FullError} and $\kappa_4$ in \Cref{tab:curvature}. The respective $\bar\kappa$'s are given by 
\begin{align} 
	\bar \kappa_2 =1.26(13)\,,\qquad \bar \kappa_4 =2.3 (2)\,.
	\label{eq:barkappas2019}
\end{align} 
\Cref{eq:barkappas2019} indicates that the $\bar\kappa$'s are not decaying as is suggested by the smallness of $\kappa_4$ but stay of the same size. We hence may take the $\mu_B/T$ value, where the first two terms have the same size as an indicative scale where physics may change qualitatively. For the $\kappa$'s from \cite{Fu:2019hdw} this entails 
\begin{align} 
	\frac{\mu_B}{T_\chi} = 7.0\,, 
\end{align}	
which agrees with the evaluation in \cite{Braun:2019aow}, see also \Cref{fig:FierzcompleteDominance}.

\bibliographystyle{utphys_mod}
\bibliography{ref-lib}

\end{document}